# Low Field Magnetotransport in Manganites

P.K. Siwach[a], H.K. Singh[b] and **O.N. Srivastava**[a]*
[a]Physics Department, Banaras Hindu University, Varanasi-221 005, INDIA
[b]National Physical Laboratory, Dr K S Krishnan Marg, New Delhi-110012, INDIA

## Abstract

The perovskite manganites of general formula $RE_{1-x}AE_xMnO_3$ (RE= rare earth, AE =Ca, Sr, Ba and Pb) have drawn considerable attention, especially following the discovery of colossal magnetoresistance (CMR). The most fundamental property of these materials is strong correlation between structure, transport and magnetic properties. They exhibit extraordinary large magnetoresistance pronounced as CMR in the vicinity of insulator-metal/paramagnetic-ferromagnetic transition at a relatively large applied magnetic fields. However, for applied aspects, occurrence of significant CMR at low applied magnetic fields would be required. This review consists of two sections: In the first section we have extensively reviewed the salient features e.g. structure, phase diagram, double exchange mechanism, Jahn Teller effect, different types of ordering and phase separation of CMR manganites. The second is devoted to an overview of experimental results on CMR and related magnetotransport characteristics at low magnetic fields for doped manganites such as polycrystalline $La_{0.67}Ca_{0.33}MnO_3$ films, Ag admixed $La_{0.67}Ca_{0.33}MnO_3$ films, polycrystalline ($La_{0.7}Ca_{0.2}Ba_{0.1}MnO_3$) and epitaxial ($La_{0.7}Ca_{0.3}MnO_3$) films on different substrates, nanophasic $La_{0.7}Ca_{0.3}MnO_3$, manganites-polymer composites ($La_{0.7}Ba_{0.2}Sr_{0.1}MnO_3$ –PMMA and $La_{0.67}Ca_{0.33}MnO_3$ –PMMA) and double layered polycrystalline bulk ($La_{1.4}Ca_{1.6-x}Ba_xMn_2O_7$) and films ($La_{1.4}Ca_{1.6}Mn_2O_7$). The results of investigations pertaining to these studies carried out in our lab will be described and discussed. Similar investigations of other workers will also be outlined. Some other potential magnetoresistive materials e.g. pyrochlores, chalcogenides, ruthenates, diluted magnetic semiconductors, magnetic tunnel junctions, nanocontacts etc have also been briefly dealt with. The review concludes with the summary of results for low field magnetotransport behaviour and prospects for application.

**Corresponding author:** Phone & Fax: +91-542-2368468 / 2307307
**E-mail:** hepons@yahoo.com, pksiwach@yahoo.com

**Contents**

# 1. Introduction

Materials science and engineering have been at the frontier of technological advancement since the bronze and iron ages. The present information age relies on the development of "smart" and "smaller" magnetic materials for memories, data storage, processing and probing. These magnetic materials can be divided into numerous categories, depending on their origin and applications. One prime example are the transition metal oxides (TMO) having perovskite ($ABO_3$ type) structure which form an important class of materials from the point of view of fundamental physics as well as technological applications [1]. They have been attracting intense attention because of their exotic properties such as ferroelectricity of titanates (doped $BaTiO_3$) [2], high temperature superconductivity of cuprates ($La_{2-x}Ba_xCuO_4$, $HgBa_2Ca_2Cu_3O_{8-\delta}$), [3], colossal magnetoresistance of manganites (doped $LaMnO_3$) [4] and unconventional p-wave superconductivity of ruthenates ($Sr_2RuO_4$) [5].

During the last decade, TMO based magnetoelectronics is fast emerging as future viable technology. In the early days, magnetic materials e.g. $Fe_2O_3$, $CrO_2$, were used only in motors and generators, as permanent magnets followed eventually by applications like magnetic media (disks and tapes) for data storage, magnetic field sensors, read heads etc. The ever-exciting field of magnetism still continues to draw attention both for scientist and industrial community [6]. R & D for magnetic recording efforts are focused on the goal of achieving higher areal density (the number of bits/unit area on a disk surface), which is possible by increasing the linear as well as track densities. Improvement in the linear density requires advancement in materials, in the recording techniques and in miniaturization of the components. Track densities can be increased by improving the magnetic media or material characteristics. The media or material developments cover a broad range of materials and processes. The recent development of deposited metal film media makes it possible to achieve still higher magnetization value having high linear as well as track densities. Therefore increased attention is being paid to metal thin films. Thin films with higher coercivity are required for higher density media. The most challenging issue towards achieving high recording density media is to produce magnetic thin films which have a large signal-to-noise ratio despite the smaller size of the bit or data being detected. The flexibility of thin film technology makes it possible to tailor the magnetic properties to meet specific design requirements [7,8].

Magnetism is still a very competitive technology not only for recording but also for other novel applications. Recently, an approach to electronics is emerging that is based on the up or down spin of the charge carriers rather than on electrons or holes as in traditional semiconductor electronics. Devices that rely on electron's spin to perform their functions form the foundation of spintronics or magnetoelectronics [9-12]. These spintronics devices are being developed for applications such as ultrasensitive magnetic sensors and magnetoresistive random access memory (MRAM). The key factor for the growth of magnetism based technology is the increase in the areal density. The areal density of the state of the art production was 700 Mbit in$^{-2}$ in 1995. In the quest to lower the cost and improve the performance, the areal density has increased more than 20 million-fold in modern disk drives and currently doubles every year (Fig.1). Nonetheless, the pursuit of higher areal densities still continues, as is evident in recent laboratory experiments of recording densities beyond 100 Gbit in$^{-2}$ [13,14] and the next big challenge now looming ahead is to achieve 1 Tbit in$^{-2}$ recording density [15].

The main limitation on the size of the smallest bit depends on the design of read-write head and the intrinsic signal-to-noise ratio of the material. Herein comes the next generation devices based on Giant and Colossal Magnetoresistive materials. Very sensitive magnetoresistive materials called giant magnetoresistive (GMR) materials and more recently the colossal magnetoresistive materials have been rediscovered in the past few years, due to the intense, new materials research. These materials exhibit a resistance change when subjected to a magnetic field and may eventually evolve into magnetoresistive (MR) heads to achieve the required areal densities. Commercialization of the GMR and CMR effects will require materials which have both high magnetization and low activation fields of the order of few tens of guass or smaller. Thus, research on magnetic materials (magnetic thin films and creating new materials system) and understanding its magnetic properties has the potential to make of significant contribution to information technology [8].

## 1.1. THE PHENOMENON OF MAGNETORESISTANCE

Magnetoresistance (MR) refers to the relative change in the electrical resistivity of a material on the application of an external magnetic field. MR is generally defined by the equation,

$$\% \text{MR} = \frac{\Delta \rho}{\rho_0} \times 100$$

$$= \left[ \frac{\rho_H - \rho_0}{\rho_0} \right] \times 100 \qquad (1)$$

where $\rho_H$ and $\rho_0$ are the resistivites at a given temperature in the applied and zero magnetic fields, respectively. MR can be positive or negative depending on the increase or decrease in resistivity, respectively.

All metals show some MR, but upto only a few percent. Nonmagnetic metals such as Au, exhibit small MR, but the magnitude is somewhat greater (upto 15%) in ferromagnetic metals such as Fe and Co. The semimetal Bi also shows ~ 18% MR in a transverse field of 0.6T, which rises to a 40-fold change at 24T [16]. Cu is more typical in that the same very powerful field (24T) gave rise to change of only ~2% at room temperature. This is the classical positive magnetoresistance that varies as $B^2$ (B = applied magnetic field) in half metallic ferromagnets such as $CrO_2$, $Fe_3O_4$ at low temperature [17]. It is absent in the free electron gas [18] but appears when the fermi surface is non-spherical (as in case of semimetal Bi) [16]. This MR originates from the impact of the Lorentz force on the moving charge carriers similar to the Hall effect. Its value is ~ 10% at 10T. A classification of magnetoresistance phenomenon is based on the distinction familiar in magnetism between intrinsic properties such as anisotropy constants, which depend only on the crystal structure, composition and purity, and extrinsic properties such as coercivity which depend on the structure on a mesoscopic or microscopic length scales [19].

**1.2. Giant Magnetoresistance (GMR)**

The GMR effect was originally discovered in Molecular Beam Epitaxy grown epitaxial (100) oriented Fe/Cr/Fe sandwitches [20] and Fe/Cr multilayers [21] but the effects were quite modest at room temperature. Shortly afterwards it was discovered that similar effects could be found in polycrystalline sputtered Fe/Cr multilayers [22] and subsequently very large room temperature MR was found in Co/Cu and related multilayers [23-25]. The GMR has also been observed in variety of inhomogeneous granular (clusters and alloys) systems predominately comprised of Fe, Co, Ni and their various alloys in Cu, Ag and Au matrices [26-29]. In granular magnetic systems, where small ferromagnetic grains (e.g., Fe, Co, Ni etc.) are embedded in an immiscible insulating matrix, the macroscopic properties depend on the metallic volume fraction 'x', the grain size and intergranular distance. When the relative orientation of grain is antiparallel, it results in a minimum in conductance. When antiparallel grains are forced to be parallel by the application of a magnetic field, conductance increases and results in large magnetoresistance [30-32]. On the other hand, in magnetic multilayers spin dependent scattering (SDS) at the interface is responsible for the GMR effect [33]. Grunberg et al. (1986) [34] have observed MR in thin film multilayers comprising of two layers of Fe, and Cr layer sandwiched between them. The report of

Grunberg went unnoticed by most researcher till Baibich et al. (1988) [21], independently observed a drop in resistivity of almost 50% in artificially engineered multilayers by application of external magnetic field and named the phenomenon as Giant Magnetoresistance (GMR). Parkin et al. (1995) [35] have found that the relative orientation of the magnetic moments of two neighbouring Co (magnetic) layers depends on the thickness of the intervening spacer Cu (non-magnetic) layer.

Fert and Campbell (1976) [36] have given a simple explanation of GMR by mapping the electric current that flows through the magnetic multilayers. They considered a trilayer with two magnetic layers separated by a nonmagnetic spacer layer. As the GMR relies on the fact that electron spin is conserved over a distance of up to several tens of nm, which is greater than the thickness of a typical multilayers, so one can assume that electric current in the trilayer flows in two channels, one corresponding to electrons with spin up projection ↑ and the other to electrons with spin down projection ↓ [37]. Since the ↑ and ↓ spin channels are independent (spin is conserved) they can be regarded as two wires connected in parallel. The second essential ingredient is that electrons with spin projection parallel and antiparallel to the magnetization of the ferromagnetic layer are scattered at different rates when they enter the ferromagnet. Thus the GMR in a trilayer can be explained qualitatively using the simple resistor model as shown in Fig. 2. This simple physical resistor model of the GMR effect is believed to be correct but needs to be converted into a quantitative theory that can explain the difference between the current-in-plane (CIP) and current-out-plane (CPP) geometries, the observed dependence of the GMR on the layer thickness and also the material dependence of the effect. The simple resistor network model discussed above predicts that the resistance of a magnetic multilayer will be higher for antiparallel alignment of the magnetic layers as compared with parallel alignment [37,38].

Several theoretical models have been developed, but most of them are based on a pioneer model of the electrical conduction in ferromagnets (FM) proposed by Mott [39]. Mott hypothesized that the electric current in FM metal is carried independently in two conduction channels that correspond predominately to the spin-up and spin-down s-p electrons. These electrons are in broad energy bands with low effective masses. This assumption is believed to be good at temperatures significantly below the magnetic ordering temperature of the magnetic material so that there is little spin mixing between two conduction channels. Mott established that the conductivity can be significantly different in the two channels which is related to the corresponding spin-up and spin-down density of empty states at the Fermi level. These states will be largely of d character, and as a result of the exchange split d bands, the ratio of spin-up to spin-down density of empty states at the Fermi level can be significantly different in ferromagnetically ordered state of Fe, Co, Ni and their alloys. Consequently this leads to the possibility of substantially different mean free paths $\lambda^{\pm}$ and spin-down (minority) electrons as compared with spin-up (majority) electrons [40-42]. An extensive review on all aspects of giant magnetoresistance effect is given by Tsymbal and Pettifor (2001) [43].

## 2. Emergence of a New Type of Magnetoresistnce: Colossal Magnetoresistance (CMR)

Another magnetoresistive material which has drawn considerable attention in the last decade are the unique intrinsically layered perovskite ($ABO_3$ type) manganites of the form $RE_{1-x}AE_xMnO_3$, where RE is a trivalent rare earth element e.g. La, Pr, Nd etc. and AE is divalent alkaline earth element e.g. Ca, Ba, Sr etc. Chahara et al. (1993) [44], von Helmholtz et al. (1993) [45] and Jin et al. (1994) [46] observed a high magnetoresistance in these doped rare earth manganites (bulk as well as thin films) in a magnetic field of several tesla [~6T] (Fig. 3).

As the physical origin of the magnetoresistance in manganites was completely different from the Giant Magnetoresistance effect (GMR), and hence the term Colossal was coined (Jin et. al.) [46] to describe the effect. Since doped perovskite manganites are the theme of the present review so the basics of colossal magnetoresistance materials is described and discussed in detail.

### 2.1. History of Manganites

About half a decade ago Jonker and Santen (1950) [47-49] described the preparation of polycrystalline samples of $(La,Ca)MnO_3$, $(La,Sr)MnO_3$ and $(La,Ba)MnO_3$ manganites and reported ferromagnetism and anomalies in the conductivity at the Curie temperature with variation in lattice parameter as a function of hole doping. Few years later Volger observed a notable decrease of resistivity for $La_{0.8}Sr_{0.2}MnO_3$ in FM state, in applied magnetic fields [50]. Soon after a significant research effort have started on the studies of low temperature measurement in manganites such as specific heat, magnetization, dc and ac resistivity, magnetoresistance, magnetostriction, I-V curves, dielectric constant, Seebeck effect and Hall effect [50,51].

After those pioneering experiments, Wollan and Koehler (1955) [52] carried out extensive neutron diffraction study to characterize and draw the first magnetic structures of $La_{1-x}Ca_xMnO_3$ in the entire composition range. [Wollan and Koehler (1955) were among the first to use the technique of neutron scattering to study magnetism in materials]. They found that in addition to FM phase many other interesting anti-ferromagnetic phases were also present in manganites (Fig.

4a). Among them the most exotic spin arrangement is the CE-type state, which following their classification is a mixture of the C-phase with the E-phase (Fig. 4b). This CE-state was the first evidence of charge-ordering and mixed phase (phase separation) tendencies in the manganites. Further progress come somewhat later when the group at Manitoba grew a high quality mm long single crystal of another interesting manganites (La,Pb)MnO$_3$, which has a Curie temperature well above room temperature [53-55]. Jirak et al. (1979) [56] and Pollert et al. (1982) [57] studied the structure and magnetic properties of another very popular manganites (Pr,Ca)MnO$_3$ by X-ray and neutron diffraction technique. They observed charge-ordering phases which are totally different from the ferromagnetic phases of other manganites.

A burst of research activity on manganites started during 1990 because of the observation of large magnetoresistance. Work by Kusters et al. (1989) [58] on bulk Nd$_{0.5}$Pb$_{0.5}$MnO$_3$ reveals the large MR effect. Another work by von Helmholtz et al. (1993) [45] on thin films of La$_{2/3}$Ba$_{1/3}$MnO$_3$ also revealed a large MR effect at room temperature. Thereafter similar conclusion was reached by Chahara et al.(1993) [44] using thin films of La$_{3/4}$Ca$_{1/4}$MnO$_3$ and Ju et al. (1994) [59] for films of La$_{1-x}$Sr$_x$MnO$_3$. They all observed MR values larger than those observed in artificially engineered multilayers (GMR) [35]. A defining moment for the field of manganites was the publication by Jin et al. (1993) [46] of results with truly Colossal Magnetoresistance (CMR). Jin et al. reported MR (($R_H-R_0$)x100/$R_H$) close to 1500% at 200K and over 100,000% at 77K for thin films of La$_{0.67}$Ca$_{0.33}$MnO$_3$. This enormous factor corresponds to thousand-fold change in resistivity with and without the field. One year later Xiong et al. (1995) [60] reported MR ratio of over 100,000% using thin films of Nd$_{0.7}$Sr$_{0.3}$MnO$_3$ near 60K and in the presence of magnetic field of 8T. These studies led to the obvious conclusion that manganites were a potential alternative for 'Giant' MR systems.

## 2.2. Salient Features of Manganites

### 2.2.1. Crystal Structure and Its Relevance

The characteristic properties of doped perovskite manganties like the CMR effect and the strong correlation between the structure and electronic-magnetic phases can all be attributed to the ratio of the Mn$^{3+}$ and Mn$^{4+}$ ions [47]. The parent compound crystallizes in AMnO$_3$ type perovskite structure having general formulas RE$_{1-x}$AE$_x$MnO$_3$, where RE stands for trivalent rare earth cation such as La, Pr, Nd, Sm, Eu, Gd, Tb, Y etc. and AE stands for divalent alkaline earth cation such as Ca, Br, Sr etc. In this perovskite like structure (RE, AE) occupies the vertices of the cubic unit cell, Mn occupies the body center and O occupies the six faces of the cube which forms

MnO$_6$ octahedra (Fig. 5) [61-63]. The (RE, AE) site (so called perovskite A-site) can in most cases form homogenous solid solution. Both the end members LaMnO$_3$ (A-type) and CaMnO$_3$ (G-type) are antiferromagnetic insulator having single valent Mn ions i.e. Mn$^{3+}$ and Mn$^{4+}$ respectively [52]. On partial doping of the trivalent RE-ion by divalent alkaline earth cation AE, leads to the formation of a mixed valence state of the Mn i.e. Mn$^{3+}$ and Mn$^{4+}$ to maintain the charge neutrality of the system (e.g. $La^{+3}_{1-x} Ca^{+2}_{x} Mn^{+3}_{1-x} Mn^{+4}_{x} La^{+3}_{1-x}$) [47]. The mixed valency of the Mn ions may also be controlled by varying the oxygen content [64,65]. This doping with some divalent cation causes the structure to get distorted due to the differences in the size of the various atoms and leads to Jahn-Teller effect, which is discussed in section 3.4 [66].

Perovskite-based structures occasionally show lattice distortion as modifications from the cubic structure due to doping. One of the possible origin in the lattice distortion is the deformation of the MnO$_6$ octahedron arising from the Jahn-Teller effect that is inherent to high-spin (S=2) Mn$^{3+}$ with double degeneracy of e$_g$ orbitals. Another lattice deformation comes from the connection pattern of the MnO$_6$ octahedra in the perovskite structure, forming rhombohedral or orthorhombic lattices. In these distorted perovskites, the MnO$_6$ octahedra show alternate buckling [67-69]. Such a lattice distortion of the perovskite AMnO$_3$ (where A=RE$_{1-x}$AE$_x$) is governed by the Goldsmith tolerance factor 't' [70,71] which measures the deviation from perfect cubic symmetry (t=1) and is defined as,

$$\left[ t = \frac{d_{A-O}}{\sqrt{2} d_{Mn-O}} = \frac{\langle r_A \rangle + r_0}{\sqrt{2}(\langle r_{Mn} \rangle + r_0)} \right] \quad (2)$$

where, d$_{A-O}$ is the distance between the A-site, where the lanthanide or alkaline earth ions are located, to the nearest oxygen ion i.e. (<r$_A$>+r$_o$) and d$_{Mn-O}$ is the Mn-O shortest distance which are calculated from the sum of the ionic radii for 12-coordianted A-site cations and 6-coordinated Mn-cations [72]. However, the tolerance factor is dependent on both temperature and pressure. The A-O bond has a larger thermal-expansion coefficient and is normally more compressible than the Mn-O bond of an AMnO$_3$ perovskite, which makes dt/dT > 0 and dt/dP < 0 [71].

Since for an undistorted cube the Mn-O-Mn bond is straight (d$_{A-O}$ =√2 d$_{Mn-O}$) that makes t=1. However sometimes the A-ions are too small to fill the space in the cubic centers and due to that the oxygen tend to move toward that center, reducing d$_{A-O}$ (d$_{Mn-O}$ also changes at the same time). For this reason the tolerance factor becomes smaller than one, t<1, as the A-site radius is reduced, and the Mn-O-Mn angle gets smaller than 180º. The hopping amplitude for carriers to move from Mn to Mn naturally decreases as θ is reduced from 180° [73]. Thus as the tolerance factor decreases the tendencies of charge localization increases due to the reduction in the carrier mobility. This has been observed experimentally and proved theoretically in doped manganites [74-76]. For the ideal cubic structure t=1, but the stable perovskite structure occurs over a range of 0.8 < t < 1.1. For lower values of 't', the cubic structure is distorted to optimize the A-O bond lengths. For values of 't' between 0.75 and 0.9, the $MnO_6$ octahedra tilts cooperatively to give an enlarged orthorhombic unit cell [73]. This distortion (reduction of Mn-O-Mn angle from 180°) affects the conduction band which appears as hydrbidization of the p-level of the oxygen and the $e_g$ levels of the Mn. The orbitals overlap decreases with decrease in tolerance factor and the relation between the bandwidth $\omega$ and $\theta$ has been estimated as $\omega \propto \cos^2\theta$ [77].

Hwang et al. (1995) [78] have carried out a detailed study of the structure-property correlation as a function of temperature and tolerance factor 't', for $RE_{0.7}AE_{0.3}MnO_3$ compond for a variety of RE (trivalent rare earth ion) and AE (divalent) ions. The typical relationship is shown in Fig. 6 and it shows the clear presence of three dominant regions: a paramagnetic insulator at high temperature, a low temperature ferromagnetic metal at large-tolerance factor and a low temperature charge-ordered ferromagnetic insulator (FMI) at small tolerance factor. Zhou et al (1999) [79] have also investigated the influence of the tolerance factor 't' and differences in the ionic radii of the A-site cations on curie temperature, resistivity, coercive field and magnetoresistance. They observed that large difference between the ionic radii of the A-site cations are detrimental for magnetotransprot properties. In another study Sun et al (1997) [80] examined a series of compounds at constant 33% dopant level $La_{2/3-x}R_xCa_{1/3}MnO_3$ (R=Pr, Nd, S, Eu, Gd, Tb, Y, Er, Tm) with x chosen to fix at t=0.911. Under the constraints of constant doping level 'x' and 't', the picture developed by Hwang et al. (1995) [79] would predict no variation of the insulator-metal transition temperature ($T_{IM}$) with rare earth. Sun et al. (1997) did observe a dependence of $T_{IM}$ on the choice of rare earth ion, which demonstates that a single, average tolerance factor 't' is inadequate for describing the behaviour of perovskite manganites where A-site components have widely different radii. They attributed discrepancy due to an inhomogenous distribution of cations on the A-site [80].

**2.2.2. Phase Diagram**

Phase diagram of doped perovskite manganites are exceptionally rich with different resistive/magnetic as well as structural phases [81-83]. The phase diagrams that have been established so far for different compounds e.g.$La_{1-x}Ca_xMnO_3$ (LCMO), $La_{1-x}Ba_xMnO_3$ (LBMO),

La$_{1-x}$Sr$_x$MnO$_3$ (LSMO) etc. are constructed from detailed measurements of macroscopic physical quantities such as resitivity ($\rho$), susceptibility ($\chi$) and magnetization (*M*) on single crystal and bulk ceramic samples [84,85]. Even though the phase diagram of each composition is different due to the variation in sizes of different atoms involved but they have some common features [86]. The Ca doped LaMnO$_3$ i.e. La$_{1-x}$Ca$_x$MnO$_3$ (LCMO) is the prototype of the intermediate bandwidth mixed valent perovsikte manganite because the ionic size of Ca (~1.16Å) is almost identical to the ionic size of La (~1.18 Å) and thus a true solid solution forms in the entire range of Ca concentration [82]. Furthermore, the structure, unlike other perovsikte manganites, remains orthorhombic below ~700K in the entire doping concentration. So, La$_{1-x}$Ca$_x$MnO$_3$ (LCMO) is a good candidate material for basic understanding and hence its phase diagram has been described in detail.

The first ever magnetic phase diagram for La$_{1-x}$Ca$_x$MnO$_3$ as a function of temperature was reported by Schiffer et al. (1995)[84] but the complete phase diagram for La$_{1-x}$Ca$_x$MnO$_3$ was given by Cheong and Hwang (2000) [85] on the basis of magnetization and resistivity data as shown in Fig. 7. Similar phase diagram have been obtained using thin films [88,89]. In Fig. 1.7 we have shown the phase diagram of intermediate bandwidth manganite i.e. La$_{1-x}$Ca$_x$MnO$_3$. At high temperature (>275K), for all doping levels, the system is a paramagnetic insulator (PI). At low temperature, LCMO undergoes the following transitions: The end compositions LaMnO$_3$ (x=0) and CaMnO$_3$ (x=1) are insulator at all temperatures and canted antiferromagnetic insulator at low temperature. For x=0.175, a complicated regime with FI (ferromagnetic insulator), CO (charge order) and CAF (canted antiferromagnetic) phases have been realized. On further Ca substitution from 0.175 to 0.50, the regime of CMR effect have been found. In this region, the material undergoes a insulator- metal transition, T$_{IM}$ which is usually very close to paramagnetic-ferromagnetic transition, T$_c$. At close to x=0.5, where the Mn$^{4+}$ to Mn$^{3+}$ ratio is about 1:1 a charge-ordered antiferromagnetic insulating phase starts to evolve at low temperature. This phase have been observed upto 87% of Ca, beyond which canted antiferromagnetic regime (mixture of ferromagnetic and antiferromagnetic regions) exists. Thus the phase diagram of La$_{1-x}$Ca$_x$MnO$_3$ consists of various phases such as canted antiferromagnetic, charge ordered, ferromagnetic metallic, paramagnetic insulating and others, which makes the physics of manganites interesting and challenging [84,85].

As can be seen from the phase diagram in Fig. 7, there are well-defined special features at the commensurate Ca concentration of x=N/8 (N=1,3,4,5 and 7) [85]. At x=3/8 (0.375) *T$_c$* becomes maximum whereas *T$_{CO}$* peaks at x =5/8 (0.625). The compound at the phase boundary with x=4/8 (0.5) undergoes first a ferromagnetic transition and then a simultaneous antiferromagnetic and charge ordering transition at low temperature. The system at x=1/8 concentration also undergoes two transition: first a ferromagnetic transition and then an antiferromagnetic transition accompanied by a charge ordering transition. There is another well-defined phase boundary at x=7/8 (0.875), and a magnetic transition with a significant ferromagnetic moment is observed for x>7/8. The phase diagram also shows that there exists pronounced electron-hole symmetry in the ground state properties of (La, Ca)MnO$_3$. First of all, the ground state is ferromagnetic metal for the hole concentration of 1/8<x<4/8, but the electron concentration 1/8<x<4/8 shows charge ordering at low temperatures. Furthermore, in the carrier concentration of 0<x<1/8, ferromagnetism is much more pronounced for the hole carrier as compared to electron carrier. In general holes in LCMO tend to induce metallicity alongwith ferromagnetism at low temperatures, while electron carriers are susceptible to charge ordering. These anomalies at the commensurate concentrations clearly indicate the importance of electron-lattice coupling (which induces charge localization) in the mixed-valent manganites [90,91].

Another important manganite, La$_{1-x}$Sr$_x$MnO$_3$ (LSMO) is widely studied as a representative of large-bandwidth Mn-oxides and has a high Curie temperature of 370K at intermediate hole doping [92]. In the LSMO compound a structural transition from orthorhombic (x>20%) to rhombohedral (x<20%) is present [81]. However, the structural phase diagram is even more rich [82,83]. In general the orthorhombic phase is stable at lower temperatures, while the rhombohedral phase requires higher temperature. Thus depending on the doping, one can thus obtain ferromagnetic or antiferromagnetic metallic phases, as well as antiferromagnetic insulating phase. Ferromagnetic insulators are less common, since the occurrence of ferromagnetism is associated with the movement of free carriers in the lattice, but can be obtained for some partial orbital ordering cases [93,94]. A large amount of theoretical as well as experimental work has been devoted to disclose the orbital effects on the magnetic phases of CMR manganites [95]. There are differences in the phase diagram for different manganites e.g. La$_{1-x}$Ca$_x$MnO$_3$ [84,85], La$_{1-x}$Sr$_x$MnO$_3$ [92,96], Pr$_{1-x}$Ca$_x$MnO$_3$ [87], Nd$_{1-x}$Sr$_x$MnO$_3$ [97] etc. however, in general they have some common features.

## 3. The Known Mechanisms: Their Salient Features and Inadequeacies

### 3.1. Crystal Field Splitting and Jahn Teller Effect

The physical properties of the doped perovskite manganite (LaMnO$_3$) involve a complex interplay between the spin, charge and orbital degree of freedoms, which strongly depends on the site of occupancy of the d-orbitals. The basic building blocks of the manganites are the MnO$_6$ octahedra. In the cubic environment of the MnO$_6$ octahedron, hybridization and electrostatic interaction with oxygen 2p electrons will create a crystal field for the outer 3d electrons in Mn$^{3+}$. As the d-orbitals are five-fold degenerate (Fig. 8) so this crystal field lifts the 5-fold degeneracy of d-electrons present in free Mn$^{3+}$ ions by splitting the energy levels and forming lower-lying triply degenerate t$_{2g}$ states and a higher doublet of e$_g$ states [98]. The low-lying t$_{2g}$ triplet consists of the d$_{xy}$, d$_{yz}$ and d$_{zx}$ orbitals. These orbitals have lobes oriented between the O$^{2-}$ ions. The higher energy e$_g$ doublet consists of the $d_{x^2-y^2}$ and $d_{3z^2-r^2}$ orbitals. Their lobes point in the direction of the O$^{2-}$ ions, which raises their energy because of the stronger Coulombic repulsion of the MnO$_6$ octahedra in doped LaMnO$_3$. The energy difference due to crystal field splitting (CFS) between t$_{2g}$ and e$_g$ levels for LaMnO$_3$ is approximately 1.5eV (Fig. 9) [99]. Due to strong intra-atomic Hund's coupling, all electrons of Mn$^{3+}$ and Mn$^{4+}$ are aligned parallel in the ground state, leading to a total spin of S=2 and S=3/2, respectively. All three outer electrons of Mn$^{4+}$ occupy the t$_{2g}$ sites, while the extra electron of Mn$^{3+}$ is situated in one of the e$_g$ levels. The t$_{2g}$ orbitals overlap relatively little with the p-orbitals of nearby oxygen atoms, so that the t$_{2g}$ electrons can be considered as forming a localized core spin (s=3/2). The e$_g$ orbitals on the other hand overlap with the p-orbitals of neighbouring oxygen atoms. Although strongly coupled ferromagnetically to the t$_{2g}$ spin, the e$_g$ electron (s=1/2) is more mobile and can hop between different Mn ions. Thus the partial degeneracy of the 3d orbitals has been removed by CFS. The remaining degeneracy is usually broken by the lattice motion. The oxygen ions surrounding the Mn$^{3+}$ ions can slightly readjust their locations, creating an asymmetry between the different directions that effectively removes the degeneracy. This lifting of degeneracy due to the orbital –lattice interaction is pronounced as Jahn-Teller cooperative effect. This effect tends to occur spontaneously because the energy penalization of the lattice distortion grows as the square of that distortion, while the energy splitting of the otherwise degenerate orbitals is linear. For this reason, it is energetically favourable to spontaneously distort the lattice, thus removing the degeneracy. As far as manganites are concerned there are 21 degree of freedom (modes of vibration) for the movement of oxygen and Mn ion [100]. Out of these only two types of distortion (modes of vibrations) are relevant for the splitting of e$_g$ doublet i.e. JT distortion: Q$_2$ and Q$_3$ [101] which are shown in Fig. 10. The Q$_3$ is a tetragonal distortion which results in elongation or contraction of MnO$_6$ octahedra. However in case of manganites the effective distortion is the basal plane distortion (called as Q$_2$ mode) in which one diagonally opposite O pair is displaced outwards and the other pair displaced inward. As Mn$^{4+}$ does not have an electron in the e$_g$ states, it will not act as JT ion.

Lattice distortion of the octahedral can be static or dynamic. When the carriers have certain mobility, the distribution of $Mn^{3+}$ and $Mn^{4+}$ ions is random and changes with time. Therefore, electron-phonon coupling arises and, in fact, Millis et al. (1995) [102] and Roder et al. (1996) [103] have claimed that it is necessary to take account of the lattice vibrations to explain the change in curvature of the resistivity close to $T_C$. Moreover, due to large Hund's coupling, magnetic polarons can be formed [104]. The localization of the carrier in lattice and/or magnetic polarons can explain the activated behaviour of the resistiivty for $T>T_C$ [89] (section 3.3). When the bandwidth is narrow, the localization induced by lattice deformations is much relevant and leads to charge/orbital ordering and stripe formation [93,94,105].

### 3.2. Double Exchange and Related Effect

Soon after Jonker and Santen (1950) [82,83] discovered the strong correlation between ferromagnetism and metallic conductivity in doped manganites, Zener (1950) [96] in a couple of seminal papers, proposed an qualitative explanation that remains at the core of our understanding (of simultaneous ferromagnetic-paramagnetic and metal-insulator transition) in manganites even today.

Zener interpreted ferromagnetism as arising from an indirect coupling between "incomplete d-shells" of $Mn^{+3}$ and $Mn^{4+}$ via "conducting electron" of oxygen as shown in Fig. 11. In the parent compound $LaMnO_3$, the Mn ion is in +3 state having electronic configuration $3d^4$ with three electrons occupy the $t_{2g}$ levels and are coupled to a core spin S = 3/2 by the strong intraatomic Hunds coupling and the fourth electron (itinerant) occupies one of the energetically degenerate $e_g$ orbitals. Zener noted that on doping with a divalent ion on rare earth site i.e. $RE_{1-x}AE_xMnO_3$, the Mn ions become mixed valent with Mn fraction 'x' in the tetravalent $Mn^{4+}$ ($3d^3$, $t_{2g}^3 e_g^0$, S = 3/2) and '1-x' in the trivalent $Mn^{3+}$ ($3d^4$, $t_g^3 e_g^1$, S = 2) state. He considered a cluster formed from oxygen and two $Mn^{3+}$ ions. The basic idea of Double Exchange is that the configurations, $\varphi_1 = Mn^{3+} - O - Mn^{4+}$ and $\varphi_2 = Mn^{4+} - O - Mn^{3+}$ are degenerate leading to a delocalization of the hole on the $Mn^{4+}$ site or electron on the $Mn^{3+}$ site.

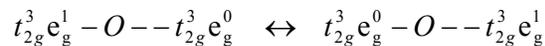

$$t_{2g}^3 e_g^1 - O -- t_{2g}^3 e_g^0 \leftrightarrow t_{2g}^3 e_g^0 - O -- t_{2g}^3 e_g^1$$

The transfer of an electron occurs simultaneously from $Mn^{3+}$ to $O^{2-}$ and from $O^{2-}$ to $Mn^{4+}$, this process is a real charge transfer process and involves overlap integral between Mn and O orbitals. Because of strong Hunds rule coupling $J_H$, the transfer-matrix element has finite value only when the core spins of the Mn ions are aligned ferromagnetically. The Hunds rule coupling of degenerate states lifts the degeneracy and the system resonates between $\varphi_1$ and $\varphi_2$ if the core spins are parallel, leading to simultaneous occurrence of metallic conductivity and ferromagnetism. Zener made a rough estimation of electrical conductivity based on Einstein's relation and the diffusion constant of a hole located at a $Mn^{4+}$ site which comes to be,

$$\sigma \approx \frac{xe^2}{ah} \cdot \left(\frac{T_c}{T}\right) \qquad (3)$$

where $\sigma$ = electrical conductivity, h = the Planck's constant, e = electronic charge, Tc=ferromagnetic transition temperature x = concentration of $Mn^{4+}$ ions and a =lattice parameter or Mn-Mn distance Zener's model was based on the assumption that the manganites are uniform and homogeneous without any form of coexisting clusters of competitive phases.

Anderson and Hasegawa (1955) [106] modified Zener's argument by treating the core spin of each Mn ion classically and the mobile electron quantum mechanically. They showed that the effective hopping matrix element is given by $t_{eff.} = t \cos(\theta/2)$ within a classical treatment of the core spins, where 't' is the transfer integral and 'θ' denotes the angle between $t_g$ spins (the core spins) located at the two Mn sites ($Mn^{3+}$ and $Mn^{4+}$) involved in the electron transfer. The exchange energy is lower when the itinerant electron's ($e_g$) spin is parallel to the total spin of the Mn cores. They also showed that the assumption of classical spins can be avoided if one replaces $\cos(\theta/2)$ by $(S_0 + 1/2)/(2S+1)$, where 'S' is total spin of the two Mn ions and the mobile electron and '$S_0$' is the core spin.

de Gennes (1960) [107] in a mean-field type description revisited the problem, treating the effect of double exchange in the presence of antiferromagnetic background. He formulated the DE problem for a lattice and derived a band model for the motion of holes. de Gennes considered a layered material with 'N' magnetic ions per unit volume, each spin 'S' coupled ferromagnetically to its 'Z'' neighbour on the same layer with exchange energy 'J'' and antiferromagnetically to 'Z' neighbours on adjacent layers with energy 'J'. de Gennes predicted that at low doping level, an antiferromagnetic superexchange interaction competes with the ferromagnetic DE interaction which leads to spin-canted state. However recent studies have shown that a strong alternative to canted states is provided by the tendency to phase separation. de Gennes further considered localization and self trapping of charge carriers which gives rise to local distortion of the spin lattice i.e. the concept of magnetic polaron [108]. Another pioneer theoretical study in manganites was carried out by Goodenough (1955) [109] regarding the charge, orbital and spin arrangements in the non-ferromagnetic regime of the phase diagram of LCMO [52]. The approach of

Goodenough was based on the notions of 'semicovalent bond' and elastic energy considerations. Semicovalent bond or semicovalency arises when the overlap of spin polarized sp orbitals of Mn ions with occupied orbitals of the oxygen allow only covalent bonds involving electrons of one spin direction [109].

Kubo and Ohata (1972) [110] considered a fully quantum mechanical approach employing mean-field theory for metallic double-exchange ferromagnets. They calculated a magnetic phase diagram, resistivity and the magnetoresistance. Their results show a ferro-to paramagnetic transition at $T_c$, accompanied by a change in the temperature dependence of resistivity, and diverging magnetoresistance at $T_c$. They have calculated low temperature resistivity due to second-order electron-magnon process and found that resistivity is proportional to $T^{9/2}$. However, neither the predicted low-temperature resistivity dependence nor the constant resistivity above $T_c$ agree with the experiments. Further, Furukawa (1995, 2000) [111,112] proposed an unconventional one-magnon scattering process in manganites using the dynamical mean field theory and found that the low temperature resistivity follows $T^3$ power law.

### 3.3. Electron-phonon Coupling and Subsequent Theories

The double exchange and subsequent theories can only explain the transport properties of manganites qualitatively. However, it overestimates the Curie temperature of most manganites, can not describe the huge magnitude of the CMR effect, underestimates the resistivity values in the paramagnetic phase by several orders, and can't account for the existence of various antiferromagnetic phases, charge/orbital ordering, phase separation scenario and strong lattice effects/anomalies seen experimentally due to its inherent limitation for several manganites. Millis et al. (1995) [102] incorporated the idea that double exchange alone does not explain the resistivity of $La_{1-x}Sr_xMnO_3$. Their argument hinges mainly on an estimate of the Curie temperature in a pure double-exchange model which comes out to be an order of magnitude larger. Moreover, Millis et al. calculated the resistivity within the double-exchange model including spin fluctuations and found that resistivity decreases below $T_c$ and a positive magnetoresistance above $T_c$, both features in contradiction to the experimental results. Millis et al. argued that the electron-phonon coupling due to the dynamic Jahn-Teller distortion plays an important role, and that a strong interplay between electron-phonon coupling, including charge localization, and Hunds coupling, generating a FM metallic phase, is responsible for the observed properties of manganites [113,122-124]. The strong e-ph coupling in manganites is mainly caused by the Jahn-Teller effect of $Mn^{3+}$. The JT effect causes local distortion of the crystal structure in which some of the Mn-O bonds become shorter and other longer. This breaks the local cubic symmetry and splits the degeneracy of the $e_g$ levels on that site. By occupying the orbital with the lowest energy, the $e_g$ electron can become effectively self-trapped to form together with the surrounding deformed lattice a quasi-particle called lattice polaron or Jahn-Teller polaron. This transport of lattice and spin distortions is also called as magnetic-polaron. Calculations by Millis et al. [104,113-115] predict the localization of charge carriers by temporal and spatial JT distortions around and above $T_c$. This would lead to the observed activated resistivity behaviour in the paramagnetic phase. Below $T_c$, the self-trapping of carriers ends leading to a relaxation of the lattice and an enhancement of the conductivity. In this theory both JT coupling and DE are needed to explain the properties in the various magnetic phases. This leads to the prediction of lower more correct $T_c$ values, and can explain the high resistivity and large CMR effect in manganites. The work of Millis et al. marked a new stage in the study of the CMR manganites by making use of the concept of static and dynamic JT instabilities [116-118], JT polarons and bipolarons [119-121], virtual phonon exchange between carriers and band insulators [103,122]. Several experimental studies have indicated self trapping of carrier as small polaron in perovskite manganite materials above $T_C$ due to Jahn-Teller induced elecron-phonon coupling, from resistivity [123], thermopower [124], Hall effect [125], optical conductivity [126], mobility [127], neutron scattering [128], volume thermal expansion [129], a large nuclear magnetic resonance [130], isotope effect [131], X-ray absorption fine structure spectroscopy [132] as well as Raman scattering [133].

Roder et al. (1996) [103,134] incorporated Jahn-Teller (electron-phonon) coupling into the double exchange model and suggested that the $e_g$ charge carrier becomes self trapped as localized lattice distortions with a spin polarization around the position of charge carrier, having a coherence length of the order of five Mn sites. These quasi-self trapped small polaron can therefore be called magnetoelastic polaron, since they are associated with spin clusters and essentially form metallic islands in a paramagnetic lattice. Ample experimental evidence regarding the existence of magnetic polaron has been reported. It has been demonstrated by neutron diffraction that for both perovskite [135] and layered manganites [136], a volume reduction and relaxation of lattice is observed as the temperature decreases below $T_c$. This is due to change in the Mn-O bonds and indicates the existence of localized $e_g$ electrons above $T_c$ which becomes delocalized in the ferromagnetic phase [137].

The close agreement between the theoretical predictions and the experimental evidence strongly indicates that the lattice effects in CMR materials are caused by the existence of small magnetic Jahn-Teller polarons above $T_c$ and "melting" of these entities below $T_c$. In this theoretical picture, the properties of manganites like the value of $T_c$, the magnitude of the CMR effect, and whether the ground state becomes metallic or stays insulating, depends on the relative strength of DE mechanism and the electron-phonon coupling, which is determined mainly by the nominal hole concentration x. There are various other theories/models e.g. Vibronic Model of Goodenough (1997) [138], Bi-polaronic Model of Alexandrov (1999) [139], Magnetoimpurity theory of Nagaev (2001) [140] and Falicov-Kimball like approach of TV Ramakrishnan (2004) [141] etc. that explains some aspect or the other of doped perovskite manganites, but a satisfactory theory of manganites is still lacking.

### 3.4. Ordering Phenomenon

A fascinating phenomenon of charge ordering (CO) is found to occur in various transition metal oxides (TMO) wherein electron becomes localized due to ordering of cations of different charges (oxidation states) on specific lattice sites. Such ordering generally localizes the electrons in the material, rendering it insulating or semiconducting. This phenomenon of Charge ordering (CO) is well known in $Fe_3O_4$ (Magnetite) which undergoes a disorder-order transition accompanied by a resistivity anomaly, popularly known as Verwey transition, at 120K [142,143]. Charge ordering (CO) has been found to occur in a few other TMO [144,145] as well but the evidence of CO in doped rare earth manganites is overwhelming due to the discovery of Colossal Magnetoresistance and other interesting properties [146]. The first evidence of CO in these rare earth doped manganites was observed by Wollan and Koehler [52] in their classic study through neutron diffraction and latter examined by Jirak et al [147]. The phase diagram of prototype $La_{1-x}Ca_xMnO_3$ material (Fig. 1.7) depicts well-defined features at the commensurate carrier concentration (Ca) of x=N/8 (N=1,3,4,5 and 7) [84,85]. CMR has been observed for Ca concentration 'x' of ~1/8-4/8. For the high doping range with x≥ 4/8, the doped charge carriers localize and order with stripe modulation at low temperatures alongwith antiferromagnetic ordering [109,148-150]. These CO with stripe modulation has a strong tendency towards a microscopic electronic phase separation when charge carriers are introduced into an antiferromagnetic, insulating background.

The general tendency of charge carrier localization and ordering in doped Mott insulators [85,151] is particularly strong in doped manganites, due to the relatively enhanced (electron/hole) carrier-lattice coupling. In addition, there exists orbital degree of freedom of the $e_g$ electrons in $Mn^{3+}$ ions. This orbital ordering can lower the electronic energy through the Jahn-Teller mechanism. Therefore, there exists orbital ordering (OO), in addition to charge ordering in mixed-valent manganites. The first direct evidence of charge ordering in $La_{0.5}Ca_{0.5}MnO_3$ (Tc ≈ 220K) was provided by electron diffraction studies reported by Chen and Cheong (1996) as shown in Fig. 12 [150]. Close to the onset of antiferromagnetism, quasi-commensurate satellite reflections were observed, with a modulation wave vector $2\pi/a$ (1/2-∈, 0,0). They interpreted these reflections as result from the coherent ordering of $Mn^{3+}O_6$ and $Mn^{4+}O_6$ octahedra, as expected for a charge ordered phase. Radaelli et al. (1997) [152] reported a detailed synchrotron X-ray and neutron diffraction investigations of 50% Ca doped $LaMnO_3$. They observed weak satellites reflection in X-ray diffraction pattern which was consistent with that of Chen and Cheong [150].

In $La_{1/3}Ca_{2/3}MnO_3$, there are twice as many $Mn^{4+}$ ($3d^3$) ions as the $Mn^{3+}$($3d^4$) ions, and the ordering of diagonal rows of $Mn^{4+}$ and $Mn^{3+}$ ions plus the orientational ordering of the $d_z^2$ orbital in $Mn^{3+}$ gives rise to the striped pattern as shown in Fig. 13. In the Fig. 13a, diagonal charge strips are evident, and their periodicity is ~16.5Å. These ~16.5 Å charge stripes from the pattern in the real space image (Fig. 14) [153] obtained from electron microscopy for x=2/3. Shown in Fig. 13b is the similar charge/orbital ordering scheme for x=0.5, where there are just as many $Mn^{4+}$ ions as $Mn^{3+}$ ions. In this case, the diagonal charge stripes adopt a wave vector δ=0.5 with a spacing of ~11 Å.

Mori et al. (1998) [153] have reported a different pattern of charge localization in the charge-ordered phase of $La_{1-x}Ca_xMnO_3$ (x≥0.5) employing transmission electron microscopy at 95K [Fig. 14]. They observed extremely stable pairs of $Mn^{3+}O_6$ stripes, with associated large lattice contraction (due to the Jahn-Teller effect), separated periodically by stripes of non-distorted $Mn^{4+}O_6$ octahedra. These periodicities, which adopt integer values between 2 and 5 times the lattice parameter of the orthorhombic unit cell, corresponds to the commensurate carrier concentrations of x=1/2, 2/3, 3/4 and 4/5; for other values of x, the pattern of charge ordering is a mixture of the two adjacent commensurate configuration. These paired Jahn-Teller stripes appear therefore to be the fundamental building blocks of the charge-ordered state in the mangnaties. The charges ordering in manganites have been accompanied by an increase in sound velocity, change in lattice parameters, anomalies in heat capacity, magnetization resistivity and the activation energy for conduction. These orbital/charge order can readily be melted to ferromagnetic metallic state by application of various impulses such as magnetic field [154], pressure [155], exposure to X-ray photons [156], high voltage (250-700V) [157], electric field [158] and visible-IR light laser pulse [159].

Charge ordering (CO) similar to that found in $La_{0.5}Ca_{0.5}MnO_3$ has been reported for $Nd_{0.5}Sr_{0.5}MnO_3$ [160] and $Pr_{0.5}Ca_{0.5}MnO_3$ [145]. Both these compounds exhibit the CE-antiferromagentic structure. However, not all $A_{0.5}A'_{0.5}MnO_3$ compounds exhibit charge ordering behaviour. For example, $Pr_{0.5}Sr_{0.5}MnO_3$ [161] has a type –A antiferromagnetic insulating ground state. It has also been demonstrated that the equal amounts of $Mn^{3+}$ and $Mn^{4+}$ is not a prerequisite to charge ordered behaviour. $Pr_{0.7}Ca_{0.3}MnO_3$ has also been reported to exhibit the CE-type antiferromagnetic structure, suggesting a similar charge ordering as in $La_{0.5}Ca_{0.5}MnO_3$ [109]. Rao et al [163] have carried out detailed investigations on the effect of average radius of the A-site cations $<r_A>$ on CO properties and concluded that *Tco* increases with decrease in $<r_A>$. The phenomenon of charge/orbital ordering in manganites is very interesting and relevant to explain various peculiar properties such as colossal magnetoresistance and phase separation [105,163-165]. Recently, Loudon et al. (2002) [166] have observed CO-FM phase in $La_{0.5}Ca_{0.5}MnO_3$ employing Lorentz electron microscopy at 90K. They observed an inhomogeneous mixture of ferromagnetic (3.4±0.2$\mu_B$ per $\mu_N$) and antiferromagnetic (zero-moment) regions which extend for several micrometers, and can span several crystallographic grains. Loudon et al. have suggested that CO occurs not only in regions with no net magnetization, but can also occur in ferromagnetic regions this is consistent with the similar coexistence in $La_{0.25}Pr_{0.375}Ca_{0.375}MnO_3$ as observed by Mori et al. (2002) [167] Very recently G. Van. Tendeloo et al. (2004) have extensively reviewed the structure and microstructural aspects of colossal magnetoresistive materials with special reference to charge ordering [168].

## 3.5. Phase Separation (PS) Scenario

The physics of solids with strongly correlated electrons such as transition metal oxides (TMO) [1,169,170] and related compounds appears to be dominated by states that are microscopically and intrinsically inhomogeneous in the most interesting range of temperatures and charge carrier (holes and electrons) densities. The most relevant examples are the cuprates at the hole densities in the underdoped region and the managanites in the regime of Colossal Magnetoresistance (CMR). In cuprates the competition occurs between antiferromagnetic insulating and superconducting or metallic phases. On the other hand in manganites the inhomogeneties arise from phase competition between ferromagnetic metallic and charge ordered insulating phases. These microscopic and intrinsic inhomogeneties lead to phase separation (PS) in manganites [171-173]. Indeed, the existence of phase separation was envisioned by Nagaev [174] in an antiferromagnetic semiconductor where the doping of electrons is expected to create ferromagnetic phase embedded in antiferromagnetic matrix. Nagaev [175,176] remarked that if the two phases have opposite charge, the coulomb forces would break the macroscopic clusters into microscopic one, typically of nanometer scale size. Percolative transport has been considered to result from the coexistence of ferromagnetic metallic and insulating phases. The tendency of PS is entirely reversible, and is generally the result of a competition between charge localization and de-localization, the two situations being associated with contrasting electronic and magnetic properties. An interesting feature of PS is that it covers a wide range of length scales anywhere between 1-200nm [177-180].

These intrinsically inhomogeneous states are more pronounced and universally accepted for manganties. These phase-separated states give rise to novel electronic and magnetic properties with Colossal Magnetoresistance (CMR) in doped perovskite manganites. CMR and related properties essentially arises from the double-exchange mechanism of electron hopping between the $Mn^{3+}$ ($t_{2g}^3 e_g^1$, JT ion) and $Mn^{4+}$ ($t_{2g}^3 e_g^0$, non JT ion) ions, which favours ferromagnetic metallic phase below $T_c$ and paramagnetic insulating state above $T_c$. In the insulating state, the Jahn-Teller distortion associated with the $Mn^{3+}$ ions localizes the electrons and favours Charge ordering (CO) of $Mn^{3+}$ and $Mn^{4+}$ ions. This CO competes with double exchange and promotes the antiferromagnetism insulating (AFI) behavior [105]. Even in many of the manganites (exhibiting CMR) which are in FMM state at low temperatures, CO clusters occurs. Thus in doped rare earth manganites the CO (AFM) and FM clusters or domains coexist, the size of which are affected by the carrier concentration or composition, average size of the A-site cations, temperature and other external factors such as magnetic and electric fields [181-184]. Phase with different charge densities and transport properties coexist as carrier-rich FM clusters or domains along with carrier-poor antiferromagnetic (AFM) phase. Such an electronic phase separation gives rise to microscopic or mesoscopic inhomogeneous distribution of electrons, results in rich phase diagrams that involve various type of magnetic structures [107]. Thus there is a clear evidence of electronic phase separation in many manganite systems. Rao et al [177,178] and Dagatto et al [185-187] have extensively reviewed all aspects of phase separation in manganites.

The phase separation (PS) scenario in manganites is somewhat complex because the transition from the metallic to insulating state is not sharp and the domains of the two phases are often sufficiently large to give rise to well defined signatures in neutron scattering or diffraction experiments. In electronic phase separation for manganites, the concentration of the charge carriers giving rise to ferromagnetic and/or metallicity in a part of the crystal causes mutual charging of the two phases. This gives rise to strong coulomb interaction, which may mix the conducting ferromagnetic and insulating antiferromagnetic phases in order to lower the coulomb energy (stabilization of microscopically charged inhomogeneous states) and gives rise to cluster of one phase embedded in another .The size of clusters depend on the competition between DE and Coulomb force. Electronic phase separation of different charge densities is generally expected to give rise to nanometer scale clusters. This is because large phase separated domain would break up into small pieces because of Coulomb interaction. Depending on the strength of interaction the shapes of these pieces could be droplet or stripes. One can visualize PS arising from disorder due to size mismatch of the A-site cations in doped manganites. Such phase separation is seen in the $(La_{1-y}Pr_y)_{1-x}Ca_xMnO_3$ (LPCMO) systems in terms of insulator-metal transition induced by disorder [189]. The size of the cluster depends on the magnitude of disorder. The smaller the disorder, the larger would be the size of the cluster. These microscopically homogenous clusters are usually of the size of 1-2nm in diameter dispersed in an insulating or charge-localized matrix. Such a phase separation scenario bridges the gap between the double exchange model and the lattice models. In the last couple of years, phase separation have been reported in several rare earth manganites and the phenomenon has been investigated by a variety of techniques [186] Keeping in view of the wide implication of this phenomenon in solid state and material science it is important as well as necessary to give an idea about the concept of phase separation in manganites.

The first evidence of phase separation in manganites was given in the pioneer neutron diffraction study of $La_{1-x}Ca_xMnO_3$, by Wollan and Koehler [52] in 1955. They reported the coexistence of ferromagnetic and A-type antiferromangetic reflections in non-stiochiometric $LaMnO_3$ (14, 18 and 20% $Mn^{4+}$) and in $La_{0.89}Ca_{0.11}MnO_3$ . The most important results that have convincingly shown the presence of coexisting clusters of metallic and insulating phases in the CMR regime of manganites was obtained by Uehera et al. (1999) in their study of $La_{5/8}Pr_YCa_{3/8}MnO_3$ using transport, magnetic and electron microscopy techniques [188]. They observed an enormous low temperature resistivity in spite of the fact that $\partial\rho/\partial T > 0$ suggests metallic behavior. By itself this shows that a homogenous picture of manganites will likely fail, since only a percolative state can produce such large but metallic resistivity. The magnetoresistance is large and increases rapidly as $T_c$ is reduced. The value of MR can be very large even at low temperature where the resistivity is flat, far from the actual ferromagnetic transition, suggesting again the mixed phase tendencies in the system. Uehara et al. (1999) [188] interpreted their results in terms of two-phase coexistence of a stable FM state at small 'y' (Pr-content) and a stable CO state in the large 'y' LPCMO compound. They proposed a percolative transition in the intermediate regime of compositon. To further strengthen their results they carried out transmission electron microscopy and found 500 nm large coexisting domains of CO insulator and FM metallic phases for Pr=0.375 at 20K. At 120K, these clusters become nanometer in size (Fig. 15). However, the experimental results for LPCMO are in excellent agreement with the ideas presented by Moreo et al. [189], where first order transitions are transformed into regions of two-phase coexistence by the intrinsic disorder of the manganites which is called as disorder induced phase transition.

Another remarkable evidence of mixed phase tendencies in $La_{0.7}Ca_{0.3}MnO_3$ single crystals and thin films has been given by Fath et al. (1999) employing scanning tunneling spectroscopy (STS) [190]. Below $T_c$, phase separation was observed where inhomogenous clusters of metallic and insulating phases coexist. The cluster size was found to be as large as a fraction of micrometer and that depends strongly on magnetic field. They believe that $T_c$ and the associated MR behaviour is caused by percolative transition. Fath et al. (STS gives real space picture at microscopic level) remarked that the presence of "clouds", which can be metallic or insulating, having size of tens to hundreds of nanometers are not at all compatible with the picture of homogenously distributed small polarons, which is competing theory to the mixed phase scenario. Other workers have also studied the mixed phase tendencies by STS [191,192], STM [193,194] and low temperature MFM [195]. The result provide an atomic scale basis for description of the manganites as mixture of electronically and structurally distinct phases, in excellent agreement with modern theoretical studies [185] and a wide range of experiments [186].

Several other experimental techniques e.g. EXAFS [196], PDF (X-ray as well neutron) [197], Neutron Scattering [198], Raman Scattering [199], Mossbauer Spectroscopy [200], Muon Spin Relaxation [201], Infrared Reflectivity [202], Photoabsorption Spectroscopy [202], Isotope Effect [204], Specific Heat [205], Thermal Expansion measurements [206], Optical studies [207], Internal Friction [208], and others give strong evidence of phase separation in a variety of manganite systems. Another evidence of phase separation in manganites comes from inhomogeneous conductivity in the vicinity of the phase transition (near the insulator- metal transition) for the noise characteristics of film and single crystal samples. The noise measurements provide direct proof of conduction in perovskite manganites dominated by a mixed phase tendencies leading to percolative process below $T_c$ [209-212]. Merithew et al. (2000) [213] and Raquet et al. (2000) [212] have carried out detailed analysis of voltage spectral density ($1/f$) for $La_{2/3}Ca_{2/3}MnO_3$ and concluded that the effect arises from fluctuations between local states with different conductivities. These fluctuations are likely located along the percolative backbone expected in manganites. The domains have a size of $10^4$ to $10^6$ unit cells. These effects are not caused by chemical inhomogenetiy but seem to be intrinsic to the material. Similar results were reported for $Pr_{2/3}Ca_{1/3}MnO_3$ by Anare et al. (2000) [214] and for $(La_{1-y}Pr_y)_{1-x}Ca_xMnO_3$ in the mixed phase regime by Podzorov et al. (2001) [209,215]. Very recently D. D. Sarma et al (2004) observed the formation of distinct electronic domains by direct spatially resolved spectroscopy [216].

There has been considerable theoretical work, motivated by experimental research on perovskite manganites, in the analysis of models for these materials. Several many body techniques for modeling strongly correlated electron systems were developed and improved during recent efforts to understand high temperature superconductors; thus, it is natural to apply some of these models to manganite systems, of particular relevance here are the computational technique that allow unbiased analysis of correlated models on finite clusters [217]. Intimately related to the concept of phase separation is the idea of percolation of insulating and metallic regions. The effective medium approach suggests that metallic and insulating regions coexist as interpenetrating clusters, also suggesting a percolative mechanism for the insulator metal transition. One such percolative model was proposed by Bastiaansen and Knops (1998) [218] based on a random registor network. A Monte Carlo simulations (MCS) of resistor networks with the 2D Ising Model formed the basis of the calculation, with unit resistors connecting aligned nearest-neighbor and next-nearest –neighbor sites and infinite resistance linking unlinked sites. The MCS is in good agreement with experimental data concerning both the temperature variation of the resisitvity and the influence of the magnetic field. Quite similar results for the resistivity of manganites by random resistor network model have been reported recently by Mayr et al. (2001) [219]. Weibe et al. (2001) [220] proposed a two-phase scenario of competing ferromagnetic metallic and insulating polaronic phase; the balance between these two states/phase can be tuned by the variation of various parameters. The magnetization exhibits a first order transition, which is consistent with the neutron scattering data of Lynn et al. (1996) [221] and the magnetization data of Mira et al. (1999) [222] and Ziese (2001) [223]. A more relativistic model was proposed by Lyuksyutov and Pokrovsky (1999) [224] which is based on Varma's theory [225] of magnetic polaron formation, modified to include Jahn-Teller effects. Magnetic polarons, which coexist with small lattice polarons, are assumed to be large, basically comprising magnetically correlated regions. As the temperature is lowered, the magnetic polaron density increases until the magnetic polaron overlaps, which defines the percolation point. The authors argue that long-range Coulomb effects render implausible suggestions that macroscopic charge separation underlines the CMR effect [172,226]. Similar ideas have been discussed by Gorkov and co-workers [227,228]. Recently Dzero et al. (2000) [229] applied percolative model to study the phase separation at low doping. Modeling the metallic phase as a two-band Fermi liquid, they arrived at a transition from the antiferromagnetic insulating to the metallic ferromagnetic state at x=0.16.

In addition a variety of mean field and variational calculations also led to phase separation. This shows that the evidence of phase separation in manganites is not restricted to computational methods only. Dynamical mean-field theory of Millis et al. (1996) [104] and simple mean-field model of Jaime et al. (1999) [230] and several other MFT's only give qualitative description of phase separation rather than the essential contributions of magnetic /conducting fluctuation. In fact, using approximate analytic techniques, Recently Ahn et al. (2004) [231] have shown that the strong coupling between the electronic and elastic degrees of freedom is essential in explaining self-organized inhomogeneties over both nanometer and micrometer scales (phase coexistence).

In short we can say that the instability towards phase separation and the formation of inhomogeneous states or competing phases (e.g. CO/AF and FM) is an intrinsic property of doped perovskite manganites. The existence of these performed clusters (inhomogeneous state or competing phase of CO/AF and FM) and their easy alignment with modest magnetic fields leads to Colossal Magnetoresistance. Thus phase separation appears to be at the heart of various magnetotransport phenomenon in manganites. So, one has to be careful, in attributing CMR and the associated insulator-metal transition to percolation of FMM domains.

## 4. Colossal Magnetoresistance at Low Magnetic Fields (≤ 10 kG)

The observation of colossal magnetoresistance (CMR) close to $T_c$ in good quality single crystals and epitaxial films of doped manganite perovskites generally requires large applied magnetic fields (>1 Tesla), which severely restricts their applicability [46,232-234]. Reducing the field scales and increasing the temperature range has been the motivation of many research groups all over the globe. Significant differences in the magnetoresistance (MR) properties of polycrystalline and single crystal material versions have been reported since the initial discovery of CMR in the manganites. In particular, initial work on bulk polycrystalline samples has shown a substantial MR at temperatures much below $T_c$, and have relatively flat temperature dependence of MR, whereas MR magnitude is usually very small in single crystals or epitaxial films of the same composition [235,236]. A number of subsequent studies, both on bulk and thin film samples, have confirmed the important role of grain boundaries on the MR behavior of manganites because of enhanced MR at small fields (low field magnetoresistance, LFMR) of the order of few oersted [235-255]. As low field magnetotransport properties of polycrystalline manganite is the theme of present review, it is important to give a brief overview of grain boundary (GB) induced low field magnetoresistance (LFMR) in doped manganites.

## 4.1. Overview of Grain Boundary Magnetoresistance in Manganites

### 4.1.1. Comparison of Magnetotransport of Single Crystals/Epitaxial Films and Polycrystalline Samples

Mahesh et al (1996) [237] were the first who have studied the effect of particle size on the transport and magnetic properties of bulk polycrystalline $La_{0.7}Ca_{0.3}MnO_3$. They prepared samples with different particle sizes (0.025–3.5 μm) by a citrate-gel route, followed by heat treatment at different temperatures in an appropriate atmosphere. The $Mn^{4+}$ concentrations were kept similar in the different samples because it is a crucial factor in controlling the transport and magnetic properties. They observed that the resistivity ($\rho$) of the LCMO samples, with similar $Mn^{4+}$ concentration, has been found to increase substantially with decreasing particle size (Fig. 16). Furthermore, the ratio of $\rho$ at 4.2 K and at the peak, corresponding closely to the $Tc$, decreased with increasing particle size. The $Tc$ has also been observed to broaden with decreasing particle size, with the 0.025 μm sample not exhibiting a well-defined transition temperature. They noted that despite these changes, the MR near the $Tc$ did not show any significant changes with the size of the particle. However, the MR at 4.2 K is found to have both low and high-field components, with the former increasing with decrease in particle size (Fig. 16). Based on this and the observation that single crystal and epitaxial films exhibit very small MR at 4.2 K, they concluded that a substantial part of the MR at low temperatures arises from the grain boundaries. This work clearly brought out the role of GBs in magnetotransport leading to low temperature and LFMR.

Another study by Hwang et al (1996) [235] further elaborated the role of GBs in manganites by direct comparison of the MR and field-dependent magnetization at low temperatures for single crystal and bulk polycrystalline samples of $La_{0.67}Sr_{0.33}MnO_3$ (LSMO). Single crystal of LSMO was grown by floating zone method, whereas polycrystalline LSMO samples are prepared through conventional solid-state reaction in air at 1300 and 1700 °C. Both the single crystal and polycrystalline samples show a sharp ferromagnetic transition at 365K.

However, at low temperatures, $\rho$ of the polycrystalline samples is found to be significantly higher than that for the single crystal as shown in Fig. 17. At 5 K, a $\rho$ of 35 $\mu\Omega$-cm is observed for the single crystal LSMO; the $\rho$ of the 1700 °C sintered polycrystalline sample is about an order of magnitude higher owing to scattering introduced by the grain boundaries. Furthermore, the $\rho$ of the 1300°C sintered sample is almost an order of magnitude larger than the 1700ºC sintered sample because of increased scattering which is due to smaller grain size. Despite the differences in $\rho$, the temperature dependence of the magnetization at 0.5T is quite similar for the three samples. A comparison of the field dependence of $\rho$ in the longitudinal geometry (magnetic field, parallel to the current direction) along with the magnetization of the single crystal and the polycrystalline samples at 5K and 280 K is shown in inset of Fig. 17.

For the single crystal there is negligible MR at low temperatures, and with increasing temperature there is increasing negative MR. Correspondingly, the magnetization shows a rapid rise because of magnetic domain rotation at low applied fields, followed by a slow approach toward saturation at higher fields. The variation in the magnetization at various temperatures for the single crystal closely tracks the MR, strongly suggesting that the suppression of magnetic fluctuations is the origin of the negative MR in the single crystal sample. Unlike the single crystal, both polycrystalline samples (sintered at 1300°C and 1700°C) exhibit a sharp drop in the resistance at low fields followed by a slower background negative MR at higher fields. The sharp drop is greatest at the lowest temperatures and decreases with increasing temperature. In contrast to the resistivity variations, the magnetization data are very similar to that for the single crystal. This suggests that the MR in the polycrystalline samples is dominated by intergrain effects, with the magnetic field associated with the sharp drop in resistance identical to that associated with magnetic domain rotation. In order to analyze the temperature ($T$) dependence of the low-field MR component, Hwang et al have back-extrapolated the high field $\rho(H)/\rho(0)$ to find the zero-field intercept and calculated the magnitude of the MR associated with magnetic domain rotation. Their result clearly showed that MR increases with decreasing temperature. Another important observation, apparent from Fig. 18, is that above 0.5 T, the MR in the polycrystalline samples appears to have the same field dependence for the entire temperature range 5–280 K. It appears that the MR has weak $H^2$ dependence in addition to the visually obvious dominant H linear term. They concluded that LFMR observed in polycrystalline LSMO is due to spin polarized tunneling between misaligned grains. It was further shown by Wang et al [238] that, phenomenologically, one has to distinguish weak and strong links between the grains. Only weakly linked grain boundaries give considerable low field magnetoresistance.

Gupta et al (1996) [239] have systematically explored the properties of epitaxial and polycrystalline films of $La_{0.67}Ca_{0.33}MnO_3$ (LCMO), $La_{0.67}Sr_{0.33}MnO_3$ (LSMO), and $La_{0.75}MnO_3$ and found that, unlike the epitaxial films, the polycrystalline films show substantial MR over a wide temperature range in all three systems (Fig. 19b). They used a wide-angle Kerr microscope to image the domains in the polycrystalline manganite films of LSMO. Because these domains have different coercivities, they are weakly coupled and orient successively in an increasing field. Fig. 19a shows a Kerr $M$–$H$ loop of a 14 μm grain size polycrystalline LSMO film and three corresponding Kerr images of the film recorded at room temperature. Image (*a*) displays the nearly uniform magnetic state of the sample at remanence. Near the coercive point, half the grains switch orientation, as seen in image (*b*). In image (*c*) most of the grains have switched as the loop nears saturation. Individual grains can be observed to switch at different fields, and the loop data are an average over the many grains contained in the images. Some of the grains are also observed to switch by wall motion. For example, in the top right corner of (*b*), a wall can be seen to cross a grain boundary. Wall motion can be impeded by surface defects such as scratches on the film. This work clearly brings out the difference between magnetization in polycrystalline and epitaxial LSMO films at microscopic level, which gives a qualitative idea about the LFMR in polycrystalline samples.

**4.1.2. Magnetotransport in a Crystal having Bicrystal Grain Boundary**

Several other studies [240-249] using polycrystalline films have probed the role of grain boundaries in samples consisting of multiple grains with different orientations over the measurement distance of the probe contacts, which gives only qualitative information because the total MR contribution is a complex convolution of the MR due to 'intrinsic' CMR contribution inside the grain and 'extrinsic' contribution coming from the GBs. Bicrystal substrates consisting of a single-grain boundary, with a well-defined misorientation between the grains, are ideal for the growth of films in order to isolate the contribution of a single-grain boundary. In order to enhance the grain boundary contribution to the total $\rho$, the number of grain boundary intersections is increased by patterning a meander track on the substrate. Reduction of the grain region that is probed will lead to further enhancement of the grain boundary signal.

Mathur et al (1997) [250] have followed such an approach to pattern thin film devices on a bicrystal substrate in order to isolate the contribution of a single-grain boundary. Epitaxial LCMO films, 2000 Å thick, are grown by pulsed laser deposition at a growth temperature of about 800°C on bicrystal substrates consisting of two $SrTiO_3$ (001) crystals misaligned by 24°. In order to study the artificial grain boundary directly, they patterned a variety of highly symmetric Wheatstone-bridge structures into the bicrystal films using optical lithography and ion milling, as shown in Fig. 20a. The symmetry of the bridge structures ensures that all resistance contributions balance to zero, except those arising from the grain boundary. Bridge structures are provided on each chip in which current is able to flow either across the grain boundary using a meander track or along it. The devices have been characterized in a liquid nitrogen cryostat in a magnetic field up to 300 mT by passing a constant current through each bridge and measuring the voltage across the output terminals. This measured voltage represents a direct measure of the resistance introduced by the grain boundary. A large bridge MR (27% at 77 K) is observed during magnetic field sweeps within ± 200 mT over a range of temperatures down to 77 K, with a strong low-field hysteresis in bridge resistance at low temperatures (Fig. 20b). At higher temperatures, the magnitude of the peaks in MR decreases, and the peaks disappear altogether at temperatures around 230 K. These results are qualitatively similar to those observed on polycrystalline films except for the magnitude of the low-temperature MR, which is about a factor of two to three higher for the single-grain boundary junctions. Interestingly, a small positive MR is observed above $T_c$, which is masked in experiments using polycrystalline samples by the contribution from the grains. Steenbeck et al (1997,1998)[251,252], Westerburg et al (1999) [253], Ziese et al (step edge junction) (1999) [254] and several others [255-261] have also reported similar field dependence of LFMR in artificially engineered grain boundaries.

The above discussions makes it evident that in the polycrystalline samples (both bulk and thin films) Colossal Magnetoresistance (CMR) can be observed at compatibly low magnetic fields (~mT). This is important since this interfaces directly with the application aspect of CMR. For application only low magnetic fields (~mT) is feasible. Keeping these aspects in view we envisaged to carry out investigations on the synthesis and growth of polycrystalline films, to monitor the structural/ microstructural characteristics and to bring out correlations between the CMR properties. In the present review we have summarize the low field magneto-transport studies carried by our group giving detailed of work carried out by other workers.

### 4.2. Low Field Magnetoresistnce (LFMR) in Various Doped Manganites

### 4.2.1. Studies on Polycrystalline $La_{0.67}Ca_{0.33}MnO_3$ (LCMO) Films

Different techniques have been used to prepare polycrystalline films of $La_{1-x}Ca_xMnO_3$ on different type of substrates. These include laser ablation [46,232-234,248,250], chemical vapour deposition [246] and spray pyrolysis [249]. Of these techniques the last one, viz., spray pyrolysis is convenient and capable of producing films similar to those produced by more sophisticated techniques like laser ablation [46,232-234,248,250]. This technique has been very successfully used to grow polycrystalline high $T_C$ superconducting films and some SQUIDs have also been fabricated on these films [262]. Since the La-manganites and the copper oxide superconductors are similar, the spray deposition can be a suitable and viable technique for the growth of polycrystalline LCMO films. Another advantage of this technique is that synthesis can be done at relatively lower temperatures because of the fact that metal nitrates used for making the spray solution have lower dissociation temperatures as well as smaller synthesis duration. As for $La_{1-x}Ca_xMnO_3$ system optimum properties such as insulator-metal transition, paramagnetic-ferromagnetic transition and maximum magnetoresistance have been observed for 33% of Ca doping. LFMR has been rather sparsely studied and all aspect of CMR related to this are not yet clearly intelligible. Therefore there is significant interest in such studies. In our laboratory the said type of studies have been carried out in detail. The following gives the brief but relevant account of these investigations.

Polycrystalline $La_{0.67}Ca_{0.33}MnO_3$ (LCMO) films have been synthesized through two step route. In the first step, the deposition of LCMO films from aqueous nitrate solution containing relevant cations in the desired ratios, that is, La/Ca/Mn = 0.67/0.33/1 has been carried out. The 0.2M nitrate solution have been prepared by dissolving lanthanum nitrate ($La(NO_3)_3.6H_2O$), calcium nitrate ($Ca(NO_3)_2.4H_2O$) and manganese nitrate ($Mn(NO_3)_3.4H_2O$) in appropriate cationic ratio in double distilled water. This solution was thoroughly homogenized with the aid of a hot plate and a magnetic stirrer. This step is crucial for stoichiometric homogeneity of the film. The solution was then sprayed on a single crystal yttrium stabilized zirconia (YSZ) substrate in (100) orientation, with the aid of an ultrasonic nebulizer. The substrate temperature was maintained at 300±5°C. After the spray was over samples were then cooled to room temperature. In the second step, the as deposited samples were transferred into an alumina boat which was then put in a silica tube provided with oxygen inlet and outlet ports. The samples were then put in a tube furnace. The furnace was programmed to rise to temperature 900°C in 3 hrs, the temperature was then kept constant at 900°C for 30 minutes and then furnace cooled to room temperature.

These as grown films were subjected to gross structural characterization by X-ray diffraction (XRD, Philips PW 1710) using CuK$_\alpha$ radiation ($\lambda$ =1.54106 Å) at room temperature in the 2$\theta$ range of 20-80° and microstructural characterization using scanning electron microscopy (SEM, Philips XL 20) in secondary electron imaging mode. The magnetotransport measurements have been carried out in four probe geometry under applied dc magnetic fields in the range 0-300 mT from room temperature to 77K (liquid nitrogen).

XRD analysis (Fig. 21) reveals that LCMO films are single phasic without any impurity phase. All films are poly-crystalline and have cubic lattice with a lattice parameter a=3.884 Å. The calculation of particle size from the XRD data using values of full width at half maximum (FWHM) and employing the Scherrer formula yields an average particle size of ~125 nm. The microstructural/surface morphological evaluation employing scanning electron microscopy reveals that the LCMO films are dense and uniform over large area. As the substrate lattice parameter (a=5.214 Å) is larger than the lattice parameter a = 3.884 Å of LCMO film and therefore the film is under strain. Consequently some microcracks due to the film/substrate lattice mismatch are also observed and these cracks are spread all over the sample. This is shown in Fig. 22(b). A determination of the grain size employing SEM shows that the film has uniform grain size of ~125 nm. It is also clear from the micrographs that the distribution of grain size is quite even. A typical SEM micrograph showing the typical representative grain distribution is shown in Fig. 22(c). The film thickness was determined by exploring the film substrate interface when this is made parallel to be electron beam and it is ~1μm.

The electrical transport behaviour has been studied by the standard four-probe technique with and without applied dc magnetic fields. An electromagnet was used for applying magnetic field and it was varied from 0 up to 3 kOe. The variation of resistance with temperature results are shown in Fig. 23(a,b) where normalized resistance, that is, resistance at any temperature divided by resistance at room temperature [R(T)/R(room temp.)] is plotted against temperature in Fig. 23(a) while the absolute values of resistance and resistivity are plotted in Fig. 23(b). It has been observed that in zero applied field, the *R-T* curve shows a peak at ~195K and this peak temperature is called as insulator-metal transition temperature. It is also clear that the peak resistivity is about ~4.5 times the resistivity at room temperature. At this temperature a decrease in resistivity takes place which goes on decreasing down to 77K, the lowest temperature at which measurements were carried out.

This semiconductor to metal transition is not very sharp as indicated by the broad peak in the *R-T* curve. The nature of variation of resistance with temperature in zero magnetic field as well as at applied fields $H_{dc}$ = 1 kOe and 3 kOe is similar except that the observed peak in the resistivity curve shifts slightly to higher temperatures at applied magnetic fields of $H_{dc}$ = 1 kOe and 3 kOe. The broadness of the electrical transition at all the applied fields can be explained in terms of presence of antiferromagnetic insulating regions near grain boundaries [246]. Since the magnetic transition depends on the ferromagnetism of the grains these insulating regions have no effect on the magnetic transition. The presence of insulating regions around grain boundaries makes intergrain conduction poor and a broad electrical transition shifted towards lower temperature is observed.

The magnetoresistance (MR) has been calculated by using the expression MR=[R($H_{dc}$)-R($H_{dc}$=0)]/R($H_{dc}$) =$\Delta R_H$/R($H_{dc}$). We have obtained a maximum MR = – 18% at 77K for $H_{dc}$ = 3.kG The temperature dependence of magnetoresistance is shown in Fig. 24(a) where the magnitude of MR is plotted against temperature. It is clear from the lMRl-T plot that below 120K the increase in MR is very rapid and this is more so at the higher applied dc magnetic field. For example at 133K the MR at 1 kG is –9% while at the lower field of 1 kG it is – 4% which is more than twice the value at higher field value. A similar trend is observed up to 190K, that is, up to this temperature the observed MR at $H_{dc}$ = 3kG is nearly twice the value at $H_{dc}$ = 1 kG. MR effects are observed up to 243K for $H_{dc}$ = 1 kOe and 248K for $H_{dc}$ = 3 kG.

The dependence of MR on magnetic field strength at different temperature has also been studied. This MR-$H_{dc}$ behaviour for two different temperatures 77K and 170K is plotted in Fig. 24(b). It is clear that the increase in MR (at constant applied dc field) is faster at lower temperatures than that at higher temperatures. Further as seen in Fig. 24 (b) the rate of increase in MR with applied magnetic field is larger at low applied fields than at high fields. As suggested by Pignard et. al. (1998) [246] such behaviour is associated with domain rotations caused by the applied field. When one by one all the domains are oriented by the increasing field and when all the electrons are parallel to the magnetic field, the scattering of polarized electrons is reduced. The low field MR is expected to be due to polarization of electrons in magnetically disordered regions near the grain boundaries.

It is therefore, clear that the spray deposited LCMO film as prepared in the investigation is polycrystalline and it has been observed to exhibit average particle size ~125 nm. This particle (grain) size is smaller than usually observed ones. The smaller particle size means that a large number of grain boundaries are present in the films grown in the present investigation. SEM studies also show that grains are loosely connected. It has been observed that samples containing large number of grain boundaries exhibit MR which goes on increasing as the temperature is reduced [235,237,246]. A similar behaviour has been observed in the present case for applied dc magnetic fields $H_{dc}$ = 1 kG and 3 kG. The large values of MR observed in this case at relatively low applied magnetic fields seems to be due to the presence of large number of grain boundaries [238,246]. As has been explained by Wang et. al. (1998) [238] when the grain size coincides with ferromagnetic domain size, the domain walls will coincide with the grain boundaries. Inter-domain interactions in each weak link region are reduced by the domain walls at the weak-link grain boundaries. This results in enhanced alignment of spins in inter-domain region even at a low dc magnetic field. Thus significant MR values are obtained even at low applied dc magnetic fields [246]. Since the observed grain size in the present case is comparable to the ferromagnetic domain size this interpretation is equally valid here also. Further, observed variations in MR dependence on applied dc magnetic fields and temperature in the present investigation are also in line with the above arguments. The rapid increase in MR at temperatures below 135K shows that the weak link behaviour is further enhanced at lower temperatures because of a strong reduction in the inter domain interaction and enhanced conduction across the grains leading to the observed rapid increase in the magnetoresistance.

### 4.2.2. Investigations on Ag Admixed $La_{0.67}Ca_{0.33}MnO_3$ (LCMO) Films

Mixed-valence perovskite manganites are under intense study since the discovery of the colossal magnetoresistance in them. CMR near $Tc$ driven by the double exchange mechanism and the dynamic Jahn Teller distortion as well as the low field MR due to grain-boundary effects [247,252] provide a fertile ground for application as magnetic sensors [263]. In mixed-valence perovskite manganite, where the hopping conductivity is strongly dependent on the oxygen stoichiometry and the tolerance factor, the influence of oxygen dynamics as an extrinsic $1/f$ source of noise cannot be neglected, especially in the high temperature range [212]. Much more relevant are the temperature dependences $1/f$ resistance fluctuations observed in CMR Mnanganite thin films in the ferromagnetic regime. In Ca and Sr doped manganites, a 3 or 4 order of magnitude increase of the noise level occurs as the metal-insulator transition is approached [209-215,264-266]. The noise peak temperature is slightly lower than that of the resistivity peak. By applying a magnetic field, a large reduction of the noise is observed in high field, comparable to the magnetoresistance. Therefore, the $1/f$ fluctuations are mostly of magnetic origin. The effect has been attributed to magnetic domain fluctuations near the transition in a time scale consistent with low frequency noise [211], or spin fluctuations coupled to the resistivity by electron-magnon scattering according to the double exchange theory [211].

There are some generic issues concerning noise in these perovskite manganites [267],

(1) The most likely source of noise in these materials is associated with motion of oxygen. This is because in most of the perovskite oxides, oxygen is a particularly mobile species and has much smaller activation energy of migration (often <1 eV). This is much less than that of the cations. Also the perovskites are defect stabilized and contain oxygen vacancies that facilitate oxygen migration. Interestingly oxides like the rare earth manganites (CMR oxides) are used as oxygen conductors.

(2) The electronic charge transport is strongly related to the presence or absence of oxygen at a lattice site. This is because an oxygen-transition metal–oxygen network (2D or 3D) determines the conduction path. As a result, a small fluctuation associated with local oxygen density can cause appreciable conductance fluctuation.

(3) The noise is a strong function of oxygen stoichiometry but is not a monotonous function. In some cases, the noise shows a low value at certain non-stoichiometric conditions.

(4) These materials are susceptible to strain which plays a very important role in deciding the charge transport. As a result, the presence of inhomogeneous strain can cause a large conductance fluctuation. It is to be noted that the strain in the grown layer (arising due to substrate quality, growth condition, lattice mismatch substrate quality, growth condition, lattice mismatch and the thickness) can be large and inhomogeneous. Therefore, the noise in these films are large and are often irreproducible even if the resistances are reproducible. Good control of the film growth, controls of strain and oxygen stoichiometry are necessary to produce low noise oxide films

(5) Devices made from artificial grain boundaries are common in these oxides. GBs (particularly high angle GBs ) are a prominent source of noise.

(6) The oxide interconnects are susceptible to electromigration of oxygen ions. This is a serious cause of lack of stability. In oxides since noise can be linked to oxygen migration, it provides a good probe of this process.

For various applications of magnetoresistance materials, a large magnetoresistance at low field is required. Another requirement is lower value of conduction noise. In addition to that the long-term stability of the manganite films is a prerequisite for technological applications. Epitaxial thin films of CMR materials exhibit large magnetoresistance at high field (a few tesla) and in a narrow temperature range, while polycrystalline films [44,46,234,239,268,269] exhibit substantial magnetoresistance at low field and at temperatures much lower than the ferromagnetic transition temperature. The presence of grain boundaries in polycrystalline CMR films generally leads to suppression of the peak temperature ($T_{IM}$), and enhancement of noise. It has also been found that oxygen deficiency in the CMR films may lead to higher resistivity, lower value of insulator-metal transition temperature ($T_{IM}$) and shift in paramagnetic-ferromagnetic transition temperature ($Tc$) to lower value [264,270,271]. These oxygen deficiencies have been produced by post annealing the films in Argon-atmosphere at higher temperatures or by varying oxygen pressure during deposition. Prokhorov et al. (2002) [272] reported that in an epitaxial film of $La_{0.8}Ca_{0.2}MnO_3$ kept at room temperature for six months, structural phase transformation occurs which changes the characteristic of the film. In this section we have investigated the effect of Ag addition in $La_{0.67}Ca_{0.33}MnO_3$ polycrystalline films deposited by spray pyrolysis and compare its magnetotransport and conduction noise properties with Ag free $La_{0.67}Ca_{0.33}MnO_3$ films.

Films of $La_{0.67}Ca_{0.33}MnO_3$ (LCMO) and $Ag-La_{0.67}Ca_{0.33}MnO_3$ (Ag-LCMO) were prepared on $SrTiO_3$ single-crystal (100) substrates using an ultrasonic spray pyrolysis technique. A two-step synthesis route [273] was employed to prepare these films which was already described in section 4.2.1. Both LCMO and Ag-LCMO films were subjected to structural characterization by x-ray diffraction (XRD) using Cu Kα radiation at room temperature in 2θ range of 20-80° and micro-structural characterization using scanning electron microscopy (SEM) employing secondary electron. The resistance of the film was measured using a standard four-probe technique. For magnetoresistance measurements, a magnetic field was applied in the plane of the film with the direction parallel to the current direction. AC susceptibility was used to measure paramagnetic-ferromagnetic transition temperature $T_C$ of both the films. The conduction noise of the film was measured using the four-probe technique. A 3 $\mu A$ current was passed through the film using a battery-operated low-noise current source. The voltage signal was dc filtered, amplified by a low-noise amplifier and measured by a dynamic signal analyser for observing the frequency spectrum of the voltage noise. All the measuring instruments were interfaced with a computer for automatic data collection. Voltage noise spectral density of the sample was measured at different temperatures from 77 to 300 K. In order to test the stability of the LCMO polycrystalline film, repeated measurements on the same film has been done over a period of one week. In every measurement, the film was cooled to 77 K and again heated to room temperature. The film was left for two days in the ambient atmosphere at room temperature before the measurement is repeated. XRD, SEM, resistivity versus temperature curve, ferromagnetic transition temperature $T_C$ and conduction noise of the LCMO film was recorded. Similar type of measurements has also been performed on Ag–LCMO film.

The structure of the LCMO and Ag-LCMO film was examined by XRD and the corresponding patterns are shown in Fig. 25. The XRD data analysis reveals that both the films are polycrystalline and have cubic unit cell with slightly different lattice parameters; $a_{LCMO}$=3.890 and $a_{Ag-LCMO}$ = 3.885 Å. Some impurity peaks are also observed; however, the amount, as indicated by the relative intensity of these peaks, is small. These impurities are possibly some unreacted calcium oxide. The presence of Ag in the Ag–LCMO film is evidenced by the occurrence of small reflections corresponding to metallic silver. The small decrease of the lattice parameter of the Ag added film seems to be due to better oxygenation of the LCMO film. This decrease in lattice constant of the LCMO film due to Ag addition has also been observed in Ag-LCMO epitaxial film prepared by the laser ablation technique [274].

In CMR materials, a decrease in lattice parameter has been found to be related to an increase in oxygen content [240]. Thus, in the present case, the $AgNO_3$/AgO in La–Ca–Mn–O film decomposes to yield metallic Ag and nascent oxygen. The metallic Ag does not interact chemically with the parent perovskite structure and diffuses into the inter-granular regions [274-278]. The nascent oxygen released from the decomposition of $AgNO_3$/AgO provides better and more efficient in-situ oxygenation. This is because the bond energy of Ag–O is much lower (220 mJ/mol.) than the O–O bond (498 mJ/mol.) [276]. In order to see the effect of thermal cycling on the structural characteristics of the LCMO and Ag–LCMO, XRD studies have been carried out after the fourth cycle. As revealed by the XRD data no changes have been observed in the structural characteristics of both the films. Fig. 26 shows SEM microgrpahs of LCMO and Ag-LCMO films. It is evident that the surface morphologies of the two films are quite different. The surface morphology of the Ag-free film shows large microcracks and the film look porous. In the case of Ag admixed LCMO films the large microcracks have disappeared and the surface looks smooth. Some spherical Ag droplets are also seen over the surface. At ~950 °C, which is the synthesis temperature in the present case, Ag melts and during the slow cooling process, silver atoms or a cluster of silver atoms may sit on the grain boundaries and can also migrate to the surface of the film. Silver also acts to reduce or moderate the thermal gradient between the film and the substrate. This leads to a decrease in the extent and number of microcracks in the Ag-LCMO films.

Figure 27 shows resistance–temperature curves for LCMO and Ag-LCMO films at zero magnetic field. The curve shows that both the films undergo a transition from a semiconducting like ($dR/dT < 0$) to a metallic-like ($dR/dT > 0$) behaviour as the temperature is decreased from room temperature. The addition of silver to the LCMO film is found to shift the peak in the $R$–$T$ curve ($T_{IM}$) to a higher temperature. In the LCMO film the peak in resistance occurs at a temperature ($T_{IM}$) of 190 K, whereas in Ag-LCMO the peak occurs at 200 K.

Figure 28 shows the variation of MR with field at 77 K. The variation of MR with the field is similar for both the films except that at a higher field ($Hdc \geq 1$ kOe) the MR in the Ag-LCMO film is slightly smaller than that of the LCMO film. The variation of MR in both the films shows a faster increase in the beginning and afterwards the increase is slow. This type of variation of MR with dc field is typical in polycrystalline CMR films [239,269]. A sharp increase in MR at low field is attributed to spin polarized tunnelling across the grain boundary in the CMR material [235]. Grain boundaries provide both a kind of thin insulating barrier for electron tunnelling and magnetic decoupling of grains. Magnetic domains associated with the grains are randomly oriented and application of a small field aligns them and thus increases the tunnelling across the grain boundaries. A slow increase in MR at larger field is attributed to the field induced reduction of magnetic disorder at the grain boundaries [269,279]. In the double exchange mechanism [61] charge transport between $Mn^{3+}$ and $Mn^{4+}$ depends upon the relative orientation. A high strain at the grain boundaries may cause the spin disorder. Application of a magnetic field aligns the spins of $Mn^{3+}$ and $Mn^{4+}$ at the grain boundaries and thus reduces the resistivity of the sample [279]. A smaller increase in the MR of the Ag-LCMO film in the higher field region indicates that the film has less magnetic disorders at the grain boundaries compared to the LCMO film.

Figure 29 shows the temperature dependence of normalized noise ($Sv/V^2$) for LCMO and Ag-LCMO films. It shows that the normalized conduction noise has reduced by more than one order due to silver addition. However, the nature of variation of ($Sv/V^2$) is similar in both the films. In the paramagnetic state ($T > Tc$), $Sv/V^2$ decreases with the decrease in temperature, whereas below $Tc$, $Sv/V^2$ shows an increase. This type of variation of $Sv/V^2$ is similar to what has been observed in LCMO epitaxial film [264]. The real cause of the observed conduction noise in CMR material is still not very clear. It may be due to an unconventional conduction mechanism in this material. The Zener double exchange mechanism associated with the mixed valance of Mn ions is believed to govern the charge transport in this material [61,280]. In this model the conductivity is proportional to the hopping rate of the electron between the Mn ions in $Mn^{3+}$–O–$Mn^{4+}$. The hopping rate is very sensitive to the Mn–O bond length and Mn–O–Mn bond angle, which are also influenced by the oxygen vacancies. The oxygen vacancy migration and spin fluctuation in the ferromagnetic state can also contribute to the noise in CMR material. In the polycrystalline CMR film, the presence of defects and oxygen vacancies at the grain boundaries will increase the conduction noise further. In the Ag-LCMO film, during the synthesis, silver segregates at the grain boundaries while the released nascent oxygen is captured in situ by the LCMO lattice in the interfacial region resulting in better oxygenation of the LCMO film as indicated by a decrease in the lattice constant of Ag added film. This will reduce oxygen vacancies at the grain boundaries which may lead to the observed reduced conduction noise. Improved surface morphology of the film due to silver addition as observed in SEM will also contribute to lowering the conduction noise. The addition of Ag in the LCMO film has also been found to improve the stability of the film. In the Ag-free LCMO film, $T_{IM}$ in the $R$–$T$ characteristic shifted to lower value and the noise of the film increased after each measurement. However, in the case of the Ag-LCMO film, the $R$–$T$ curve and noise of the film did not change, even after several measurement cycles.

In order to test the stability of the LCMO polycrystalline film, repeated measurements on the same film has been done over a period of one week. In every measurement, the film was cooled to 77 K and again heated to room temperature. The film was left for two days in the ambient atmosphere at room temperature before the measurement is repeated. XRD, SEM, resistivity versus temperature curve, ferromagnetic transition temperature $T_C$ and conduction noise of the LCMO film was recorded. Similar type of measurements has also been performed on Ag–LCMO film. The paramagnetic to ferromagnetic phase transition was studied by ac susceptibility measurement. The Curie temperature ($Tc$) of the LCMO and Ag–LCMO films have been found to be 230K and 250 K, respectively. The higher $Tc$ of the Ag–LCMO film is due to the better oxygenation brought about by $AgNO_3$ admixture. The $Tc$ of both the films was found to remain constant in all the subsequent thermal cycles. The $T_{IM}$ of the LCMO film is found to be 190 K while that of the Ag–LCMO film, it is 200K. The resistivity of the Ag–LCMO film is lower than that of the LCMO film. The lowering of resistivity and improvement in $T_{IM}$ of the Ag– LCMO film is due to better oxygenation induced by $AgNO_3$ supplied nascent oxygen and improved grain connectivity provided by metallic silver. Fig. 30a shows resistivity vs temperature curves of the LCMO film as recorded in the first, second, third and fourth measurements. It is evident that thermal cycling of the film from room temperature to 77 K leads to an increase in the resistance of the film. A shifting in the peak temperature of the resistivity vs temperature curve ($T_{IM}$) towards lower side is also noticed in the subsequent measurements. In a similar type of measurements for Ag–LCMO film, it has been found that resistivity vs temperature curves are reproducible in all the four measurements (Fig. 30b). Right hand panel of Fig. 30c shows measured values of $T_{IM}$ and $Tc$ of the LCMO and Ag–LCMO film for different measurements.

It is interesting to note that in LCMO film although $T_{IM}$ is shifting to lower temperature but $Tc$ remains same in all the measurements. However, in case of Ag–LCMO film, both $T_{IM}$ and

*Tc* did not change even after several thermal cycles from room temperature to 77 K. This result indicates that addition of Ag in LCMO film increases the stability of the film. The decrease in $T_{IM}$ is related to the deficiency of oxygen in CMR films [234,268]. The deficiency of oxygen in CMR films also affects the *Tc* of the films. In the polycrystalline films $T_{IM}$ depends on the properties of both the grain as well as the grain boundaries whereas the *Tc* is the bulk property depending mainly on the grain. In the present study of LCMO film only $T_{IM}$ is found to shift to lower value while *Tc* remain the same with the thermal cycling. This indicates that the thermal cycling is bringing change only at the grain boundaries and not inside the grain. This change seems to be related to the creation of oxygen deficiency or more disorders at the grain boundaries during the thermal cycling. Fig. 31a shows the temperature dependence of conduction noise of LCMO film for the first, second, third and fourth measurements. The conduction noise of the film increases in the subsequent measurement for the entire temperature range. Fig. 31b shows the temperature dependence of conduction noise for Ag–LCMO film. It shows that the conduction noise for Ag–LCMO film did not change in the subsequent measurements. In order to have better comparison of the conduction noise data normalized conduction noise $S_v/V^2$ has been calculated.

Figure 32 shows the temperature dependence of $S_v/V^2$ for LCMO and Ag–LCMO films. Here also it is found that the normalized noise of the LCMO film is increasing in the subsequent measurements whereas the normalized conduction noise for Ag–LCMO film did not change. Increase in conduction noise of LCMO film at all temperature indicates that thermal cycling results in the creation of more defects. In order to confirm that the decrease in $T_{IM}$ and increase in conduction noise of LCMO film after thermal cycling is due to creation of oxygen vacancies, the LCMO film is post-annealed at 600°C for 2 h in oxygen atmosphere. After the post annealing treatment the initial value of $T_{IM}$ and conduction noise is recovered. During the synthesis of the Ag added films, silver segregates at the grain boundaries [240,274-278]. It seems that the silver segregated at the grain boundaries in the Ag–LCMO film not only increases conductivity of the grain boundaries but also reduces the effect of thermal strain at the grain boundaries of the LCMO film and hence Ag-LCMO films are stable for long time periods.

### 4.2.3. Lattice Strain Effect on Magnetotransport Properties: Some Recent Investigations

The last decade has seen the emergence of epitaxial transition metal oxide (TMO) films as one of the most attractive and overwhelming subject for the material science community not only because of fundamental physics but also due to their application potential [281]. The resurgence of such interest was primarily stimulated by the discovery of high temperature superconductivity

(HTSC) in doped cuprates and more recently by the advent of colossal magnetoresistance (CMR) in doped manganites. Since most technological applications such as ferroelectric field effect [282], bolometric [283], optical [284] and spintronic devices [285] require manganite thin films having good magnetic and electrical properties on suitable substrates, thus the ability to prepare such high quality single crystal epitaxial films and understand their properties is of prime importance. The synthesis (growth optimization) has become possible for manganite thin films by benefiting from studies of high temperature superconducting thin films in late 1980s. Another reason for this quick transfer of growth technology is that, these oxides crystallize in the same perovskite structure like HTSCs and in several respects, they are quite similar [286,287]. Different growth techniques, such as sputtering [288,289], molecular beam epitaxy (MBE) [290,291] electron beam/thermal coevaporation [292], chemical vapor deposition (CVD) [293], metal-organic chemical vapour deposition (MOCVD) [246,294] pulse laser deposition (PLD) [232,234] nebulized spray pyrolysis [273,295] and the sol-gel technique [296] have been extensively used for deposition of manganite thin films.

It is known that the physical properties of the manganites are governed by Double Exchange interaction between Mn ions spins via $Mn^{3+}$-O-$Mn^{4+}$ networks [61]. Therefore, even a small perturbation in $Mn^{3+}$-O-$Mn^{4+}$ network affects CMR properties significantly. In addition, the strong electron-phonon coupling was found to play an important role in the mechanism of CMR [113]. In particular, the Jahn-Teller (JT) effect affects both the magnetic and transport properties in manganite system [102,104,113]. There are two types of distortions of the ideal perovskite structure in bulk samples. One is the rotational distortions of the $MnO_6$ octahedra due to the mismatch between the averaged ionic radii of the A-site species $<r_A>$ and the ionic radius of Mn $<r_{Mn}>$. It is a strain effect induced by chemical pressure [297,298]. The other is JT distortion of $Mn^{3+}O_6$ octahedra. However, for mangantie thin films in addition to doping level and average size of dopant, the substrate induced lattice strain (caused by the lattice mismatch between the film and substrates) plays a key role [299,300]. This kind of biaxial strain is different from the strain induced by hydrostatic [301] or chemical pressure [297,298], since an in plane strain generally accompanies an out of plane strain with a different sign, which can cause electronic behaviour not found in bulk materials of the same chemical composition [299]. Substrate-induced lattice strain has been reported to change a wide variety of properties, with examples including the crystal symmetry [302-305] transport [306,307], and magnetic anisotropics [308], the magnitude of ferromagnetic $T_c$ [309] and charge ordering melting field [310,311] the spin and orbital order structure [312] and the tendency towards phase separation [231,299]. In most cases, these changes are interpreted in terms of substrate-induced strain, which relaxes with increase in thickness [313-316,232,234]. The lattice mismatch δ along the interface is defined by δ = ($a_{p\ substrate}$ - $a_{p\ bulk}$)/$a_{p\ substrate}$. When the film is grown on a substrate whose lattice parameter is smaller or larger than that of the bulk material, the epitaxial strain is expected to be compressive (the cell is elongated along the growth direction and compressed in the film's plane) or tensile (the cell is elongated in the film's plane and compressed along the out-plane growth direction), respectively. Compressive strain usually reduces the resistivity and shifts $T_C$ towards higher temperature. These effects have been confirmed in $La_{0.7}Ca_{0.3}MnO_3$ films [317,318] and $La_{0.7}Sr_{0.3}MnO_3$ films [319-323] grown on various substrates.

For the case of CMR films, the study of strain dependence of the film properties is important from the physics standpoint as well as from the point of view of applications, because most devices are based on supported thin film configurations. A biaxial strain is induced by lattice mismatch, and in some cases by a large thermal expansion coefficient mismatch, with the substrate. Jin et al. (1995) found that insulator-metal transition is suppressed in very thin films because of strain [324]. A theoretical analysis has suggested that ferromagnetic $T_C$ is extremely sensitive to biaxial strain that can have an impact on both the Mn–O bond distance and the Mn–O–Mn bond angle [309]. Konishi et al. (1999) [325] have shown that thin films of LSMO under tensile strain are metallic while under compressive strain they are insulating. They attributed the difference due to strain-induced orbital polarization on the magnetic state of the films. Strain effects on magnetoresistance, magnetic anisotropies and magnetic domain structure have also been studied [326,327]. Duo et al. (2003) have studied the magnetotransport properties of the LSMO epitaxial thin film with varied oxygen background pressure and thickness in relation to lattice strain. They observed that $T_C$, $T_{IM}$ and $T_{MR}$ are more sensitive to oxygen pressure than film thickness and concluded that the oxygen partial pressure is associated with the oxygen defects that leads to magnetic and structural inhomgeneities and hence affect the magnetotransport [328]. Nelson et al. (2004) observed that substrate induced lattice strain for $Pr_{0.6}Ca_{0.4}MnO_3$ thin films have dramatic effect on the crystal symmetry, low-temperature charge and orbital ordering, transport and magnetization behavior. They found that low temperature ordering to be more robust under compressive strain as compared to tensile strain and suggested the importance of Mn-O-Mn bond angle in the formation of low temperature charge and orbital ordering [329]. Wu et al. (2003) have investigated $La_{0.67-x}Pr_xCa_{0.33}MnO_3$ (x=0.13, 0.20, 0.27) thin films under both internal chemical pressure from Pr doping and external strain from lattice mismatch with substrates. They found that lattice strain not only produces a widespan of insulator-metal transition temperatures, but also induces the tendencies towards multiphase coexistence in films. Large tensile strain eliminates the metallic behaviour altogether and produces a ferromagnetic-insulating phase separated mixture [330]. Dale et al. (2003) have tuned the strain state of epitaxial $La_{1/2}Sr_{1/2}MnO_3$ thin films on $BaTiO_3$. As BTO undergoes numerous phase transitions as a function of temperature so the BTO surface lattice can be dynamically changed in an attempt to significantly alter the strain state of the epitaxial film. They correlated the temperature dependence of the structure with that of the fractional changes in magnetization and magnetoresistance [331]. S. Valencia et al. (2003) have grown 50nm thick fully strained epitaxial films of $La_{2/3}Ca_{1/3}MnO_3$ on $SrTiO_3$ (001) substrates. After detailed analysis of the structural and magnetotransport properties, of high temperature annealed film, they have found a progressive increase of out-of plane cell parameter, as the annealing temperature rises but no change in the in plane cell parameters are observed. Simultaneously, the magnetic transition temperature $T_C$ and saturation magnetization, $Ms$, substantially increases. They argued that high temperature annealing promotes an enhancement of the itinerant charge carriers likely to be due to an oxygen uptake accompanying the lattice relaxation [332]. Kanki et al. (2003) have deposited $La_{0.8}Ba_{0.2}MnO_3$ (LBMO) thin films on $SrTiO_3$ substrate by PLD having atomically flat surface (width of atomically flat surface was 150 nm with step height of 0.4 nm, corresponding to one unit cell of LBMO). They found room temperature ferromagnetism down to 5 nm thick films. The 5 nm thick film shows $T_C$ at 290K. They also observed several tens of nanoscale ferromagnetic (local magnetization) domains at room temperature by using noncontact magnetic force microscopy (NC-MFM). The observation of nanoscale magnetic domain behaviour makes it possible to create workable spintronic devices using a perovskite system, and that NC-MFM techniques will also be applicable to decide the magnetic state of nanostructural devices [333]. Andres et al. (2003) have studied structural and magnetic properties of $La_{0.7}Ca_{0.3}MnO_3$ epitaxial films with varying thickness from 2.4 to 80 nm on $SrTiO_3$ substrate. By extensive X-ray diffraction study employing synchrotron radiation they demonstrated that for fully strained films (<12nm), the origin of $T_C$ reduction is due to limitation of the spin fluctuation by the film thickness [334]. Only recently, it has been realized that because of phase separation, different lattice parameters associated with various phases can result in large-scale inhomogeneous strain in the interfacial regions of supported thin films [335], but the relation between strain, magnetic properties and electronic phase separation in manganite films remains unclear.

The interplay between substrate and film allows the modification of properties and can even enhance the magnetoresistive effect [45,46]. There are two mechanisms by which substrates modify film properties. The first one is associated with the lattice distortion of epitaxially grown films. The substrate induces both bulk and biaxial strains in the film, which alters the physical properties of films [336-339]. The other one is associated with the dynamics of film growth and the manner of strain release [340], which induce phase separation and inhomogeneities in films [341,342]. These suggest that the interesting phenomena may be observed in highly strained thin films of manganite. The influence that lattice strain appears to exhibit over so many properties of perovskite film suggests that it could be used advantageously in order to enhance the property of interest in a given material. That is substrate induced lattice strain could be used to tune/tailored the behaviour of the films. But before such tuning becomes more than a possibility, the role of substrate-induced strain in these materials needs to be more clearly understood. These strain effects have been evaluated by the dependence of properties on the thickness and lattice matching between the films and substrates. Although consistent behaviors have been reported concerning the thickness dependence of $T_C$ of the manganite films [314,343,344] disagreement exists concerning the origin of the observed phenomena. Some investigators have argued that the difference in oxygen content is the most important factor responsible for the $T_C$ variation in manganites and that strain has less effect [320,344], while others have claimed that a change in structure, which is strongly coupled with the electronic system, must account for the origin of the behaviors observed [314,345]. Thus, the strain effect in manganite films has not become completely intelligible. Keeping these varied views we have carried out detail investigations on the effect of lattice strain on magnetotransport properties for polycrystalline films prepared by spray pyrolysis and epitaxial films by pulsed laser ablation.

### 4.2.3. (i) Example of Effect of Substrate on CMR Properties of Polycrystalline $La_{0.7}Ca_{0.2}Ba_{0.1}MnO_3$ Manganite

Polycrystalline films of CMR materials exhibit significant magnetoresistance (MR) at low fields over a wide range of temperature whereas the epitaxial films show high MR at large magnetic fields (~ several tesla) and only in the vicinity of the Curie temperature ($T_C$) [232-235]. Recently several reports have been published on the effect of film–substrate lattice mismatch strain on the properties of CMR epitaxial films [304-350,230-232]. All major magnetotransport properties, such as, the Curie temperature ($T_C$), insulator-metal transition temperature ($T_{IM}$), resistivity, and MR have been observed to be profoundly affected by the film-substrate mismatch strain [304-350,230-232]. Since polycrystalline films show the enhanced low field magnetoresistance (LFMR), it would be interesting to explore the effect of strain due to lattice mismatch of the polycrystalline film on various substrates. Till date, a detailed study of the magnetotransport properties of polycrystalline CMR films grown on various substrates having different lattice parameters and hence corresponding to different extent of lattice mismatch, is not available. Therefore, a study of polycrystalline thin films having different lattice mismatch between the film and the substrate seems to be pertinent and timely. We have carried out detailed investigations on the effect of substrate on magnetotransport behaviour of $La_{0.7}Ca_{0.2}Ba_{0.1}MnO_3$ films. In the following the work carried out in our lab will be described and discussed.

Polycrystalline films of $La_{0.7}Ca_{0.2}Ba_{0.1}MnO_3$ (abbreviated as LCBMO hereafter) were prepared on LAO (cubic, 100, a=3.821Å), STO (cubic, 100, a=3.905Å), ALO (hexagonal, $10\bar{1}0$, a=4.758Å, c=12.99Å) and YSZ (cubic, 100, a=5.14Å) single crystal substrates by chemical spray pyrolysis. The substrate temperature was maintained at 300±5°C during the deposition. A two-step synthesis route was employed to prepare these films which was already described in section 4.2.1a. The deposition and annealing conditions for all the four films were kept identical. The LCBMO films deposited on LAO, STO, ALO and YSZ will be referred hereafter as films F1, F2, F3 and F4 respectively. We have also synthesized bulk LCBMO sample by standard solid state reaction route to compare the magnetoresistive properties with the films. Stoichiometric weight of $La_2O_3$, CaO, $Ba(NO_3)_2.4H_2O$ and $MnCO_3$ have been mixed thoroughly and calcined in air at 900°C for 12 hrs. This calcined powder was again reground at 1000-1100°C for 24 hrs with two intermediate grinding. Finally the pellet has been sintered at 1300°C for 24 hr to get desired LCBMO bulk pellet. The structure / microstructure characterization has been carried out by X-ray diffraction (XRD, Philips PW1710, Cu $K_\alpha$) and scanning electron microscopy (SEM, Philips XL 20). The transport measurements were carried out by standard four-probe technique and magnetic characterization has been done by ac susceptibility measurements.

The observed X-ray diffraction patterns for LCBMO bulk and films are shown in Fig. 33. The XRD data reveals that all the samples (bulk as well as films) are single phasic and polycrystalline in nature. The structure of the bulk LCBMO sample has been found to be cubic (a=3.892Å), whereas all the films crystallize in an orthorhombic unit cells. The lattice parameters for bulk and various films are listed in Table 1 and the lattice parameters of the films deposited on various substrates do not vary significantly. The lattice strain have been calculated following the conventional method [74] as given by,

Lattice Strain (%) = ( $a_{bulk}$ - $a_{substrate}$ ) x 100/ $a_{substrate}$

Thus for films F1 (LAO/LCBMO), F2 (STO/LCBMO), F3 (ALO/LCBMO) and F4 (YSZ/LCBMO) the strain comes out to be +1.858, -0.333, -18.201 and -24.280 percent respectively. As our films have thickness of 500 nm thus the lattice mismatch induced strain does not persist but gets relaxed. This relaxation of strain may lead to creation of disorders of various degrees in different films [76]. One such possible disorder is the increase in the grain boundary density. In order to study the effect of varying degree of lattice mismatch on the surface morphology and the grain size, we have investigated the surface microstructure through scanning electron microscopy (SEM). The representative micrographs taken from low lattice mismatch film F2 on STO and high lattice mismatch film F4 on YSZ are shown in Fig. 34. The SEM investigations reveal that the higher strain (as for example -24.280 % for film F4) leads to higher degree of disorder and the grains are smaller in size as shown in Fig. 34b. On the other hand when the strain is smaller (as for example -0.333 % in case of film F2 deposited on STO) it leads to a smaller degree of disorder and consequently the grain size is larger as shown in Fig. 34a. It may be mentioned that smaller grain size will imply higher grain boundary density and the reverse will be true for larger grain films.

The magnetic transition, that is, the paramagnetic to ferromagnetic transition in polycrystalline LCBMO films was studied by employing ac susceptibility measurements to determine the Curie temperature ($T_C$). The $T_C$ values as measured for films F1, F2, F3 and F4 are 283K, 285K, 254K and 243K respectively while the $T_C$ of the bulk sample taken as a reference material in the present study was found to be 287K (Table 2). It is evident that $T_C$ values of all the films are smaller than that of the reference bulk sample. Since in the present study all the films were grown together under identical conditions, the large $T_C$ variation observed in case of films F3 and F4 as compared to films F1 and F2 can not be attributed to intrinsic factors such as oxygen stoichiometry etc. This observed variation in $T_C$ can, however, be explained by taking into account the disorders created due to relaxation of strains originating from the lattice mismatch between substrates and the films. The observed trend in the $T_C$ values of the LCBMO films shows that as the LCBMO-substrate lattice mismatch increases, the disorder resulting from strain relaxation also increases and consequently the $T_C$ gets lowered. This is clearly manifested for films F3 and F4 deposited on ALO and YSZ respectively which have higher disorder density as compared to the films F1 and F2 on LAO and STO respectively.

Apparently the occurrence of these disorders may result in depression of $T_C$ as observed in the present study. Thus the larger the lattice mismatch the higher the disorder (as for example for the case of YSZ/LCBMO film F4), the greater is the reduction in $T_C$. The temperature dependence of resistance (*R-T*) and magnetoresistance (MR-T) was measured by the four-probe technique in the temperature range 300-77K and applied dc magnetic fields up to 300 mT. The low field magnetoresistance (LFMR) was calculated using the relation LFMR=($[R_0-R_H]\times100)/R_0$. The *R-T* behaviour at zero field for all the films is shown in Fig. 35 where the normalized resistance is plotted against temperature. It is evident from the plot that the temperature dependence of resistance for films F1 and F2, which have small lattice mismatch with the substrate and therefore lower disorder, is almost similar. As the temperature is decreased, the resistance increases and insulator-metal transitions ($T_{IM}$) occurs. The $T_{IM}$ for films F1 and F2 are 259K and 264K respectively (Table 5.2). Like the Curie transition temperatures, the insulator to metal transition temperatures ($T_{IM}$'s) of films having smaller disorder are close to each other. The *R-T* of films having high density of disorder (films F3 and F4) is markedly different. In these films, the resistance is higher by at least one order of magnitude and $T_{IM}$ is depressed to much lower temperatures. As for example $T_{IM}$ for films F3 and F4 are 160K and 167K respectively. The width of insulator-metal transition is broad showing higher disorder induced by large lattice mismatch. It is well known that the grain boundaries have dominant role in determining the electrical transport in doped manganites [273,333,338]. In case of films having small disorder e.g. films F1 and F2 the grain boundary density are not affected significantly. The *R-T* behaviour in such films is also not affected much and resembles that of an epitaxial film. But in case of films possessing higher disorder as reflected through increased density of grain boundaries (Fig 34b) the $T_{IM}$ gets significantly affected and it gets substantially reduced.

When a small dc magnetic field is applied, the resistance is lowered but the broad *R-T* features remain the same except that the $T_{IM}$'s go up by 2-4K. The variation of LFMR with the dc magnetic field in the range 0-300 mT at 77K is shown in Fig. 36. The LFMR-H plot shows that the field dependence of LFMR is different for films having different degrees of disorder. The LFMR-H behavior of films F1 and F2 which have comparatively smaller lattice mismatch of +1.858 % and -0.333 % respectively and hence lower density of disorder, lower MR (~2 to 3%) has been observed. On the other hand the films F3 and F4 with larger lattice mismatch of -18.201% and -24.28% respectively and hence higher density of disorder have larger LFMR (~7 to 9%). At low fields (H< 0.6 kOe), the increase in LFMR in films F3 and F4 is much steeper than films F1 and F2 and even at higher fields the rate of increase of LFMR in former films is greater. As pointed out earlier the films having larger lattice mismatch, as for example, films F3 and F4 have smaller grain size. A smaller size implies larger grain boundary density in these films. The increased number of grain boundaries may have larger number of disordered $Mn^{3+}$ and $Mn^{4+}$ ions (which in absence of magnetic field lead to higher resistance) than the films having smaller grain boundary density. Consequently, when a magnetic field is applied larger number of such disordered spins are affected leading to comparatively larger decrease in the resistance and hence higher LFMR. The variation of LFMR with temperature (measured at $H_{dc}$=1.5 kOe) is shown in Fig. 37. Here again two different types of temperature dependence are observed. First, the films grown on LAO (F1) and STO (F2) have almost similar temperature dependence of LFMR in which there are two distinct regions. The lower temperature region from ~77K to ~245K, in which the LFMR shows a very weak temperature dependence and the high temperature region around $T_{IM}$ where LFMR attains maximum value. Beyond this LFMR rapidly drops to zero. The peak in MR is attributed to the strong coupling between the conduction electrons and the local magnetic moments through the double exchange and it has intragrain origin.

This also shows that in polycrystalline thin films having low density of grain boundaries, there is an intragrain contribution to MR even at low magnetic fields. The nearly temperature independent LFMR below $T_{IM}$ may be due to the fact that the films on LAO and STO have relatively smaller disorder. The LFMR-T behaviour of films grown on ALO (F3) and YSZ (F4) is markedly different and the LFMR goes on increasing as the temperature is reduced. At 77K the MR values are ~9% and ~7% for films F3 and F4 respectively. A small hump is, however, observed in LFMR-H curve of film F3 on ALO but vanishes in case of film F4 on YSZ. This hump, as discussed earlier is due the double exchange controlled MR of intragrain origin. The decrease in this hump with increase in the grain boundary density implies that the intragrain contribution to MR which is present even at low fields becomes smaller and the intergrain contribution due to spin polarized tunneling across the grains starts dominating. As discussed earlier, in polycrystalline thin films having large lattice mismatch with the substrate the grains boundaries are disordered and more $Mn^{3+}$ and $Mn^{4+}$ ions having disordered spins are available there. For films F3 and F4, spin polarized tunneling occurring across grain boundaries gives rise to large LFMR, which increases, as the temperature is lowered [338].

We have also measured *R-T* and temperature dependence of magnetoresistance (at 150 mT) of bulk polycrystalline $La_{0.7}Ca_{0.2}Ba_{0.1}MnO_3$ (LCBMO). As the temperature is increased the resistance increases with metal like behavior and a insulator-metal transition is observed at $T_{IM}$ ~283K (Fig. 38) above which the resistance shows a sharp decrease. The observed value of $T_{IM}$ is very close to the Curie transition temperature (*Tc*~287K) of the bulk sample and higher than that of the LCBMO films. The nature of the temperature dependence of MR in LCBMO bulk is similar to the films deposited on $Al_2O_3$ and YSZ substrates except that the peak in MR at 283K is more prominent. The value of MR for bulk LCBMO is comparable to the MR of the LCBMO film on $Al_2O_3$. The low field MR in the low temperature region (*T<<Tc*) can arise due to grain boundary or due to structural effect. In the present case the value of MR in bulk LCBMO (cubic) is similar to the film on $Al_2O_3$ (orthorhombic), which indicates that low field MR is not dominated by structural effect but by the grain boundaries.

Our study reveals that polycrystalline thick films behave differently to lattice strain and they affect low field magnetotransport by producing disorders depending on the amount of strain relaxation. A larger film substrate lattice mismatch (as in case of films deposited on ALO and YSZ) results in higher disorder density that reduces $T_C$, gives rise to larger zero field resistance and higher negative MR. The $T_C$'s of the films deposited on ALO and YSZ are 254K and 243K respectively and corresponding MR ($H_{dc}$=1.5 kOe, 77K) values are ~9% and ~7% respectively (Table 2).

### 4.2.3. (ii) Example of Effect of Substrate on CMR Properties of Epitaxial $La_{0.7}Ca_{0.3}MnO_3$ films

It is known that lattice strain due to film-substrate mismatch affect the properties for thin films (< 100 nm) and it influences the magnetotransport properties ($T_{IM}$, $T_C$ and MR) significantly. However, for thicker films (> 100 nm) lattice strain gets relaxed and produces various kinds of imperfections such as stacking faults, dislocations etc. In strain relaxed epitaxial thin films of $La_{0.7}Ca_{0.3}MnO_3$ significant low field MR has been observed and this is supposed to be closely related to the degree of lattice relaxation and hence density of defects that disrupt the long range order [349,351]. In this section we work out the possible correlation between the degree of strain relaxation and the various physical properties such as $T_{IM}$, $T_C$ and low field MR for relatively thicker films (~ 200 nm) of $La_{0.7}Ca_{0.3}MnO_3$ which were deposited by Pulsed Laser Deposition in our lab.

For the preparation of the film by Pulsed Laser Deposition (PLD), first a stoichiometric ceramic target of $La_{0.7}Ca_{0.3}MnO_3$ has been synthesized by standard solid-state reaction as described in section 4.2.3(i). For the depostion of LCMO films, all three cleaned substrates (for present study $LaAlO_3$, $SrTiO_3$ and MgO; all in (100) orientation) were glued side-by-side on the heater with silver paste to ensure good thermal contact and to have identical synthesis condition. A typical temperature for film deposition is 700-800 °C at a oxygen pressure of 200-400 bars. The laser beam ($\lambda$ = 248 nm) is focused on a target with an energy density of ~2 J/cm$^2$ in a spot of 3-4 mm$^2$ with repetition rate (pulse frequency) of 2 Hz. During the entire ablation process the deposition chamber is kept at a dynamic oxygen pressure of 200- 400 bar. The heater, carrying the substrate, is held at a distance of ~4 cm from the target. Films presented in this study are all deposited in on-axis geometry, i.e. perpendicular incidence of plume to substrate. The film thickness is monitored by the number of ablating laser pulses during deposition. The typical growth rate for the manganites at the parameters used is around 10 nm per minute. After the desired deposition time (typically around 20 minutes yielding a film thickness of around 200 nm), the laser is switched off, and the sample is kept in an oxygen atmosphere (usually 1 bar) at the deposition temperature (for ≥ 1 h) and then slowly cooled (typically 5° /min) down to room temperature. After that films have been post annealed in flowing oxygen at 1000°C for 8 hrs to ensure full oxygen content.

The crystal structure and orientation of all films were characterized by X-ray diffraction using CuK$_\alpha$ radiation ($\lambda$ =1.54106 Å) at room temperature in the 2θ range of 20-80°. X-ray rocking curve analysis has been done to check the crystallinity of the films. The magnetization measurements have been performed in vibrating sample magnetometer at an applied magnetic field of 153 Oe in the range of 77-300 K. The electrical resistance has been measured in the range of 77-300 K by four-probe technique. The MR was calculated by using the resistance data with and without applied field ($H_{dc}$ = 1.0 T), which is applied parallel to film plane.

XRD patterns of La$_{0.7}$Ca$_{0.3}$MnO$_3$ (LCMO) films fabricated on SrTiO$_3$ (001), LaAlO$_3$ have been found to match with the (00$l$) reflections expected from the bulk. The pattern confirms that the films are indeed single phase without any impurity and grown epitaxially on the c-axis as shown in Fig. 39. In order to study the effect of the substrate on the structure of the deposited films, the (002) peaks in the XRD pattern are analyzed. The out of plane lattice parameters of the LCMO thin films deposited on STO, LAO and MgO single crystal substrates have been measured to be 3.878 Å, 3.889 Å and 3.899 Å respectively and the same for the bulk LCMO target is 3.92 Å. The in plane lattice parameters are found to be 3.890 Å, 3.911 Å and 3.912 Å respectively on STO, LAO and MgO substrates. Thus there is no appreciable difference between in plane and out of plane lattice parameters. The near equality of in plane and out of plane lattice parameters is due to the fact that the thickness of these films (~200 nm) is much larger than the critical thickness up to which the biaxial strain originating due to the difference between the lattice parameters of the substrate and LCMO can be effective. Since the film thickness is larger the lattice strain will get relaxed. At film thicknesses larger than the critical thickness, the lattice relaxation defined as L = [(a$_{film}$−a$_{substrate}$) x 100]/a$_{substrate}$ , will play the dominant role. In fact it has been shown that at larger thickness such as ~200 nm, the lattice strain relaxes leading to generation of variety of defects [349]. The epitaxial nature of the films has also been confirmed by the rocking curve analysis of the films and rocking curves corresponding to the (002) reflection is shown in Fig. 40. As seen in the figure, in the case of LCMO/STO film the intensity is highest and decreases for LCMO/LAO and LCMO/MgO films in that order. Similarly the FWHM is observed to be the minimum for the LCMO/STO film and increases for LCMO/LAO and LCMO/MgO in that order. The FWHM for these films is 0.39º, 0.52º and 1.24º respectively.

The above observed broadening may be due to one or more of the following factors (i) the mosaic spread, (ii) reduction of the long-range order in the lattice and (iii) increased static disorder due to imperfect growth. The long range order can be disrupted by the increased density of the stacking faults and tilting out of the in plane grains. Static disorders may lead to the effects such as electron localization etc. In the present case the magnitude of lattice relaxation has been found to be 0.384 3.057 and 6.411 percent for films deposited on STO, LAO and MgO single crystal substrates respectively [349]. Obviously as the difference between the lattice parameters of the LCMO film and substrates increases the magnitude of lattice relaxation also increases. At the same time the FWHM of the rocking curves is found to increase with the lattice relaxation. Increasing FWHM as evidenced by the broadened rocking curves with increasing lattice relaxation shows that defects are being generated as a consequence of this relaxation. As mentioned earlier the lattice relaxation indeed gives rise to extrinsic distortions/defects such as dislocations, grain-boundaries, stacking faults, cationic vacancies etc. The lattice relaxation, may not appreciably affect the more fundamental features such as the magnetic exchange interactions, but it may affect Jahn Teller distortions and hence the electron lattice coupling. Thus the effects of lattice relaxation will get reflected in the magneto-transport characteristics [304-352,232-234].

The paramagnetic–ferromagnetic phase transition ($T_C$) was studied by measuring the magnetization ($M$) in the temperature range 300-77 K and the $M$-$T$ data is plotted in Fig. 41a. The PM – FM transition temperature or the Curie temperature ($T_C$) was determined from the maxima in the magnitude of the $dM/dT$. The observed $T_C$ values are ~ 244 K, 218 K and 175 K respectively for LCMO films deposited on STO, LAO and MgO. The $T_C$ of the bulk target has been measured to be ~ 245 K and the corresponding $M$ – T data is plotted in Fig. 41b. The $T_C$ of the LCMO bulk target and the LCMO thin film on STO substrate are nearly equal while that of the LCMO films on LAO and MgO substrates are smaller than that of the bulk LCMO target. In fact, the $M$ – $T$ data shows that the transitions in the LCMO thin films on LAO and MgO substrates are not as pronounced as in case of the LCMO bulk and thin film on STO. The broadness of the transition suggests towards either a percolative regime or a fluctuation regime having its origin in competition between related interactions around $T_C$. There are host of factors that can account for a $T_C$ depression as observed in the LCMO thin films grown on LAO and

MgO substrates. One of such possible factors is the biaxial lattice strain, but as the film thickness is around ~ 200 nm, the lattice strain is expected to be almost completely relaxed. Therefore the contribution from this factor will be negligibly small [304-352,232-234]. The second factor leading to $T_C$ depression is the oxygen deficiency and it may well be one of the reasons leading to $T_C$ depression in the present case, especially in the case of films deposited on MgO substrates. Another reason and perhaps the dominant one in the present study might well be the structural disorders originating due the lattice strain relaxation. In fact as can be seen in Table 3, in case of LCMO film on STO where the magnitude of lattice relaxation is very small (0.384 %), the $T_C$ of the film is nearly same as that of the bulk.

However, as the magnitude of the lattice relaxation increases to 3.057 and 6.411 percent for LCMO films on LAO and MgO, the $T_C$ decreases quite rapidly to 218 K and 186 K, respectively. The $T_C$ depression can be accounted as in the following: As pointed out and discussed earlier, due to the difference between the lattice parameters of the substrate and film, lattice relaxation occurs. This may lead to generation of various types of defects, such as mosaic spread and decrease in the long range order due to increased density of stacking faults, in plane tilting of grains, creation of oxygen vacancies etc. It is expected that the density of such disorders would increase as the magnitude of the lattice relaxation increases. Induced disorders that disrupt the long range order in the lattice, such as the stacking faults may produce modifications in the Jahn Teller distortion of the $MnO_6$ octahedra leading to electron localization effects and hence may increase the polaronic component to the transport in the paramagnetic phase. The larger value of the lattice relaxation in LCMO/MgO film thus results in the higher density of defects such as the stacking faults, dislocations, etc. The increased density of these lattice defects may lead to larger distortion of the local lattice environment such as the $MnO_6$ octahedra.

The temperature dependence of resistivity of all the LCMO thin films was measured between 300K and 77K and the corresponding data are plotted in Fig. 42. At room temperature (~300K) the resistivity of LCMO films on STO, LAO and MgO are measured to be 36 mΩ-cm, 31 mΩ-cm and 38 mΩ-cm. Thus at room temperature all the films, irrespective of the substrate have nearly the same resistivity value. However, as the temperature is reduced resistivities are observed to increase and at temperature well below room temperature insulator-metal transition ($T_{IM}$) is observed. The $T_{IM}$s of LCMO films on STO, LAO and MgO are found to be 243K, 217K and 191K and the corresponding resistivities at $T_{IM}$ are 72 mΩ-cm, 113 mΩ-cm and 275 mΩ-cm, respectively. Thus compared to the room temperature values the resistivity of LCMO films on STO, LAO and MgO undergo a 2-fold, 3.65-fold and 7.25-fold while the $T_{IM}$ for the LCMO film on MgO is higher than the $T_C$. At 77 K the resistivity of all the films have been measured to be less than 2 mΩ-cm. Thus the observed variation of the resistivity of the LCMO films shows that increased lattice relaxation enhances the resistivity at $T_{IM}$ while the room temperature as well as the low temperature resistivity remain nearly independent of the nature of the substrate and hence the lattice relaxation. The resistivity enhancement around $T_{IM}$ and in the paramagnetic phase in case of LCMO film having larger lattice relaxation suggests increased in carrier localization.

As outlined in the above, disorders are invariably present in the present films, the $\rho$-$T$ data above $T_C$ was analyzed in the framework of the Mott's variable range hopping (VRH) of polarons which is given by [353],

$$\rho(T) = \rho_0 \exp\left(\frac{T_0}{T}\right)^{1/4} \qquad (4)$$

Where $\rho_0$ is the residual resistivity and $T_0$ is the characteristic VRH temperature. It has been observed that in all LCMO films the data above $T_C$ fits well to the above Mott's VRH equation. This is amply clear in the ln ($\rho$) - $T^{-0.25}$ plots for the LCMO films as depicted in Fig.43 (a,b,c).

The value of the constant ($T_0$) at room temperature has been evaluated from the linear fit of the resistivity data and for LCMO films an STO, LAO and MgO the values of $T_0$ are found to be 1.24 x $10^7$ K, 2.25 x $10^7$ K and 3.83 x $10^7$ K respectively. Thus with increased lattice relaxation the characteristic VRH temperature ($T_0$) also increases. Since the value of the parameter $T_0$ is a measure of the strength of the Jahn-Teller distortion and is inversely related to the extent of the localized states, therefore, the increasing value of $T_0$ suggests that increased lattice relaxation decreases the localization length which in turn reduces the average hopping distance.

The localization length (1/$\alpha$) has been calculated using the modified formula proposed by Viret et al. (1997) [354]. They have proposed that in case of manganites with 30% divalent cation doping at the rare earth site, the carrier localization above $T_C$ is caused by a random potential of magnetic origin. This potential is due to the Hund's rule coupling -$J_H$ $\vec{S}_i \cdot \vec{S}_j$ between localized Mn $t_{2g}$ ion cores (S=3/2) and spin s of $e_g$ electrons in the conduction band. The modified formula for localization length (1/$\alpha$) in this model is given by,

$$\frac{1}{\alpha} = \left(\frac{171 \, U_m V}{K_B T_0}\right)^{1/3} \qquad (5)$$

Here $U_m = 3\dfrac{J_H}{2}$ is the splitting between the spin up and spin down $e_g$ bands and its value has been found from optical spectra to be 2eV [355]. V is the unit cell volume per Mn ion and $T_0$ is the characteristic VRH temperature and $k_B$ is the Boltzmann constant. Substituting the value of $U_m$ and $k_B$ above expression becomes,

$$\frac{1}{\alpha} = \left(\frac{3.965 \times 10^6 \, V}{T_o}\right)^{1/3} \quad (6)$$

Thus the average nearest-neighbor hopping (R) distance is given by

$$R = \left(\frac{9}{8\pi\alpha \, N(E) K_B T}\right)^{1/4} \quad (7)$$

where $N(E)$ is the density of available states in the random potential regime of Viret et al.[355] and its value estimated by them for LCMO is $9 \times 10^{26}$ m$^{-3}$eV$^{-1}$. Using this value of $N(E)$ and the Boltzmann constant the above expression transform to,

$$R = \left(\frac{4.6112 \times 10^{-24}}{\alpha T}\right)^{1/4} \quad (8)$$

Using the above expressions the value of the localization length $1/\alpha$ has been calculated to be 2.65Å, 2.18Å and 1.83Å for LCMO films on STO, LAO and MgO respectively. Similarly, the average hopping distance R has been found to be 14.21Å, 13.53Å and 12.98Å for LCMO films grown on STO, LAO and MgO, respectively. From these data of localization length ($1/\alpha$) and average hopping distance (R), it becomes evident that as the lattice relaxation increases in the sequence STO (-0.384)→LAO (3.057) →MgO (-6.411) the localization length sequentially decreases, that is, the carriers becomes more and more localized resulting a decreases in the average hopping distance. This also suggests that the lattice relaxation indeed leads to generation of defects such as stacking faults, dislocations and other defects. The decrease in the localization length ($1/\alpha$) and average hopping distance (R) is quite significant at 4.8 % and 8.7 % as compared to the same for the LCMO thin film on STO.

The low field magnetoresistance (LFMR) of the LCMO thin films grown of different substrates was measured in the temperature range 300 – 77 K and at magnetic fields $H_{dc} \leq 10$ kOe and the corresponding data are plotted in Fig. 44a and Fig. 44b. It is observed that at $H_{dc} = 5$ kOe as well as $H_{dc} = 10$ kOe the MR becomes significant around $T_C$ while the peak in the MR occurs at a lower temperature. This feature, as also pointed out by de Andres et al. [351], is in sharp contrast to the high field MR that peaks around $T_C$. For the LCMO thin film grown on STO the low field MR becomes significant at $T_C$ while the peak MR of ~7 % and ~18 % respectively at $H_{dc}$ = 5kOe and 10 kOe occurs at a lower temperature around T ~ 220 K. The same behaviour is observed in case of films on LAO and MgO although the MR values in this case are more than double the corresponding values measured for LCMO film on STO. However, apart from the magnitude, the temperature spread of MR around the respective peak MR values also increases as one moves from film on STO (low lattice relaxation) to film on MgO (high lattice relaxation). The LCMO film on STO by virtue of low lattice relaxation (~ 0.4 %) is expected to have relatively smaller density of extrinsic defects while the films on LAO (Lattice relaxation ~ 3.06 %) and MgO (Lattice relaxation ~ 6.411 %) have much higher density of disorders and defects. It has been discussed by de Andres et al. [351] that in epitaxial thin films the low field MR around $T_C$ or $T_{IM}$ is related to the lattice distortions that could result as a consequence of the lattice relaxation induced defects such as the stacking faults etc. In the present investigation it has been shown from the rocking curve analysis as well as the temperature dependence of resistivity in the paramagnetic state that lattice distortions indeed increase with increasing lattice relaxation. It is known that around $T_C$ polaron clusters form in the FM phase of the thirty percent doped LCMO. This polaron cluster formation is due to a drastic reduction in the bandwidth due to the lattice distortions around the defects. Since in the present investigation the density of defects increases as the lattice relaxation increases, therefore the polaraonic density is also expected to increase as one move from STO to LAO and then to MgO. These polarons are weakly bound in the FM phase and consequently can be delocalized by a relatively weak magnetic field. Thus the low field MR in epitaxial thin films around $T_C$ or $T_{IM}$ has its origin in the delocalization of the small polaron clusters formed around the $T_C$ or $T_{IM}$ in the FM phase.

The contribution of the defects/disorders is more or less confined to the temperature regime around $T_C$ and $T_{IM}$. In the LCMO films, polaron clusters form in the ferromagnetic phase around $T_C$ by a drastic reduction of the bandwidth due to the lattice distortions around the defects. The decrease in the bandwidth caused by lattice distortion is given the Narimanov-Varma [356] formula

$$W = W_0 \cos\frac{\theta}{2} \exp\left(-\frac{k(\delta u)^2}{\hbar \omega_0}\right) \qquad (9)$$

Here $\theta$ is the angle between the two Mn spins and $\delta u$ are lattice distortions that can be dynamic (e. g. phonons), intrinsic (related to La substitution by a divalent cation) or due to lattice defects around impurities. The lattice distortions around the defects favour the localization of polarons ($T_0$ increases). As proposed by de Andres et al. [351] these polarons are rather weakly bond in the ferromagnetic phase and therefore can be delocalized by a relatively smaller dc magnetic field. These small polarons can only be formed near $T_C$ when the carrier bandwidth is reduced either by spin disorder or by lattice fluctuations. In the present case this is quite obvious from the decrease in the $T_{IM}$ values with increasing lattice relaxation and also from the fact that the value of the parameter $T_0$ that eventually determines that localization length increases as the magnitude of the lattice relaxation increases in the LCMO films on STO, LAO and MgO.

In summary, epitaxial and single phasic $La_{0.7}Ca_{0.3}MnO_3$ films have been deposited on $LaAlO_3$, $SrTiO_3$ and MgO substrates As the films investigated in the present case are thicker (~200 nm), the substrate induced strain does not persist but it relaxes and these relaxation gives rise to extrinsic distortions/defects such as dislocations, grain-boundaries, stacking faults, cationic vacancies etc. The lattice relaxation has been found to be 0.384 3.057 and 6.411 percent for films deposited on STO, LAO and MgO. The observed $T_C$ values are ~ 244 K, 218 K and 186 K respectively for LCMO films deposited on STO, LAO and MgO. The $T_{IM}$ s of LCMO films on STO, LAO and MgO are found to be ~ 243K, 217K and 191K. The decrease in $T_{IM}$ and $T_C$ for MgO films have been explained on the basis lattice strain relaxation. Higher degree of relaxation creates more defects which affect the magnetotransport properties and $T_{IM}$ ($T_C$) decreases. Keeping in view the variable presence of disorder in the present films, we have analyzed the transport above $T_C$ through Mott`s VRH model. Based on this model the increase in lattice relaxation will produce defects, which will result in decrease of the tendency of charge localization (hopping distance also decreases) and so will $T_{IM}$ and $T_C$ decreases. As already described in section 2.4.3 these are in keeping with our experimental results.

### 4.2.4 Nanophasic $La_{0.7}Ca_{0.3}MnO_3$ (LCMO) Manganite

During the last decade nanocrystalline form of various materials have drawn considerable attention because they typically exhibit physical and chemical properties that are distinct from their bulk counterparts. The physical properties of manganites are also expected to depend on material size due to both the nanoscale phase inhomogeneities inherent to these materials and additional surface effects. Hwang et al [235] pointed out that the large low field magnetoresistance (LFMR) of the polycrystalline samples is dominated by spin polarized tunneling between neighbouring grains. This is quite different from their single crystals and epitaxial films counterparts where the double exchange mechanism is prominent. The bulk samples of manganites were usually synthesized by the conventional ceramic methods that need higher sintering temperature and long annealing time to obtain homogenous composition and desired structure. These methods are not appropriate for many advanced applications, due to formation of large particles, agglomerates, poor homogeneity, undesirable phases, abnormal grain growth and an imprecise stoichiometric control of cations. However, the sol gel process has potential advantage over the other methods not only for achieving homogenous mixing of the components on the atomic scale, but also for the possibility of forming desired shapes which are of technological importance. Other advantages of the sol-gel route are lower processing temperatures, short annealing times, high purity of materials, good control of size and shape of the particles and particle size well below 100nm at the lower processing temperature. It may be opportune to mention that there are several reports on the synthesis of nanophasic manganites by sol gel based synthesis methods but none of them seems to have carried out extensive studies on the effect of sintering temperature on sol-gel synthesized manganites in relation to low field magneto-transport properties [237,357-366]. In this work, we have described and discussed the synthesized of $La_{0.7}Ca_{0.3}MnO_3$ perovskite manganite by low temperature polymeric precursor route and studied the effect of sintering temperature on microstructure and low field magneto-transport properties which we have carried out in our lab.

We have adopted the sol-gel based polymeric precursor route to synthesize $La_{0.7}Ca_{0.3}MnO_3$ (LCMO) samples having nano size particles at a significantly lower sintering temperature compared to the conventional solid-state procedure. In this technique, aqueous solution of high purity $La(NO_3)_3.6H_2O$, $Ca(NO_3)_2.4H_2O$ and $Mn(NO_3)_2.4H_2O$ have been taken in the desired stoichiometric proportions. An equal amount of ethylene glycol has been added to this solution with continuous stirring. This solution is then heated on a hot plate at a temperature of ~ 100-140$^0$C till a dry thick brown colour sol is formed. At this temperature ethylene glycol polymerizes into polyethylene glycol, which disperses the cations homogeneously forming cation-polymer network. This has been further decomposed in an oven at a temperature of ~ 300$^0$C to get polymeric precursor in the form of black resin like material. The polymeric precursor thus obtained is then sintered at different temperatures ranging from 500 to 1000$^0$C. Several time periods ranging between 2 to 6 hrs were employed. It was found that sintering for ~4 hrs gave optimum results for all temperatures. Phase pure completely crystalline samples have been obtained at the temperature as low as $600^0$C. The LCMO samples sintered at $600^o$C, $700^o$C, $800^o$C, $900^o$C and $1000^o$C will hereafter be referred to as S6, S7, S8, S9 and S10 respectively.

All the synthesized samples have been subjected to gross structural characterizations using powder X-ray diffractometer [XRD, Philips PW 1710] using Cu K$\alpha$ radiation at room temperature and microstructural characterization by scanning electron microscopic technique [SEM, Philips XL20]. The magnetotransport measurements have been done by standard dc four-probe technique in the temperature range of 300-80K and in applied magnetic field in the range of 10 kOe. The magnetic characterizations have been carried out by ac susceptibility measurements.

Sintering temperature is one of the key factors that influence the crystallization and microstructure of the perovskite $La_{0.7}Ca_{0.3}MnO_3$ (LCMO) samples, which in turn affect the magneto-transport properties. The crystallinity and phase analysis of all the synthesized samples (S6, S7, S8, S9 and S 10) were determined by the powder X-ray diffraction and the corresponding pattern are shown in Fig.45a. All the samples are orthorhombic and single phasic without any detectable secondary phase. In the present polymeric precursor method, the characteristic perovskite phase formation starts at a significant low temperature of $600^o$C as compared to other conventional methods. The lattice parameters (orthorhombic unit cell parameters a, b, c and the unit cell volume V= abc) decrease in a systematic fashion with increasing sintering temperature as shown in Table1. The intensity of the X-ray peaks for the LCMO perovskite phase increases as the sintering temperature increases from $600^o$C to $1000^o$C indicating that the crystallinity of $La_{0.7}Ca_{0.3}MnO_3$ becomes better with higher sintering temperature. Fig. 45b shows the (200) reflection of S6, S7, S8, S9 and S10 samples. It is clear from inset that as the sintering temperature increases there is a decrease in the full width at half maximum (FWHM) and hence the crystallite size increases. The average crystallite sizes of the samples are obtained by the X-ray line width using Scherrer formula P.S. $\simeq$ k$\lambda$ /$\beta$ Cos $\theta$; where k $\simeq$ 0.89 is shape factor, $\lambda$ is wavelength of X-rays, $\beta$ is the difference of width of the half maximum of the peaks between the sample and the standard silicon used to calibrate the intrinsic width associated with the equipment and $\theta$ is the angle of diffraction. The average crystallite size of a sample sintered at $700^o$C (S7) is calculated to be ~32nm. The same increases to ~ 35nm, 40nm and 62 nm for sample sintered at $800^o$C (S8), $900^o$C (S9) and $1000^o$C (S10) respectively. The crystallite sizes, lattice parameters and unit cell volumes obtained for the different samples are listed in Table 4. Fig. 46 (a-d) shows the representative images elucidation surface morphology for the samples S10, S9, S8 and S7 respectively employing scanning electron microscope in secondary electron imaging mode. SEM observation reveals that there is a distribution of particle size for all samples and as the sintering temperature increases, the particle size increases and the porosity decreases. The highest temperature (1000°C) sintered sample (S10) has well connected particles whereas as we go down to lower temperature sintered sample (S7), the particle connectivity becomes poor. The average particle size is ~ 50, 75, 125 and 200 nm respectively for samples S7, S8, S9 and S10. Fig.47 shows the variation of crystallite sizes (CS) obtained from the width of the X-ray diffraction peaks and the particle sizes (PS) obtained from SEM. Both crystallite as well as particle size increase as the sintering temperature is increased due to congregation effect. However, it has been observed that there is a difference between CS and PS at all sintering temperature and is more pronounced at higher sintering temperature. For example, CS = 32 nm and PS = 50 nm for S7 and for S10 it is ~ 62 and ~ 200nm respectively. This difference is due to the fact that particles are composed of several crystallites, probably due to the internal stress or defects in the structure [102].

The paramagnetic-ferromagnetic transition temperature ($T_C$) has been measured by measuring the ac susceptibility ($\chi$) in the temperature range of 300-80K. The variation of $\chi$ with temperature for the samples S6, S7, S8, S9 and S10 is shown in Fig.48, which depicts a ferromagnetic ordering transition for all the samples. We have observed only a slight variation in $T_C$ for the samples sintered at different temperatures, which have been examined by the peaks in ($d\chi/dT$). The values of $T_C$ are given in Table 5. As is clear from this table $T_C$ shows an increase from 256K for the sample S6 to 274K for the sample S10. It has also been observed that as the sintering temperature decreases the width of transition broadens which suggests that at low sintering temperatures grains are loosely connected as also visible in the scanning electron micrograph shown in Fig.46. One of the main advantages of polymeric precursor method, as we have stated earlier, is enhancement in $T_C$ as compared to standard solid-state method, which is seen here up to ~30K. Also Fig.48 indicates that the magnetization of the samples increases as the sintering temperature increases which is same as found in earlier results [187].

The dc resistivity ($\rho$) was measured in the temperature range of 300-80 K by four probe technique with and without magnetic field and the data for resistivity without applied magnetic field for samples S6, S7, S8, S9 and S10 are plotted in Fig.49. At room temperature (~300 K), resistivity values are ~ 309.15, 221.48, 89.18, 19.25 and 2.42 Ω-cm, respectively for S6, S7, S8, S9 and S10 samples. Thus even at room temperature large enhancement, by more than two orders of magnitude, in the resistivity is observed for S6 as compared to S10 as a consequence of lower $T_S$ and hence smaller particle size. This increase in resistivity is believed to be caused mainly due to enhanced scattering of the charge carriers by the higher density of magnetic disorder in GBs at smaller particle size. On increasing $T_S$ the particle size increases leading to decrease in the GBs and the associated disorder. This results in decrease in scattering of the carriers expressed by a decrease in the resistivity. All the samples show an increase in the resistivity on lowering temperature and at a characteristic temperature, which is lower than the corresponding $T_C$, an insulator ($d\rho/dT < 0$) to metal ($d\rho/dT > 0$) like transition is observed. The measured characteristic insulator-metal transition temperatures ($T_{IM}$) are ~267, 240, 210, 179 and 138K for S10, S9, S8, S7 and S6 respectively. The corresponding peak resistivity values are ~ 3.91, 33.9, 241.55, 1013.51 and 3748.27 $\Omega$-cm respectively. The sol-gel prepared samples show a large difference between $T_C$ and $T_{IM}$ and the difference increases as we lower the sintering temperature. This variation of $T_C$ and $T_{IM}$ with the sintering temperature is plotted in Fig.50. The large difference in the $T_C$ and $T_{IM}$ for all the LCMO samples is thought to be due to the existence of the disorder and is in fact a common feature of the polycrystalline manganites [235]. The $T_C$ being an intrinsic characteristic does not show significant change as a function of the sintering temperature. On the other hand $T_{IM}$ is an extrinsic property that strongly depends on the synthesis conditions and microstructure (e.g. grain boundary density). Thus the $T_{IM}$ goes down by 129K on lowering the sintering temperature from 1000°C to 600°C whereas Tc undergoes a small decrease with a concomitant transition broadening. The strong suppression of the $T_{IM}$ as compared to $T_C$ is caused by the induced disorders and also by the increase in the non-magnetic phase fraction, which is due to enhanced grain boundary densities as a consequence of lower sintering temperature. This also causes the increase in the carrier scattering leading to a corresponding enhancement in the resistivity. Thus lowering of sintering temperature reduces the metallic transition temperature and hence the concomitant increase in resistivity. When a magnetic field is applied, the FM clusters grow in size and the interfacial Mn spin disorder is suppressed resulting in the improved connectivity and consequently a decrease in the resistivity has been observed.

The temperature dependence of MR for S7, S8, S9 and S10 samples measured in the range 80-300 K at 3 kG and 10 kG are shown in Fig.51. All the samples show a sequential increase in low temperature MR with decreasing sintering temperature. The appearance of peak in the MR-T curve around $T_C$ depicts that in all the samples there is a contribution of the intrinsic component of MR, which arises due to the double exchange mechanism around $T_C$. However, around $T_C$ the peak in the MR-T curve of the sample S10 is significantly higher at all applied magnetic fields in comparison to other samples. The magnitude of MR peak around $T_C$ decreases as we lower the sintering temperatures as well as we reduce the magnitude of the applied magnetic field (Fig.51).

The peak MR values are ~13.07 and 10.34 % at 10 kG applied field for samples S10 and S9 whereas for sample S8 and S7 there is a hump in the MR variation around $T_C$. At 80 K, the MR values are measured to be ~12.87, 13.66, 13.85 and 15.46 for S10, S9, S8 and S7 respectively at the field of 3 kG. Thus, decreasing crystallite/grain size leads to the enhancement in low field MR at lower temperatures while the MR in the higher temperature regime is suppressed. The disappearance of the high temperature MR can be explained by weakening of the double exchange (DE) mechanism around the respective PM-FM transition temperatures due to decrease in particle size which results from low sintering temperature. It has been found that all the samples are showing significant MR at low field. The low field MR (LFMR) increases as the sintering temperature and hence particle size decreases. This is consistent with previous studies [244,364].

The magnetic field dependence of the low field MR at 80 K and 150K for all the samples are given in Figure 52. The low field MR of all the samples is observed to increase with increasing magnetic field. The MR-H curve shows two different slopes, the one below H ~ 1.5 kG is steeper while the other above H ~ 3 kG is rather weak. As the sintering temperature is lowered the MR increases and the slope of the MR-H curve in both the temperature regimes becomes steeper. For example in the S10 sample, the low field MR at H = 1.5 kG and 12 kG is measured to be ~ 6.37 % and 16.4 %, respectively at 150K whereas the same increases to ~ 12.16 % and 22.56% at 80K. The LFMR values at 150K and 80K for all the LCMO samples are listed in Table 2. It should also be noted that the variation of MR does not show any saturation in MR even up to 12 kG fields. The enhanced slope of the 80K MR-H curves in the low field regime as compared to those taken at 150K suggests that in the lower temperature regime spin polarized transport may the MR causing mechanism. These studies clearly depict that polymeric precursor technique is an effective method to enhance ferromagnetic ordering temperature along with enhancement of low field magnetoresistance.

### 4.2.5. Perovskite Manganite-PMMA Composites

The extrinsic low field magneto-resistance (MR) effect observed in polycrystalline materials is caused either due to spin polarized tunneling (SPT) [235,256] or spin dependent scattering (SDS) [367] as the conduction electrons traverse the grain boundaries (GBs). The dilution of GBs in manganites with insulating materials, such as a polymer, adjusts the barrier layer influencing the tunneling/scattering process which takes place across the interface/GBs and also influences the degree of magnetic disorder present therein. Enhancement of MR at low temperature and low magnetic field due to the alignment of neighboring FM grains is caused by the extraneous effects acting as pinning centers in the demagnetization by domain wall displacement. That is why a spin misorientation of the magnetically virgin state of the system is crucial to achieve and enhanced MR at low fields which is more useful for device application.

As is known that composite materials represent new material variant synthesized from two (or more) materials so that the individuality of the components is maintained in the composite. However, the properties of the composite materials are better (improved) than the properties of any of the individual components. In the case of magnetoresistance materials, the central aim is to synthesize a material which shows large magnetoresistance at low fields and near room temperature. Any single CMR material does not meet all the requisites. Even though one or two of the above criterions are satisfied by a single colossal magneto resistance material. Therefore, recent efforts are being directed to synthesize composite materials consisting of two (or more) different CMR or CMR and polymer/insulator materials which may show large magnetoresistance at low fields and at near room temperatures [358-372]. Several workers have attempted to enhance the low temperature–low field MR (LFMR) or the room temperature MR by making a composite of these CMR oxides with a secondary phase like an insulating oxide or a hard FM material or a polymer [368–382]. Most commonly used polymer for making CMR-polymer composite is polyparaphenylene (PPP) [370,373]. However, no study seems to be available in literature on making the composite with poly (methyl methaacrylate) (PMMA). Here we present the studies of LFMR for CMR-PMMA composites which we have carried out in our group with reference to $La_{0.7}Ba_{0.2}Sr_{0.1}MnO_3$ (LBSMO) and $La_{0.7}Ca_{0.3}MnO_3$ (LCMO) systems. In the present investigation we have synthesized the nano-sized grains of $(LCMO)_{1-x}(PMMA)_x$ (x = 0.0, 10, 20, 35 and 50 wt %) and LBSMO-x (PMMA) (x = 0, 2, 6, 10 wt% of PMMA) composites by a sol-gel/polymeric precursor route. The polymeric precursor/sol gel technique has efficiently been used to synthesize high quality nano-crystalline manganites [362,363,383]. The uniqueness of this method is that the PMMA is mixed with the CMR precursor before the grain growth, that is, the CMR nanocrystals are grown in the presence of PMMA and the maximum temperature (Ts) used is ~ 500 $^0$C.

### (i) Investigations on $La_{0.7}Ba_{0.2}Sr_{0.1}MnO_3$ –PMMA Composites:

The LBSMO sample was prepared by the solid state reaction route by using $La_2O_3$, $Ba(NO)_3$, $Sr(NO)_3$ and $MnO_2$ powders of 3N purity. These powders were mixed in the requisite ratio, ground and then calcined at 1193K for 48 h. The calcined mixture was then reground and mixed properly and pressed in the form of a pellet and sintered at 1273K for 48 h. The LBSMO–x wt% PMMA (where x = 0, 2, 6 and 10) samples were synthesized by mixing LBSMO powders with the requisite amounts of PMMA by grinding the mixture in the presence of chloroform. Since PMMA is soluble in chloroform, its addition facilitates a more homogeneous mixture of the CMR and PMMA. The resultant mixture was then dried in an oven at 373K for 6 h and pressed in the form of rectangular pellets and sintered at 673K for 12 h. Temperature of 673K was found to be safe because no decomposition of PMMA was observed. The resulting LBSMO–PMMA composites (having x wt% of PMMA, where x = 0, 2, 6, 10) designated as P0, P2, P6 and P10, respectively, were then characterized by employing X-ray diffraction (XRD), scanning electron microscopy (SEM), low field magneto-resistance (LFMR) and resistivity ($\rho$) measurements in the temperature range 77–300 K, and AC susceptibility ($\chi$) measurement in the temperature range 77–330 K.

In order to check the phase purity of the BSMO (P0) and to evaluate any structural modifications in P2, P6 and P10, the XRD of all the LBSMO–PMMA samples was carried out. The virgin LBSMO sample has an orthorhombic symmetry with unit cell parameters; a = 5.534 Å ,b = 5.498 Å , c = 7.802 Å. Within the accuracy limit of the diffractometer, the lattice parameters of BSMO in the LBSMO–PMMA composites do not change. This shows that LBSMO maintains its identity and that no intrinsic physico-chemical transformation has occurred as a consequence of MMA admixture. The XRD data (results not shown) do not reveal any signature of PMMA in he composites. The FWHM of the XRD profiles of all the samples do not reveal any change in the article size. To make the microstructural/morphological features more explicit, all the samples ere subjected to scanning electron microscopy SEM). The pure LBSMO sample has quite uniform particle size (~125 nm) of sub micron dimension (Fig. 53(a)). The particle size almost emains the same in all the composites. Fig. (a)–(c) depicts the representative SEM micrographs of samples P0, P2 and P10 corresponding to 0%, 2% and 10% PMMA admixed LBSMO. Amongst the composites, the PMMA is easily observable although its distribution within the BSMO matrix is not very uniform and consequently partially scattered PMMA clusters are also observed in sample P10. A comparison of Fig. 53(b) and (c) shows that the density of PMMA in the composites is observed to increase with the polymer concentration. In higher concentration composites (P10), some LBSMO grains also appear to be coated with the PMMA.

The PM to FM phase transition was studied by AC susceptibility ($\chi$–T) measurements in the temperature range 77–330K and the data corresponding to all the samples is plotted in Fig. 54.

The onset of PM to FM transition ($T_{C\text{-onset}}$) as well as transition temperature ($T_C$) both appears to be affected by the increase in the PMMA concentration. The $T_{C\text{-onset}}$ of pure LBSMO sample is measured to be ~320K which gradually decreases to ~300K in P10 while the transition temperatures determined from the peak in the $d\chi/dT$–T curve are found to be ~303, 290, 270 and 243 K, respectively. The transition width which is ~20K in the virgin sample increases quite appreciably with increase in the PMMA concentration. The observed lower value of $T_C$ and the transition broadening can be explained as follows. In the virgin sample, the transition is broadened by the strong magnetic disorder induced by the small grain size due to synthesis temperature (~1273 K) lower than commonly employed (1573–1673 K) incase of manganites. Some disorders like oxygen vacancies and very thin insulating oxides layer in the grain boundaries may also be present. PMMA admixture further broadens the transition and reduces the $T_C$ values successively to 243 K. Since PMMA is an insulator, its admixture to highly disordered LBSMO results in introduction f additional insulating regions into the LBSMO matrix. Hence PMMA segregates into the grain boundaries/interfacial regions and the intergranular separation is bound to increase with increasing PMMA concentration. This interfacial/grain boundary segregation of PMMA also affects the DE and this effect is more pronounced in the FM domain walls [370]. This consequently leads to suppression of PF–FM transition temperature to the lower side with increase in PMMA concentration. In fact the broadening of the PM–FM transition resulting in lower values of mid-point $T_C$ values suggests towards the suppression of the DE mechanism. In fact it has been shown by Yan et al. [370] that addition of polymer (PPP) leads to the dilution of the magnetization as well as yields additional magnetic disorders.

The temperature dependence of resistivity of all the samples measured by the four probe technique at zero magnetic field is shown in Fig. 55 (a). The virgin LBSMO sample (P0) exhibits the usual behavior of a fine grained disordered manganites i.e. the resistivity increases slowly with decreasing temperature and broad transition to the metal-like state taking place at a temperature; $T_P$~150K which is much smaller than the values observed in case of single crystals, epitaxial thin films or large grained polycrystals. For LBSMO, the PM to FM phase transition temperature ($T_C$) has been measured to be ~303K. Thus the metallic transition temperature ($T_P$) is drastically reduced as compared to the $T_C$ of the LBSMO sample. This observed discrepancy in the $T_C$ and $T_P$ can be attributed to the strong disorder due to small grain size in LBSMO synthesized at lower synthesis temperature (Ts)~1273 K. These disorders may include oxygen deficiency, high density of blocked Mn spins at the grain boundaries, increased misalignment of the neighboring FM domains, etc. Further, in fine-grained manganites, it has been shown that even in the FM regime below TC, the magnetic clusters consist of a mixture of FM and PM phases [373]. As the temperature decreases, the FM clusters grow at the cost of the PM phase and when the FM phase fraction crosses the percolative threshold, the metallic transition is induced. In the PMMA admixed samples there is huge enhancement in the resistivity and also the resistive transition to the metal-like state is found to disappear from the temperature range of the present measurement. In fact the resistivity at 77K is found to increase by almost four orders of magnitude from the virgin LBSMO (P0) to P10 composite whereas increase in the resistivity at 293K (room temperature) is nearly three orders of magnitude. The variation of resistivity with PMMA concentration at 293 and 77K is shown in Fig. 55(b). In addition, the temperature dependence of resistivity is also modified, and in fact as the polymer concentration is increased, the temperature dependence shows a transition from semiconductor like behavior for sample P0 to an insulator like behavior for sample P10. The reason for this observed resistive transformation seems to be the breaking up of the percolative network as well as breaking of the DE mechanism with increasing PMMA concentration. In fact the admixture of non-magnetic and insulating PMMA separates the FM metallic clusters and as the polymer concentration increases the spatial separation of these grains/clusters further increases. This leads to the disappearance of the metallic transition and concomitant enhancement in resistance. When a magnetic field is applied, the FM clusters grow in size resulting in improved connectivity and consequently the resistance decreases and the samples exhibit significant low field magneto-resistance.

The temperature dependence of the LFMR measured at H =3.6 kOe is shown in Fig. 56. It is evident from Fig. 56 that the temperature dependence has two distinct regimes. In the higher temperature regime (T>125 K), the LFMR of the virgin LBSMO (P0) sample is larger than that of LBSMO–PMMA composite samples (P2, P6 and P10) while in the lower temperature regime (T <125 K), the same is higher for the LBSMO–PMMA composites. The disappearance of the high temperature MR can be explained by taking into account the dilution of the FM magnetic phase and the DE mechanism around the respective FM transition temperatures as a consequence of increasing PMMA concentration. In polycrystalline samples, the low field MR (H~3 kOe) has its origin in the spin polarized tunneling across the grains. In the present study, the increase in MR at low temperature, although small, seems to be due to slightly enhanced intergrain tunneling because of the addition of PMMA. The magnetic field dependence of MR at 77K is shown in Fig. 57. The steep slope of the MR–H curve which increases slightly in PMMA admixed samples suggests that spin polarized tunneling across the FM grains may be the possible transport mechanism in all the samples [235]. The inset of Fig. 58 shows the variation of MR with PMMA concentration at 77 K. The smaller increase in MR in the present investigation may be due to the high resistivity of PMMA that seems to be encapsulating/separating the LBSMO grains. At 77K and H =1.2 kOe, the observed MR for virgin LBSMO sample is 12% and it increases to 14% for 10 wt% PMMA admixed LBSMO composite. The observed increase in MR with increasing PMMA concentration is rather small (~17%). This is due to the fact that even the virgin LBSMO sample is highly disordered due to smaller grain size and presence of oxygen vacancies, impurity segregation into the grain boundaries, etc. Since MR at lower fields ( ~3 kOe) is due to the spin polarized tunneling [235] and is controlled by the thickness of the disordered inter-granular insulating layer, the optimum thickness of this layer is the one that facilitates spin conserved tunneling process across the FM grains. Due to the lower synthesis temperature, the thickness of the insulating layer in the grain boundary region is near optimum and consequently the low field MR has already reached the saturation level. The thickness of the insulating layer between the two FM grains increases with increasing PMMA concentration and this leads to creation of some additional spin disorders and when a magnetic field is applied, the spin disorder is suppressed resulting in slightly higher MR especially above a magnetic field H~1.5 kOe. Thus the overall positive change in MR at lower temperatures is due to the slightly enhanced spin disorder in the intragranular regions which leads to enhanced spin polarized tunneling. Recently, in polycrystalline LCMO–PMMA composites the enhancement in MR ~ 30–35% is observed [374].

**(ii) Studies on $La_{0.7}Ca_{0.3}MnO_3$–PMMA composites**

In another work LCMO-PMMA composites were prepared by polymeric precursor route [362,363]. In the first step of synthesis, the relevant metal nitrates viz., La(NO3)36H2O, Ca(NO3)24H2O and Mn(NO3)24H2O (purity > 99 %) were dissolved in deionized water in the desired cationic ratio La : Ca : Mn = 0.7 : 0.3 : 1. After the proper homogenization of the aqueous solution, an equal amount of ethylene glycol was added. The resulting solution was thoroughly homogenized and then continuously stirrered at ~1500 C. The heat treatment was continued till the solvents dried and a resin was formed. This residual resin was further decomposed by heat treatment at ~2500 C. The decomposition of the resin yields slightly blackish foam like substance which acts as a precursor for the growth of polycrystalline LCMO and LCMO-PMMA composites. This foam is then crushed and pressed in form of pellets which is heat treated in air at ~500 °C for 8 hrs to get the nano-crystalline LCMO. The LCMO-PMMA composites were prepared by the following process. PMMA was weighed in the desired amounts (10, 20, 35 and 50 wt %) and dissolved in chloform. Then the LCMO precursor was added to the chloroform-PMMA solution and mixed thoroughly till the whole of the chloroform evaporated. The resulting powder mixture was dried at ~150 °C and then pressed in the form of rectangular pellets of 10x5x1 mm3 dimension. These pellets were then heated in air at ~5000 C for 8 hrs to get the LCMO-PMMA nano-crystalline composites. LCMO phase in the composite is grown in the presence of PMMA, the homogeneity of the composites is expected to be better, because of the amorphous nature of the LCMO precursor foam obtained at ~250 °C. The LCMO samples having polymer concentration x = 0.0, 10, 20, 35 and 50 wt % will hereafter be referred to as LCMP0, LCMP10, LCMP20, LCMP35 and LCMP50 respectively. These LCMO-PMMA composite samples were characterized by powder Xray diffraction (XRD), scanning electron microscopy (SEM), resistivity ($\rho$); low field magneto-resistance (LFMR) and ac susceptibility ($\chi$) measurements in the temperature range 77-300 K.

Figure 58 shows the X-ray diffraction pattern of LCMP0, LCMP10, LCMP20, LCMP35 and LCMP50, respectively. This figure reveals that all the samples are polycrystalline and have orthorhombic unit cell. The lattice parameters of the virgin LCMO are; a = 5.436 Å, b = 5.482 Å and c = 7.675 Å and a small increase is observed in PMMA admixed samples which, however, saturates at higher PMMA concentrations. The reflected intensity is observed to decrease with increasing PMMA concentration with a concomitant increase in the full width at half maximum (FWHM) of the corresponding reflections. Figure 59 depicts the close up of the (110) reflection of LCMP0, LCMP20 and LCMP50 samples. The increase in FWHM suggests a slight decrease in the LCMO grain size in the composite or may be suggestive of the presence of PMMA in the background. As expected, this suggests a decrease in the degree of crystalline nature with increasing PMMA concentration. As evidenced by shifting of the (110) reflection of LCMP20 and LCMP50 towards lower 2θ, the lattice parameters of the orthorhombic unit cell show small increase with PMMA concentration beyond 10 wt %. The orthorhombicity of the system usually defined by the expression Or(%) = [(b-a)/(b+a)]*100, is observed to be 0.421 % for LCMP0 and is almost not affected in the composites. The average grain diameter/grain size (p) was evaluated indirectly from the XRD data by applying the Scherrer formula p = $0.9\lambda / [\beta \cos\theta]^{-1}$ using the FWHM (full width at half maximum) of the (110) reflection. The same was also determined directly by employing scanning electron microscopy (SEM) to image the surface microstructure of the LCMO-PMMA composite samples. Figure 60 (a-d) shows the scanning electron micrographs of LCMP0, LCMP20, LCMP35 and LCMP50. The micrographs reveal a small variation in the grain size with increase in the PMMA concentration. The average grain size is found to be ~ 40 nm in the virgin LCMO sample and size does show a small decrease as the

PMMA concentration is increased. This feature is clearly observed in the SEM micrograph of LCMP50 (Figure 60 (d)).

The PM-FM phase transition studied by temperature variation of ac susceptibility ($\chi$-T) measurement in the temperature range 77-300 K also reveals some interesting aspects of LCMO-PMMA composites. The $\chi$-T data of all the samples is shown in Figure 61. The $T_C$ of the samples was determined from the peak in the $d\chi/dT$ curve (results not shown) and is representative of the mid point of the transition. The virgin LCMO sample (LCMP0) undergoes a PM-FM transition at $T_C \sim 261$ K while all the composite samples have successively lower $T_C$ values. The $T_C$ of LCMP10, LCMP20, LCMP35 and LCMP50 are $\sim$ 256, 255, 244 and 236 K, respectively. In addition to the successive lowering of the $T_C$ with increasing PMMA concentration, the transition width is also observed to increase appreciably. This suggests that despite the non-magnetic nature of PMMA, its admixture to LCMO results in partial decoherence of the FM domains. Since PMMA cannot be incorporated into the LCMO matrix it diffuses and segregates into the GBs/interfacial regions. These GBs/interfacial regions are known to have both magnetic and structural disorder. The most common disorder being Mn spins blocked at GBs, increased anisotropy in the interfacial regions and misalignment of the magnetic moments of the neighboring FM domains. The addition of PMMA, despite its nonmagnetic character, is expected to further increase the density of these disorders. The increased density of these disorders means the widening of the GB regions and thus the system can be considered a phase separated one, having a mixture of FM (metallic) phase and the non-magnetic (insulating) phase. As the PMMA concentration is increased, the overall density of the non-magnetic/insulating region in the bulk increases, resulting in the partial decoherence of the long range FM order. Since even at 50 wt % PMMA the FM transition is still as high as 236 K, it is conjectured that the fundamental/intrinsic mechanism of the FM transition, viz., the double exchange, is not modified appreciably due to the PMMA admixture. The increased density of the disorders described above is expected to lead to large enhancement in the dc resistivity of the LCMO-PMMA composites with increasing polymer concentration.

The dc resistivity ($\rho$) was measured in the temperature range 77-300 K by four probe technique and the data are plotted in Figure 62. At room temperature ($\sim$298 K), resistivity values are $\sim$3.75, 6.05, 34.94, 49.35 and 90.13 $\Omega$-cm, respectively for LCMP0, LCMP10, LCMP20, LCMP35 and LCMP50. Thus even at room temperature large enhancement in the resistivity is observed as a consequence of PMMA admixture. This resistivity enhancement is caused mainly by the scattering of the charge carriers by the magnetic disorder in the GBs and as the PMMA concentration increases more and more scatterers are produced resulting in the further increase in the resistivity. The lowering of the temperature towards the $T_C$ causes an increase in the resistivity and at a certain temperature which is lower than $T_C$ an insulator ($d\rho/dT < 0$) to metal (($d\rho/dT > 0$) like transition is observed in all the samples. The measured insulator-metal transition temperature ($T_{IM}$) are ~218, 213, 173, 153 and 108 K, respectively for LCMP0, LCMP10, LCMP20, LCMP35 and LCMP50 and the corresponding ρ values are ~15.92, 26.56, 227.97, 432.25 and 3832.51 Ω-cm, respectively. The variation of $T_C$ and $T_{IM}$ with the PMMA concentration is plotted in Figure 63 where as the resistivity dependence on PMMA concentration at different temperatures is depicted in Figure 64. The large difference in the $T_C$ and $T_{IM}$ for the virgin LCMO sample (~ 43 K) is due to the existence of the disorder and is in fact a common feature of the polycrystalline manganites having small grain size [235]. The strong suppression of the $T_{IM}$ as compared to $T_C$ is caused by the PMMA induced disorders and also by the increase in the non-magnetic phase fraction. This also causes the increase in the carrier scattering leading to a corresponding enhancement in the resistivity. In fact, the admixture of non-magnetic and insulating PMMA separates the FM metallic clusters and as the polymer concentration increases the spatial separation of these grains/clusters further increases. This leads to the lowering of the metallic transition temperature and concomitant enhancement in the resistivity. When a magnetic field is applied, the FM clusters grow in size and the interfacial Mn spin disorder is suppressed resulting in the improved connectivity and consequently a decrease in the resistivity.

The temperature dependence of MR measured in the range 77-300 K at 2 kOe is shown in Figure 65. The nearly monotonic temperature dependence of MR below $T_C$ suggests that in LCMP0 and LCMP10 samples, there is no contribution of the intrinsic component of MR which arises due to the double exchange mechanism around $T_C$. However, around $T_C$ the jump in the MR-T curve of the LCMP0 as well in LCMP10 shows small contribution from the intrinsic component to the total low field MR. In the PMMA admixed LCMO samples the jump in the MR around $T_C$ shifts towards lower temperatures, (for example - LCMP10 and LCMP20) and finally it disappears when the PMMA concentration is increased above 20 wt %. This is due to the fact that as the PMMA concentration is increased the intrinsic contribution to the low field MR is suppressed and as a consequence of this the MR of all composite samples is lower than that of the virgin LCMO sample till a certain lower crossover temperature is reached. The MR crossover temperatures are ~150, 144 and 120 K for LCMP20, LCMP35 and LCMP50, respectively. Below these temperatures, the LCMO-PMMA composite samples exhibit larger MR than the virgin

LCMO sample. At 77 K, the MR values are measured to be ~ 13.1, 13.7, 14.8, 15.5 and 16.4 % for LCMP0, LCMP10, LCMP20, LCMP35 and LCMP50, respectively. Thus, increasing the PMMA concentration leads to the enhancement in low field MR at lower temperatures while the MR in the higher temperature regime is suppressed. The disappearance of the MR in the higher temperature regime can be explained by taking into account the successive dilution of the FM magnetic phase and the double exchange (DE) mechanism around the respective FM transition temperatures as a consequence of increasing PMMA concentration.

The magnetic field dependence of the low field MR at 77 K for all the samples are given in Figure 66. The low field MR of virgin LCMO as well as LCMO-PMMA composite samples is observed to increase with increasing magnetic field. The MR-H curve shows two different slopes, the one below H~1 kOe is steeper while the other above H~1.5 kOe is rather weak. As the PMMA concentration is increased MR increases and the slope of the MR-H curve in both the temperature regimes becomes steeper. In the virgin LCMO sample, the low field MR at H = 0.8 kOe and 3.6 kOe is measured to be ~ 9 % and ~15 %, respectively and the same in the case of LCMP50 increases to ~11 % and ~20 %, respectively. Thus as the polymer concentration is increased to 50 wt % PMMA, the relative enhancement in MR shows an increase of ~ 20 % in the lower field regime and ~35 % in the higher field regime. It may also be noted that as the PMMA concentration increases, the saturation of MR with magnetic field disappears and hence the slope of the MR-H curve in the higher magnetic field regime (H>1.5 kOe) increases.

The variation in MR as a function of PMMA concentration can be explained by taking into account the increased disorder in the inter-granular regions. The grain boundaries in polycrystalline manganites mimic the role of the thin insulating layer sandwiched between two FM grains and therefore the electron hopping across the grain boundaries depends essentially on these GBs themselves and the spin states in the neighboring grains. It is known that in polycrystalline manganites the interfacial/inter- granular regions are magnetically as well as structurally disordered and hence have disordered Mn spins. In addition, certain degree of spin canting may also be present. When a non-magnetic impurity such as PMMA is added these disorders are further increased. As the spin disorder increases, there is a concomitant increase in the carrier scattering and consequently the resistivity of the samples increases. Such variation has indeed been observed in the present investigation and the resistivity has been observed to increase with increasing PMMA concentration (increased magnetic disorder is proportional to PMMA concentration). The Mn spin disorder is suppressed under the influence of an external magnetic field causing a decrease in the resistivity and hence increase in MR. The increase in the slope of the MR-H curve in the higher field regime as a function of PMMA concentration also suggests that the spin disorder created (due to increasing PMMA concentration) has not been fully suppressed by a magnetic field H = 3.6 kOe. Consequently no MR saturation (as observed in virgin LCMO (LCMP0) and 10 wt % PMMA admixed LCMO (LCMP10)) is seen in other composite samples having higher PMMA concentration. In fact as Mn spin disorder density increases due to PMMA admixture, the carrier scattering increases leading to further enhancement in the resistivity and the magnetic interaction energy also increases and hence a higher magnetic field is needed to suppress them.

### 4.2.6. New Doped Manganite: $La_{0.7}Hg_{0.3}MnO_3$ (LHMO)

Rare earth manganites, $RE_{1-x}AE_xMnO_3$ (RE=La, Pr, Nd and AE=Ca, Ba, Sr, etc.) exhibits CMR over a wide range of temperatures and compositions but the magnetic field required is very high [45,46,84]. The compositional range is also very important and it is different for different dopants, e.g. $La_{1-x}Ca_xMnO_3$ exhibits a significant CMR in the range 0.2 < $x$ 0.5. The Curie temperature ($Tc$) increases with increase in the average cationic radius $<r_D>$. Increasing $<r_D>$ mimics an increasing hydrostatic pressure and it increases the Mn–O–Mn bond angle. If there is a large mismatch in different D site cations, the $Tc$ does not show any increase with $<r_D>$ [105]. For practical applications the CMR perovskite manganites should exhibit significant MRs at low magnetic fields, say 1 kOe. Several efforts have been made to enhance the low field MR in these magnetic materials [235-238,246,273,384,385]. In the present study, we have investigated the effect of partial replacement of $La^{3+}$ (1.22 Å) by $Hg^{2+}$ (1.12 Å). The ionic radius of $Hg^{2+}$ (1.12 Å) is larger than that of $Ca^{2+}$ (1.06 Å) but much smaller than that of $Ba^{2+}$ (1.43 Å) and $Sr^{2+}$ (1.27 Å) the most widely used dopants [72]. The radius mismatch between La and Hg is smaller than that for La and Ca and this may result in reduced intrinsic lattice strains. The substitution of Hg for La will lead to a new magnetoresistance material $La_{1-x}Hg_xMnO_3$.

The ceramic samples of $La_{0.7}Hg_{0.3}MnO_3$ (LHMO) were prepared by the standard solid state reaction route. The whole procedure consisted of two steps. In the first step, $La_2O_3$ and MnO powders (all 2 N purity) were thoroughly mixed in an appropriate cationic ratio (La /Mn=0.70 / 1.00). As $La_2O_3$ is hygroscopic so prior to weighing it was calcined at 1000° C for 10 hrs. The stoichiometric oxide mixture was then heated at 1100 °C for 36 h with two intermediate grindings to get La–Mn–O precursor. An appropriate amount of HgO (yellow powder, 2 N purity) was added to the Hg free La–Mn–O precursor so that the cationic ratio was La/Hg/Mn=0.70/ 0.35 / 1.00. Some excess of HgO was taken to maintain sufficient vapour pressure of Hg for better reaction with the La–Mn–O precursor. The mixture was thoroughly mixed and then pressed into rectangular pellets of length $l=1$ cm, width $w=0.6$ cm and thickness $t=0.1$ cm. This pellet was put in a platinum box which was then tightly closed and sealed in an evacuated ($10^{-5}$ Torr) quartz tube (length $L=5$ cm, outer diameter O.D.=1.1 cm, inner diameter I.D.=0.9 cm) (1 Torr=133.322 Pa). Since HgO decomposes above ~500°C to yield Hg and $O_2$, this sealed chamber route is unavoidable. In case of open sintering all the Hg by virtue of its high vapour pressure will escape without reacting with the La–Mn–O precursor as in the case of Hg-based HTS materials [388]. Finally, the sealed quartz tube was put in a steel tube, which was filled with sand and tightly closed. The steel tube was then heated slowly in a Heraeus tube type furnace to 850°C. The temperature was maintained at 850°C for 40 h and then cooled to room temperature. These as grown samples following the above procedure were subjected to structural characterization /phase identification using powder X-ray diffraction (Philips, PW1710) employing CuKα radiation. Scanning electron microscopy (SEM; Philips, XL20) in the secondary electron emission mode was used to evaluate the surface microstructure. The cationic composition was investigated by energy dispersive X-ray analysis (JEOL, EDSA JXA 8600 MS). The temperature dependence of the resistivity and the magnetoresistance was measured by the standard four-probe technique (Keithley Hall Setup) in applied dc magnetic fields in the range 0–300 mT.

The cationic composition as investigated by energy dispersive X-ray analysis reveals some loss of Hg which is obviously due to the high vapour pressure of Hg at elevated temperatures, e.g. at 850°C. This is the rationale behind the excess of Hg in the starting composition. The overall cationic stoichiometry is quite close to the envisaged one. The XRD data reveals that the $La_{0.7}Hg_{0.3}MnO_3$ samples have an orthorhombic unit cell having lattice parameters $a=5.5183$ Å, $b=5.6383$ Å and $c=7.5368$ Å as determined by least square fitting. The cell volume is $V=239.75$ Å. In order to confirm that Hg doped $LaMnO_3$ represents a distinct giant magnetoresistance material, we have also synthesized La-deficient $LaMnO_3$, i.e. $LaMnO_3$ (LMO) sample to compare it with 30% Hg doped LaMnO, i.e. LHMO. The XRD pattern of 30% Hg doped $LaMnO_3$ and $La_{0.7}MnO_3$ is shown in Fig. 67. It is important to note that the lattice parameters of LHMO is different than that of LMO as given in Table 6.

In particular there is significant contraction along $c$ parameter (~ 3.435%). This is expected in view of the smaller ionic size of Hg (1.12 Å) as compared to La (1.22 Å). This lead to reduction in unit cell volume of LHMO by about 1.525% (Table B.1). An approximate evaluation of the particle size from full width at half maximum (FWHM) employing the Scherrer equation reveals that samples are micro-crystalline having grain size in the range 0.5–2.0 mm. This result is further confirmed by the SEM observations. A representative SEM micrograph showing surface microstructure is given in Fig. 68. It is clear that the average grain size is indeed around 1 µm and it is also clear that there is a distribution of grain size.

The electrical resistivity ($\rho$) in the temperature range between room temperature and 77 K was measured by the standard four-probe technique in zero magnetic field and in an applied d.c. magnetic field $H_{dc}$ upto 300 mT. The dc four-probe electrical transport result is shown in Fig. 69a where $\rho$ is plotted against temperature at $H_{dc}$=100 mT and 150 mT. It is clear that as temperature is decreased, the zero field resistivity increases from its room temperature (~304 K) value 0.148 to 0.418 $\Omega$ cm at 227.13 K (~227 K) and then again it starts decreasing. The zero field resistivity decreases to 0.0831 $\Omega$ cm at 77 K. In the beginning the $\rho$–$T$ curve shows insulating behaviour ($d\rho/dt$ <0) down to 227 K and below 227K the $\rho$–$T$ behaviour is metal like ($d\rho/dt > 0$) 0). The behaviour of the $\rho$–$T$ curve in an applied magnetic field $H_{dc}$=150 mT is quite similar. The in-field dc resistivities at 304, 227 and 77 K are 0.148, 0.404, and 0.075 $\Omega$ cm, respectively. The variation of MR with temperature is shown in Fig. 69b. The MR increases with decreasing temperature. At room temperature (304 K in this case) MR is very small ~0.06% in an applied field of $H_{dc}$=150 mT. As the temperature decreases MR increases to 3.41% at 227 K and it further goes up to ~9.05% at 77 K and $H_{dc}$ =150 mT. As the field is increased the MR has been observed to increase at all temperatures, e.g. at 77 K the MR is 7.27% at 60 mT, 9.05% at 150 mT and 9.94% at 300 mT. A plot depicting the variation of the MR with $H_{dc}$ is shown in Fig. 69c. It is clear that in lower fields (<120 mT) the MR rises rapidly and above this field the increase becomes slower, although saturation like behaviour is still not visible. Below $H_{dc}$ =120 mT the MR– $H_{dc}$ behaviour is logarithmic while above 150 mT it is linear. In order to evaluate the Curie temperature both LHMO and LMO were subjected to ac susceptibility ($\chi$) measurement. The $\chi$–$T$ plot for LHMO is shown in Fig. 69d.

It is opportune to mention that the LHMO samples exhibit contrasting behaviour when compared to the LMO samples. The comparative value of different parameters the LMO and LHMO samples is tabulated in Table 6. As pointed out earlier, to show that LHMO is indeed a distinct CMR material, LMO samples were also characterized. As shown in Fig. 70a, the LMO samples show sharp insulator to metallic transition at 246.76 K (~247 K) with a second broad maxima in the resistivity at 228.64 K (~229 K). This second broad maxima in the resistivity is supposed to be due to unusual scattering of conduction electrons. In keeping with the known systems this may be taken to correspond to the scattering on critical fluctuations of the spin systems which extend rather far below the transition temperature for strongly correlated electron systems [386,387]. The overall behaviour of $\rho$–$T$ for LMO is similar to that of LHMO but the resistivity and the insulator-metal transition temperature ($T_{IM}$) are noticeably higher than in the LHMO sample. The MR of the LMO sample shows a peak around 246.76 K (MR= 4.055% at 150 mT) which was found to be close to $T_C$. This is shown in Fig. 70b. In keeping with known results on CMR in LCMO materials, this peak can be taken to be the Curie temperature [257]. This is also in agreement with the $\chi$–$T$ plot for LMO (Fig. 70c), which yields a $Tc \sim 245.61$ K.

### 4.2.6 Double Layered Manganites

The colossal magnetoresistance (CMR) effect in naturally layered perovskite manganites has attracted great attention from physicists, chemists, and materials scientists in regard to their fundamental study and technological applications. The cubic (or pseudocubic) perovskites $RE_{1-x}AE_xMnO_3$ (RE being rare-earth ions and AE divalent cations) such as $La_{1-x}Sr_xMnO_3$, with three-dimensional Mn-O networks (isotropic $MnO_6$ octahedra), are known to become ferromagnetic metal at hole doping of x > 0.2 and exhibit CMR effects [4,44-46,78,84]. In addition to the CMR effect they exhibits coupled exotic phenomenon such as insulator-metal transitions in the presence of a concomitant para- to ferromagnetic transition, enormous structural rearrangements, electronic phase separation and charge ordering. Meanwhile, it has become clear that the physics in this material group cannot be described solely in terms of the classical Double Exchange model [61] but the complex interplay of the various microscopic (spin, charge, structural, and orbital) degrees of freedom and inhomogeneties induced phase separation has to be taken into account to describe the rich phenomenology [4,186]. Most of the studies have so far been carried out on doped $LaMnO_3$ type perovskite structures. However, very recently Moritomo et al. (1996) [389] discovered CMR properties in the layered $La_{1.2}Sr_{1.8}Mn_2O_7$ system (n=2) akin to their 3D counterparts (n = ∞), thereafter followed a flurry of activity on these layered manganites especially with the n = 2 structure [390,391]. A very large CMR effect has been observed for the n = 2 compound above the Curie temperature. At 129 K, the MR of 200% at 0.3 T has been observed which is, significantly higher than the equivalent 3D Sr-based compound. The CMR at few Teslas is observed in a very broad temperature range from 100 K to close to room temperature. In another study Kimura et al. (1996) [390] reported a large MR effect in $La_{2-2x}Sr_{1+2x}Mn_2O_7$ system. There studies also revealed that the magnetotransport behavior of the composition Sr = 0.3 is very different from that of Sr = 0.4. In fact, at Sr = 0.3 the resistivity in-plane and out-of-plane behave qualitatively differently. These bilayered manganites undergoes a charge-ordering transition at ~ 210 K. The charge-ordered state is of the $(3x^2 - r^2)/(3y^2 - r^2)$ type, as in the CE-state of cubic perovskites. The x = 0.5 compound also exhibits a substantial MR effect, although the charge-ordered state could not be melted completely at a magnetic field as high as 7T [392]. In addition to CMR effect these bilayered manganites also show phase transitions from a ferromagnetic metal to either an antiferromagnetic insulator or to a paramagnetic insulator. In some compounds, the strong competition between the ferromagnetic (FM) and the antiferromagnetic (AFM) phases, below *Tc*, drives the system to a host of other new magnetic orders, such as a cluster glass, spin glass like state or to a charge or orbital ordered state. They also show an extremely rich variety of magnetic structures as a function of doping which allow for the study of dimensionality effects on the charge transport and magnetic properties in these bilayered manganites [392-402]. The reduced dimensionality of these structures is anticipated to enhance the magnetic and electronic fluctuations in the critical temperature region just above $T_C$ and in this region where external fields can harness these fluctuations to generate ordered phases and the CMR effect [392-402]. Keeping in view the dimensionality effect of doped double-layered perovskite manganites we have carried out detail studies on low field magnetotransport behaviour of polycrystalline $La_{1.4}Ca_{1.6-x}Ba_xMn_2O_7$ bulk and $La_{1.4}Ca_{1.6}Mn_2O_7$ films.

### 4.2.6 (i) Investigation on Polycrystalline $La_{1.4}Ca_{1.6-x}Ba_xMn_2O_7$ (x=0.0, 0.2, 0.4, 0.6) bulk

The doped double layered perovskite manganites are a stack of FM metal sheets composed of $MnO_2$ bilayers which are separated by the insulating $(RE,AE)_2O_2$ layers and thus form a natural array of ferromagnetic-insulator- ferromagnetic (FM-I-FM) junctions [389,395]. It has been observed indeed that the incorporation of the $MnO_2$-$(RE,AE)_2O_2$-$MnO_2$ junctions in the structure naturally leads to higher magnetoresistance especially the low field magnetoresistance (LFMR) in these layered manganites than the infinite layer simple perovskite counterparts [389,394,395,403,404]. The individual bilayers consisting of the FM-I-FM [ $MnO_2$-$(RE,AE)_2O_2$-$MnO_2$] layers are themselves weakly coupled along the c-axis resulting in a quasi two dimensional FM order in these materials and anisotropic exchange interaction. Around room temperature the double layer compounds with x=0.3 and 0.4 are paramagnets and around T~270K a short range FM order due to in-plane spin coherence evolves and the long range FM order corresponding to, both, in plane and out of plane spin coherence evolves at a much lower temperature [389,395,405,406].

The double layer manganites consist of two building blocks, viz., the perovskite block incorporating $MnO_6$ octahedra and the rock salt separation layer containing $(RE,AE)_2O_2$. Thus the RE(AE) ions are located in the perovskite block as well as in the rock salt layer and the distribution of the various RE(AE) cations between these two building blocks is dependant on the dopant ion size. Consequently, the magnetotransport properties are expected to show a rather complicated correlation with the average RE(AE) site radius. It is well known that the average or effective cationic radius of the rare earth site plays a crucial role in determining the magnetotransport properties of the colossal magnetoresistance infinite layer manganites [407,408] and some studies in this direction has also been carried out on the double layer manganites [403,404,409,410]. Shen et. al. (1999) [403] have studied the effect of $Ca^{2+}$ (1.18 Å) substitution in $La_{1.2}Sr_{1.8}Mn_2O_7$ in which the average La site radius is 1.272 Å and have observed a gradual decrease in the paramagnetic to ferromagnetic transition temperature $T_C$ from 135 K for Ca=0.0 to 82 K for Ca=0.6. This observed decrement in $T_C$ has been attributed to the contraction of the lattice unit cell due to substitution of smaller $Ca^{2+}$ cations in place of larger $Sr^{2+}$ (1.31 Å) cations. Some what similar results have been observed by Chi et.al.(1999) for $La_{1.4}Sr_{1.6}Mn_2O_7$ (<R>=1.266 Å) [409]. They report that the $T_C$ and the insulator-metal temperature ($T_{IM}$) both decrease when $Sr^{2+}$ is partially substituted by either a bigger cation like $Ba^{2+}$ (r=1.47 Å) or a smaller cation $Ca^{2+}$ (r=1.18 Å) [409]. Chi et. al. (1999) suggest that a preferential distribution of various cations between the rock salt and the perovskite blocks is responsible for the observed trends in the $T_C$ and $T_{IM}$. However, this study leaves a fair amount of ambiguity regarding the correlation between the average La site cationic radius and the various transport parameters. Recently Zhu et. al. (2002) have studied the compound $La_{1.4}Sr_{1.6-x}Ba_xMn_2O_7$ and found that with increasing the Ba content the long range or 3D FM order diminishes and finally disappears beyond x=0.3 and on further increasing the Ba content the 2D or in plane FM phase fraction increases and induces a IM transition at the percolative threshold [406,411]. Keeping this in view we have carried out detail investigations on the effect of Ba doping on the low field magnetotransport properties of $La_{1.4}Ca_{1.6-x}Ba_xMn_2O_7$ (x=0.0, 0.2, 0.4, 0.6) in our group. We have chosen this particular composition because the average La site radius is 1.197 Å which is smaller than the previously studied Ba doped compositions [406,409-411] and therefore increase in the La site radius by partial substitution of Ba at La/Ca sites is expected to lead to improvement in magnetotransport properties.

The bulk polycrystalline samples of $La_{1.4}Ca_{1.6-x}Ba_xMn_2O_7$ (x=0.0, 0.2, 0.4, 0.6) were prepared by the standard solid state reaction route. 3N pure $La_2O_3$, $CaCO_3$, $Ba(NO_3)_2$ and $MnO_2$ were weighed and mixed in appropriate cationic ratio. $La_2O_3$ is hygroscopic in nature so before weighing it has been calcined at 1000°C for 10 hrs. These stoichimetric mixtures were ground for 30 minutes and then calcined at 950°C for 96 hours with three intermediate grindings. After calcination the powders were again ground and pressed into pellets and sintered at 1200°C for 24 hours. These pellets were again ground and palletized and then sintered at 1300°C for 24 hours. The intermediate grindings and prolonged calcinations have been observed to be necessary to maintain the compositional homogeneity of the material. The phase analysis and structural characterization was carried out by X-ray diffraction employing CuK$\alpha$ radiation in the $2\theta$ range 10-80° at room temperature. The ac susceptibility ($\chi$) was measured in the temperature range 300-20 K and the four probe electrical transport measurements were done in the temperature range 300-77 K under varying magnetic fields.

The cationic composition of the samples was evaluated by energy dispersive X-ray and results obtained from several places show that the average cationic composition is very close to the nominal one. The X-ray diffraction patterns of $La_{1.4}Ca_{1.6-x}Ba_xMn_2O_7$ (x=0.0, 0.2, 0.4, 0.6) are plotted in Fig. 71. The analysis of the observed data reveals that all the samples are polycrystalline and within the limit of the resolution of our diffractometer no second phase impurities have been observed. Long term calcinations (96 hours in the present case) have been found to be conducive in single phase growth of the material and also eliminate any unreacted or other oxide impurities. The unit cell structure of all the samples has been found to be tetragonal with $Sr_3Ti_2O_7$ type structure and as shown in Fig. 71 the three most intense reflections correspond to the (110), (125) and (200) planes in that order. The characteristics parameters for Ba doped double layered manganite is shown in Table 7.

The XRD data was further refined to obtain the accurate lattice parameters and these are plotted as a function of the average La site cationic radius and Ba content in Fig. 72. The average La site radius has been calculated by the formula It seen that as the Ba content (that is x) and hence the average La site cationic ratio is increased both the lattice parameters "a" and "c" increase progressively. At the same time it is interesting to note that as the Ba content x is increased, the ratio c/a decreases thereby indicating that the "a" parameter increases at a faster rate than the "c" parameter, that is, the lattice expands more in the a-b plane directions than in the out of plane c-direction. This increase in the lattice parameters is attributed to the substitution of smaller $Ca^{2+}$(r= 1.18 Å) cations by larger $Ba^{2+}$ (r=1.47 Å) cations. As the Ba content is increased the average La site cationic radius increases causing the unit cell to expand. But the observed expansion of the unit cell is anisotropic and this can be accounted for as follows. It has been shown by Chi et al (1999) [409] that when two or more cations are substituted simultaneously at the La site the smaller cation tends to occupy the site in the rock salt layer while the bigger cation prefers the sites in the perovskite block [409]. In the present case $Ba^{2+}$ is the largest cation and therefore it will occupy the site in the perovskite block while the smaller cations $La^{3+}/Ca^{2+}$ go preferentially to the rock salt layer. The partial substitution of smaller cations $La^{3+}/Ca^{2+}$ in the perovskite block by larger cation $Ba^{2+}$ relaxes the internal pressure in the lattice and at the same time also stabilizes the perovskite structure. In the present case when Ba content x=0.0 the average La site cationic radius is $<R> = 1.1977$ Å which is much smaller than the radius that would yield an ideal or less distorted perovskite structure. As the Ba content is increased more and more $Ba^{2+}$ cations are incorporated into the perovskite block and average La site cationic radius $<R>$ ($<R> =1.2557$ Å for Ba = 0.6) increases towards the ideal value. This leads to reduction in the distortion of the $MnO_6$ octhedra in the perovskite block which consequently leads to a reduction in the electron-lattice coupling via weakening of the cooperative Jahn-Teller distortion. This weakening of the Jahn-Teller distortion can have appreciable effect on the magneto-transport properties of the layered manganites. The above mentioned preferential distribution of the three cations would lead to unequal changes in the various in plane and out of plane bond lengths and hence an anisotropic expansion in the unit cell is observed. In the present investigation it has been observed that the metal-oxygen distances (La(Ca,Ba)-O) and the Mn-O distances in the rock salt layer slightly vary with the increasing Ba concentration. The Mn-O distances in the rock salt layer are found to decrease slightly with the increasing Ba concentration while the metal ion distances (assuming that the three metal cations $La^{3+}$, $Ca^{2+}$ and $Ba^{2+}$ are distributed randomly) in the same rock salt layer show a slight increment with the increasing Ba concentration.

The magnetic characterization of the $La_{1.4}Ca_{1.6-x}Ba_xMn_2O_7$ (x=0.0, 0.2, 0.4, 0.6) polycrystals was carried out by employing ac susceptibility measurements at $H_{ac}$ = 2 mT and $f_{ac}$ =647 Hz in the temperature range 300-20 K. The temperature dependence of the real part of the ac susceptibility ($\chi$-T) for all the $La_{1.4}Ca_{1.6-x}Ba_xMn_2O_7$ is shown in Fig. 73. The Curie temperature was evaluated from the $\rho$-T data by determining the maxima in the magnitude of $d\chi/dT$ – T curves. The $T_C$ values are found to be 135, 144, 163, and 185 K respectively for Ba content (average La site cationic radius $<R>$) x=0.0 (1.1977Å), 0.2 (1.2171Å), 0.4(1.2364Å) and 0.6 (1.2557Å). As seen in Fig. 73, the x=0.0 sample has a very broad transition and as x is increased the transition becomes sharper and the transition temperature progressively shifts to higher values. The increase in the $T_C$ can be accounted for by the following quantitative explanation. As discussed in the previous section the increase in the Ba content x leads to an increase in the average La site cationic radius <R> and decrease in the c/a ratio due to unequal changes in the in plane and out of plane lattice parameters. The substitution of larger $Ba^{2+}$ cations in the perovskite block also leads to the reduction of the cooperative Jahn-Teller distortion thereby weakening the electron-lattice coupling. Its known that the $T_C$ of the manganites is determined by the relative strength of the various competing interactions such as the double exchange, electron-lattice (phonon) coupling etc and hence in the present case the weakening of the electron-lattice (phonon) coupling and reduction of the cooperative Jahn-Teller distortion may result in the enhancement of the ferromagnetic phase fraction at the cost of the paramagnetic phase and hence higher $T_C$ values. Further $Ba^{2+}$ substitution also leads to the relaxation of internal pressure in the lattice and this may also contribute to the enhancement of $T_C$ observed in the present case.

The temperature dependence of resistivity in all the samples was measured by the four probe technique in the temperature range 300-77 K and measured resistivity data is plotted in Fig 74. As seen the plot all the samples undergo a very sharp insulator-metal transition ($T_{IM}$) within a very close temperature range and the respective transition temperatures ($T_{IM}$'s) are 125, 131, 133 and 136 K for Ba contents x=0.0, 0.2, 0.4 and 0.6. The variation of the $T_{IM}$ and also $T_C$ as a function of the Ba content and <R> is shown in Fig. 75. The increase in the $T_{IM}$ with the Ba content in very small as compared to the observed $T_C$ enhancement with Ba content. The drastic difference in the $T_C$ and $T_{IM}$ is a known feature of the layered manganite and has been attributed to the intrinsic anisotropy in the magnetic exchange interaction [389,394,395,403,404]. Another factor responsible for very small increase in $T_{IM}$ may be the increased cationic disorder at the La site. Above the $T_C$ the temperature dependence of resistivity in each sample has been found to be of the of the type $\rho = \rho_0 \exp(T_0/T)^{1/4}$ thus indicating that in this temperature regime the current transport is of Mott's variable range hopping (VRH) type. This model was originally proposed by Mott [353] for magnetic semiconductors but has successfully been applied to explain the current transport in the CMR materials in the temperature regime $T > T_C$, that is, in the paramagnetic regime. In order to correlate the transport mechanism above $T_C$ with the observed variation in $T_C$, resistivity etc. we have analyzed the $\rho - T$ data in the framework of the modified VRH model proposed by Viret et. al [354]. The $\ln(\rho) - T^{-0.25}$ plot for all the samples is shown in Fig. 76.

It has been observed that the $\ln(\rho) - T^{-0.25}$ data above $T_C$ is best fitted by straight lines. The value of the parameter $T_0$ has been found to be 1.248 x $10^8$, 1.044 x $10^8$, 9.931 x $10^7$ and 7.829 x $10^7$ K respectively for Ba contents x=0.0, 0.2, 0.4 and 0.6. The value of $T_0$ is thus found to decrease with increase in the Ba content or the average La site cationic radius. Since $T_0$ is inversely related to the extent of the localized states [353,354], therefore decreasing value shows that as the Ba content is increased the localization length and hence the hopping distance also increases. The conjecture regarding the localization length and the hopping distance suggests that with increase in the Ba content (hence <R>), that is, reduced distortion of the $MnO_6$ octahedron, the size of the magnetic polaron and the hole mobility both increase. This expresses itself as an enhancement in $T_C$ of the Ba doped samples.

The MR of all the samples, that is, $La_{1.4}Ca_{1.6-x}Ba_xMn_2O_7$ (x=0.0, 0.2, 0.4, 0.6) was measure at applied dc magnetic fields in the range 0-300 mT in the temperature range 300-77 K ( Fig 77) Fig. 77 shows the variation of the low field magnetoresistance (LFMR) with temperature for all the samples at 0.6 kOe. The LFMR increases as we increase the Ba content and maximum LFMR of ~13.2 % has been observed at 77K. The low field magnetoresistance (LFMR) at 77K was measured in the field range 0-3 kOe in the steps of 30 mT. At 77 K and 0.3 kOe the LFMR values are measured to ~ 6.5, 6.8, 8.1 and 10.5 % respectively for x=0.0, 0.2, 0.4 and 0.6 and increase to ~ 10, 10.6, 11.3 and 13.2 % when the field is increased to 0.6 kOe. In the field dependence of LFMR at 77 K it is noticed that when $H_{dc}$< 1 kOe the growth of LFMR is much faster and retards down considerably beyond that. Thus at 77 K and 3 kOe LFMR values are found to be ~ 18.5, 19.1, 20.7 and 23.8 respectively for Ba content x=0.0, 0.2, 0.4, 0.6. The field evolution of LFMR with the applied dc magnetic field is shown in Fig. 78. It is evident that at smaller fields, e.g., 0.3 kOe the Ba doped samples have much larger MR than the virgin sample. It must be mentioned here that LFMR-H curves do not show any saturation like features as seen usually in the infinite layered manganites and such a behaviour may be attributed to to the intrinsic contribution to the LFMR by the stacking of the $La(Ca,Ba)_2O_2 - MnO_2 - La(Ca,Ba)_2O_2$ layers in the unit cell of the layered manganites.

### 4.2.6. (ii) Studies on Polycrystalline $La_{1.4}Ca_{1.6}Mn_2O_7$ film

The $La_{2-2x}Sr_{1+2x}Mn_2O_7$ compound was the first to be investigated and it has been observed to undergo a 3D FM transition at Curie temperature $T_C \approx 126K$ for doping level x=0.4 [389,395]. The epitaxial films of this compound have also been observed to exhibit anisotropy in transport properties and high field magnetoresistance in excess of 99 % [412,413]. The Ca doped double layer compound, viz., $La_{2-2x}Ca_{1+2x}Mn_2O_7$ has been reported to exhibit higher $T_C$ values, e.g., $La_{1.6}Ca_{1.4}Mn_2O_7$ (x=0.2) has a $Tc \approx$ 170K which rises to 215K for a doping level x=0.25 [392,394,405]. The epitaxial thin films of the Ca doped double layer manganites have also been studied and have been observed to exhibit $T_C \sim$ 130K and significant LFMR below $T_C$ [392-394]. The observed $T_C$ in these films are lower than the corresponding bulk values.

From application point of view thin polycrystalline films having significant magneto-resistance at low magnetic field are important. The technique of spray pyrolysis has emerged as quite effective technique to grow good quality polycrystalline thin films having smaller particle size and infinite layer manganite thin films have been successfully deposited and have been found to exhibit significant magnetoresistance both at low as well as high magnetic fields [249,273,414]. In this section we report the preparation, structural / microstructural and magnetotransport characterization of $La_{1.4}Ca_{1.6}Mn_2O_7$ polycrystalline films prepared by nebulized spray pyrolysis on single crystal $LaAlO_3$ substrates in our lab.

The $La_{1.4}Ca_{1.6}Mn_2O_7$ (DLCMO) films were prepared by spray pyrolysis technique employing a two-step route (section 4.2.1). Firstly, well-homogenized aqueous solution (Molarity 0.2 M) was prepared by dissolving high purity (3N) La, Ca and Mn nitrates in appropriate cationic ratio (La/Ca/Mn=1.4/1.6/2) which then sprayed on single crystal $LaAlO_3$ (LAO) substrates. A nebulizer with proper modifications was used to spray the solution and during deposition the substrate temperature was maintained at $\sim 300 \pm 5\ °C$. After the deposition, the films were cooled to room temperature and then annealed in flowing oxygen at $950°C$ for 2 hours. The thickness of the films is ~700 nm. X-ray diffraction (XRD - Philips PW1710 $CuK_\alpha$) and scanning electron microscope (SEM- Philips XL 20) were used for structural / microstructural characterization. The ac susceptibility ($\chi$) measurement and four-probe electrical transport measurement were carried out for magnetotransport characterization.

The structural characterization of the DLCMO film by XRD reveals that the films are polycrystalline and single phase having a tetragonal unit cell having lattice parameters a = 3.877 Å and c = 19.254 Å. The corresponding XRD pattern is shown in Fig. 79. The crystallite size was also calculated employing the Scherrer formula $0.9\lambda/\beta\sin\theta$ and using the full width at half maximum (FWHM) and was found to be ~ 50 nm. The surface morphology and the particle size were also evaluated by SEM. The surface morphology consists of quite densely packed nearly spherical particles of sizes distributed in the range 70-100 nm.

The magnetic transition of the DLCMO film was studied by ac susceptibility measurement. The temperature dependence of $\rho$ was measured in the temperature range 300-16 K at an ac field of $H_{ac}$ = 2 mT and signal frequency $f_{ac}$ = 647 Hz. The measured $\chi$ - $T$ is plotted in Fig. 80. As the temperature is lowered from room temperature the susceptibility starts increasing and slope of the $\chi$ - $T$ curve changes at T~125 K. This is signature of the onset of the paramagnetic to ferromagnetic transition the midpoint of which as determined from the $d\chi/dT$ is the Curie temperature $T_C$ = 107 K. From the $\chi$ – $T$ data it can be conjectured that at temperatures above 125 K the DLCMO film has a weak FM character due to the usual in plane or short range FM spin coherence observed in case of the double layered manganites [393]. The Mn spins and hence the $MnO_2$ planes couple strongly along the out of plane c-direction only below 125 K and a complete or long range FM ordering is established only below $T_C$ ~ 107 K. On further lowering the temperature a drop in $\chi$ - $T$ curve is observed at $T_{CA}$ ~ 30 K suggesting the occurrence of a spin canted state. It has been observed that in layered manganites the magnetic frustration originating due to the competition between the FM double exchange involving the itinerant $e_g$ electrons and the anti-ferromagnetic (AFM) super-exchange involving localized $t_{2g}$ electrons results in the spin canted state at low temperatures [413]. In the layered manganites the inherent anisotropy of the exchange interaction seems to further modulate this effect.

The temperature dependence of the resistivity of the DLCMO film was measured by the four-probe technique in the temperature range 300-16 K and the $\rho$-$T$ data is plotted in Fig. 81. At room temperature the resistivity is measured to be 0.278 m$\Omega$-cm. This low value of $\rho$ also shows that despite smaller particle size (~70 - 100 nm) and polycrystalline nature, the film is of reasonably good quality. A metallic transition is observed at $T_{IM}$ ~ 55 K and the resistivity at the peak of the $\rho$ - $T$ curve is 397.5 m$\Omega$-cm. Thus the resistivity is found to increase by more than three orders of magnitude as the temperature is lowered from room temperature to the $T_{IM}$, the peak resistivity temperature. At a further lower temperature the $\rho$ - $T$ curve shows an upturn that almost coincides with the $T_{CA}$ ~ 30 K, the spin canting temperature. This shows that there is close correlation between the resistivity reversal observed at $T_R$ ~ 28 K and the magnetically frustrated spin canted state. In fact, in the spin canted regime the scattering of the conduction electrons is increased due to the canting of Mn spins and consequently the resistivity shows an upturn [413,415,416].

In order to further elaborate the transport mechanism, especially in the spin canted regime we have also studied the current-voltage characteristics (IVCs) of the DLCMO film in the temperature range 300 K – 5 K measured in a liquid He cryostat. The IVCs are linear in the range 300 K – $T_C$ and the non-linearity slowly builds up below $T_C$ and at T < 80 K the IVCs become appreciably non linear. The IVCs taken at three different temperatures below $T_C$ are shown in Fig 82 and as seen the non-linearity is especially strong below the spin caning temperature $T_{CA}$. This strong non-linear character may be attributed to the occurrence of the magnetically frustrated spin canted state that leads to an increased scattering of the conduction $e_g$ electrons and this feature has also been observed in the epitaxial thin films of the layer manganite by Philipp et al. (2002) [413]. However, in case of polycrystalline films partial contributions from the grain boundaries related disorders cannot be ruled out.

In order to have an idea of the conduction mechanism in semi-conducting regime above $T_C$ where the IVCS are linear we have analyzed the temperature dependence of resistivity above the Curie temperature $T_C \sim 107$ K. In the infinite layer simple perovskite manganites the current transport in the semi-conducting regime above $T_C$ has been explained using the Mott variable range hopping of localized electrons [353]. In the present study we have found that in the temperature regime $T_C$ - 300 K the resistivity follows the Mott's variable range hopping (VRH) model expressed by the equation:

$$\rho(T) = \rho_\infty \exp\left(T_0/T\right)^{0.25} \tag{10}$$

In the conventional Mott regime the parameter $T_0$ is related to the extent of the localized states and in fact the localization length $\lambda_0 \propto T_0^{-1/3}$ [354]. The plot of $\ln(\rho)$ and $T^{-0.25}$ is shown in Fig 83. The linear fit to the observed data is also shown in the figure by solid line. The value of the parameter $T_0$ corresponding to the best fit is 3.838 x $10^7$ K and under the applied magnetic field this value is found to decrease slightly.

The MR measurements were carried out in the temperature range 300 K – 77 K and magnetic field range H = 0 – 3 kOe. The magnetic field dependence of the low field MR measured at 77 K is shown in Fig 84 where two distinct slopes can easily be distinguished in the MR – H curve. The first slope is rather steep and is observed at $H_{dc} \leq 0.6$ kOe and above this the rate of MR increment with the field slows down but still there is no sign of saturation up to $H_{dc}$ = 3 kOe as observed in case of infinite layer manganites [273,414]. The larger value of $d(MR)/dH_{dc}$ at $H_{dc} \leq 60$ mT suggests that the tunneling component of the MR is dominant in this regime. The temperature dependence of the low field MR measured at $H_{dc}$= 1.5 kOe is shown in Fig. 85. The MR is observed to decrease monotonically as the temperature is reduced and becomes less than 2 % at 185 K and finally it vanishes around 230 K. The observed field and temperature dependence of the MR shows that the grain boundary tunneling contribution or the extrinsic component dominates at these fields and temperatures.

**5. Recent Observations of Magnetoresistance in Other Materials and Relevant Mechanism:**

The discovery of Colossal Magnetoresistance (CMR) in doped perovskite manganites revolutionized material science and a lot of effort has been carried out (and still going on) to find out materials with improved magnetoresistance around room temperature at low magnetic fields. The analogies with manganites and other magnetoresistive materials are based on the presence of a large magnetoresistance, a competition between FM and AF states and universally accepted inhomogeneities. Based on above analogies, the magnetoresistive family consists of large number of materials having different crystalline structures. These materials are as following:

**(i) Mn-based Pyrochlores:** $Tl_2Mn_2O_7$ show metal-insulator transition and large intrinsic magnetoresistance at a Curie temperature of 140K with a saturation magnetic moment of $3\mu_B$ per formula unit corresponding to a ferromagnetic ordering of the $Mn^{4+}$ ions [417-420]. The pyrochlores are interesting in comparison with manganites as their large magnetoresistance does not arise from a conventional double exchange mechanism [421]. Instead the ferromagnetism is predominantly dominated by superexchange, and the conduction electron likely arise from the Tl 6s band [422]. Thus in $Tl_2Mn_2O_7$ there are two separate electronic systems, a magnetic sublattice of Mn-O and a conducting sublattice of Tl-O indirectly coupled to it. Because of the lack of $Mn^{3+}$, there is not the strong electron-lattice coupling driven by the Jahn-Teller energy gains. In the pyrochlore, a large degree of spin-polarization appears to be due to the extremely small number of carriers [418]. The value of Tc, the magnitude of resistivity peak near Tc for $Tl_2Mn_2O_7$ and the corresponding MR are generally altered by isovalent (Sc, Bi, Ru) or aliovalent substitutions [423-426]. Alonso et al. observed room temperature MR and cluster glass behaviour in $Tl_{2-x}Bi_xMn_2O_7$ [427].

**(ii) Cr-based Chalocogenides:** Ramirez et al. [428] have demonstrated moderately large high-field magnetoresistance near the Curie temperature in the Cr-based chalocogenide spinels i.e. $Fe_{1-x}Cu_xCr_2S_4$. The spinels $ACr_2Ch_4$ is a tetrahedrally coordinated cation where A=Fe, Cu, Cd and Ch is a chalcogen i.e. S, Se and Te [429]. However, unlike the perovskite manganites they do not possess heterovalency, distortion-inducing ions, manganese, oxygen or a perovskite structure. The transport in these mateials is due to electron hopping among $d^5$ states above a valence band comprised of chalcogenide p-levels [430]. Theoretical studies of the electronic structure indicate a half metallic character with a gap in the minority density of states [431].

**(iii) Ordered Double Perovskite:** Another promising compound of CMR family are the double perovskite $Sr_2FeMoO_6$ [432] and $Sr_2FeReO_6$ [433] with comparatively high Curie temperature of 420K and 400K, respectively. The structure is obtained by doubling the perovskite unit cell e.g.

$Sr_2FeMoO_6$ has alternate stacking of $SrFeO_3$ and $SrMoO_3$ to form an ordered double perovskite structure. The highest Curie temperature was reported for $Ca_2FeReO_6$ with Tc~540K [434]. These materials are half metallic in nature. Single crystals [435] do not show significant MR but a substantial low field magnetoresistance (~5% at 300K and ~20% at 5K for H~1T) [436,437] often appears in polycrystalline samples that are likely to be of extrinsic origin from grain boundary or cation-disorder scattering similar to that of the grain boundary MR observed in manganites [438]. Westerburg et al. [439] observed 5% MR under 8T field at Curie temperature in $Sr_2FeMoO_6$.

**(iv) Hexaborides:** $EuB_6$ also shows a very large MR but that are completely different from the manganites. $EuB_6$ is a ferromangeitc semimetal and consequently, the effective mass and the number of carriers are small [440,441]. It shows two magnetic transition at 12.5K and 15K [442]. A spin-flip Raman scattering shows the existence of magnetic polarons [443]. Doped $EuB_6$ e.g. $Eu_{1-x}La_xB_6$ [444] and $EuB_{6-x}C_x$ [445] also show substantial MR. Another Eu based material i.e. EuSe [446] and $Eu_{1-z}Gd_xSe$ also show MR similar to manganites.

**(v) Ruthenates:** Ruthenates are materials that are also receiving considerable attention currently [447]. In a single layer material $Ca_{1-x}La_xRuO_4$ a dramatic decrease in the resistivity is observed upon La doping and the mott insulator ($CaRuO_4$) eventually reaching a metallic state [448]. The resitivity vs temperature curve are similar in shape to those found in manganites. Both manganites and ruthenates present metallic and insulating states that can compete with each other. Ruthenates also show orbital ordering similar to manganites [449]. Recently CaO et al. [450] discovered tunneling magnetoresistance, due in part to the magnetic-valve effect, with a drop in the resitivity of the bilayer $Ca_3Ru_2O_7$ by three order by magnitude. Very recently Ohmichi et al. [451] have demonstrated the existence of two types of field induced transitions at 6 and 15T in $Ca_3Ru_2O_7$. They attributed these behviour due to strong coupling between spin, charge and lattice degree of freedoms in $Ca_3Ru_2O_7$, which can be controlled by in plane field orientation. Another ruthenate, $SrRuO_3$ is a strongly correlated ferromagnetic d-band metal having orthorhombic structure. The Curie temperature in the bulk is 165K; for thin films reduced Curie temperatures of 150K were observed possibly due to strain effect [452,453]. Near the Curie temperature a maximum in the MR was observed for $SrRuO_3$ that does not saturate in magnetic fields upto 8T. The value of peak MR depends on the current and field direction with values between –2 and –11%. Klein et al. (1998) [454] interpreted this MR peak as arising from an increase of the magnetization and corresponding reduction of spin-disorder scattering.

**(vi) Magnetite:** $Fe_3O_4$ is a ferrimagnetic oxide crystallizing in the inverse spinel structure and has the highest Curie temperature of 858K among all magnetoresistive materials [455]. At room temperature, in this structure large O ions are located on a closed-packed face-centred cubic lattice whereas the Fe ions occupy interstitial sites. There are two kinds of cation sites, namely the tetrahedrally coordinated A-site occupied only by $Fe^{3+}$ ions and the octahedrally coordinated B-site occupied by both $Fe^{2+}$ and $Fe^{3+}$ ions. The A-and B-site sublattice are ferrimagneticially aligned such that the net moment is equal to the magnetic moment $\mu=4\mu_B$ of the $Fe^{2+}$ ($3d^6$) ion [456]. Recently Ziese et al. [457] observed few percent MR in $Fe_3O_4$ single crystals at the Verwey transitions and explained it on the basis of shift of charge ordering/verwey transition on application of magnetic field. The temperature dependence of resistivity is quite complex, changing from semiconducting to metallic behaviour slightly above room temperature and back to semiconducting behaviour near the Curie temperature [458]. Band structure calculations indicate a half metallic nature with a gap in the major density of states [459]. Recently Liu et al [460] observed ~7.4% MR at room temperature in polycrystalline $Fe_3O_4$ films which they ascribed due to spin dependent tunneling through the antiferromagnetically coupled grain boundaries.

**(vii) Chromium Oxide:** $CrO_2$ is the only stochiometric binary oxide that is ferromagnetic metal having rutile structure. It is the simplest and best studied half metal [461]. Band structure calculation of $CrO_2$ predicted 100% spin polarization at fermi level [462] and spin-polarized photoemission and vacuum tunneling experiments showed nearby complete spin polarization 2ev below $E_F$ [463,464]. The resistivity of $CrO_2$ varies widely, from semiconducting to metallic, and ranges over 5 orders of magnitude at low temperature [465]. Change in the slope of the resitivity is discernible near the Curie temperature of 390K. The MR effect in $CrO_2$ are associated with transport of spin-polarized electrons from one FM region to another with a different direction of magnetization. These regions are not usually separated by a domain wall but by a grain boundary. The MR effect at low fields and low temperature can reach 50% in pressed $CrO_2$ powder [466,467] and several 100% in planar mangantie tunnel junctions [468] whereas small MR have been observed in $CrO_2$ tunnel junctions [469,470].

**(viii) Diluted Magnetic Semiconductors:** At present Diluted Magnetic Semiconductors (DMS) are the hottest candidate for spintronic devices [471]. Most of the past work on DMS has focused on (Ga, Mn)As and (In, Mn)As [472,473]. But the problem with these DMS is that they have low Curie temperature. Dietl et al. [474] theoretically predicted that GaN and ZnO would exhibit ferromagnetism above room temperature on doping with Mn, provided that the hole density is sufficient high. Presently number of excellent reviews are available which covers, experimental as well as theoretical aspects of all types of Diluted Magnetic Semiconductors including oxide based DMS e.g. ZnO, TiO$_2$ and SnO$_2$ doped with Mn, Ni, Co, Fe, V etc [475-478]. Since then a great deal of efforts have been focused on semiconductors e.g. ZnO, TiO$_2$ and SnO$_2$ doped with ferromagnetic elements (Mn, Fe, Co, V, Ni etc.) [479-492]. Among them ZnO based DMS would be very promising because of its wide spread application in electronic devices, such as transparent conductors, gas sensors, varistors, ultraviolet laser sources and detectors [483,484]

Kim et al. [485] measured the isothermal MR of $Zn_{1-x}Co_xO$: Al thin films for x=0.02, 0.06, 0.10 and 0.15 at various temperature. They observed three different types of MR behaviour below 20K depending on the Co content but above 50K all samples exhibit a very small negative MR as observed in a ZnO:Al film without magnetic impurity. A large positive MR of~60% at 5K under 2T was observed in $Zr_{0.94}Fe_{0.05}Cu_{0.01}O$ below 100K [486]. Despite non-magnetic elements, ZnO:S films also show large MR of ~26% at 3K [487]. Similarly Co doped TiO$_2$ films also show MR under a 8T magnetic field at 3K which increases from ~6% in an undoped TiO$_2$ film to ~40% in 2% Co doped TiO$_2$ film ($Ti_{0.98}Co_{0.02}O_{2-\delta}$) [488].

Prellier et al. [489] have recently reviewed the oxide based DMS. They conclude that most of the Co doped ZnO films exhibit FM above room temperature and that in the case of Mn doped ZnO thin films, a definitive $T_c$ is not found by some workers, but Sharma et al. [490] recently reported room temperature ferromagnetism in Mn doped ZnO bulk as well as thin films ($Zn_{1-x}Mn_xO$; x=0, 0.01, 0.0 and 0.1). Thus the ferromagnetic behaviour of Mn-and Co-doped ZnO have considerable doubt. Recently Rao et al. [491] extensivley studied the ferromagnetism of Mn- and Co-doped ZnO and pointed out that presence of additional charge carriers are responsible for ferromagnetism rather than the mere doping of Mn- and Co in ZnO, which is in agreement with the recent work of Spaldin [492].

**(ix) Doped Silver Chalcogenides:** Recently, an anomalously large MR was observed in two doped silver chalocogenides, $Ag_{2+\delta}Se$ and $Ag_{2+\delta}Te$, where the resistance displayed a positive linear dependence on the magnetic field over a temperature range 4.5 to 300K without any sign of saturation at fields as high as 60T. At room temperature and in a magnetic field of ~55KOe, $Ag_2Se$ and $Ag_2Te$ show ~200% increase in resistance, which are close with CMR materials [493,494]

**(x) Magnetic Tunnel Junctions:** Magnetic tunnel junctions (MTJs) shows tunneling MR that is drawing considerable attention due to the advent of sophisticated thin film junction preparation techniques [495]. MTJ consists of two ferromagnetic metallic layers separated by a thin insulating barrier layer. The insulating layer is so thin (a few nanometer or less) that electron can tunnel through the barrier if a bias voltage is applied between the two metal electrodes through the insulator. The most important property of a MTJ is that the tunneling current depends on the relative orientation of the magnetization of the two FM layers, which can be changed by an applied magnetic field. This phenomenon is called TMR or JMR (Junction Magnetoresistance). TMR is a consequence of spin-dependent tunneling (SDT) due to an imbalance in the electric current carried by up and down spin electrons through a tunneling barrier. SDT was discovered in pioneering experiments by Tedrow and Meservey (1994) [496]. The relationship between SDT and TMR was explained by Julliere (1975) [497] within a simple model that quantifies the magnitude of TMR in terms of the spin polarization (SP) of the ferromagnetic electrodes as measured in the experiments on superconductors [496]. As per Julliere's model the TMR is given as,

$$\text{TMR} = \frac{2\rho_1\rho_2}{1 - \rho_1\rho_2} \tag{11}$$

where $\rho_i = \frac{\rho_i^\uparrow - \rho_i^\downarrow}{\rho_i^\uparrow + \rho_i^\downarrow}$ (i = 1,2) is the effective spin polarization of two electrodes. For the case of two identical ferromagnets, the TMR is always −ve; it diverges for two half-metallic electrodes. It can be both +ve and −ve.

A few years ago, however, Miyazaki and Tezuka (1995) [498] demonstrated the possibility of large values of TMR in MTJ's with $Al_2O_3$ insulating layers. Moodera et al. (1998) [499] reported TMR values for a $Co/Al_2O_3/Ni_{80}Fe_{20}$ junction of 20.2%, 27.1% and 27.3% at 295, 77 and 4.2K, respectively. Various workers have reported TMR values in excess of 100% at 4.2K for ferromagnetic oxide tunneling junctions based on manganite electrodes [500-511]. Recently, extremely large TMR values upto 1800% were obtained by Bowen et al. (2003) [512] in LSMO/STO ($t_{STO}$ = 2.8nm)/LSMO/Co/Au structure. The extremely large TMR response (1800%) at 4K leads to a spin polarization of the LSMO at the interface with STO of at least 95%. Moreover, the temperature dependence of the TMR in this optimally etched junction vanishes only at T=280K ($T/T_c$=0.7). Strains as well as the mixed valency of Mn ions at the interface are the deciding factors to have large TMR response, as deduced from EELS measurements coupled with HREM observations by Pailloux et al.(2002) [513]. Very recetnly Yamada et al. (2004) [514] have atomically engineered interface of LSMO with STO and LMO [LSMO(0.4)/LMO(2uc)]/STO(2nm)/ [LMO(2uc)/LSMO(0.4)] and observed an improved performance as compared to the direct-interface junction (LSMO/LMO/LSMO; LSMO/STO/LSMO).

**(xi) Nanocontacts:** Another type of material which is making news in last couple of years are nanocontacts which show Ballastic Magnetoresistance (BMR). When the size of the contact between two FM electrodes is of the order of nano or atomic-scale then the spin polarized electron on its way through the contact, will have less of a chance to accommodate itself to the spin regime of the secondary electrode (if it is different from that of the first electrode) and will consequently scatter more prominently, translating into a Ballistic MR effect [515,516]. Recently, H. D. Chopra and S. Z. Hua (2002) of Sunny Buffalo reported remarkably large room temperature BMR of 3150 % in fields of 500 Oe at room temperature in electrodeposited Ni nanocontacts [517]. In subsequent study they observed 10,000% BMR in fields of 3000 Oe at room temperature for stable Ni nanocontacts which were made using only mechanically pulled Ni wires to eliminate the possibility of any extraneous chemical layer being present [518]. The BMR is formulated by Tartara et al. (1999) [519] and is given by,

$$\text{BMR} = \left(\frac{2\rho^2}{1-\rho^2}\right) F \tag{12}$$

where $\rho = \left(\frac{D_\uparrow - D_\downarrow}{D_\uparrow + D_\downarrow}\right)$ describes the spin polarization, $D_\uparrow$ and $D_\downarrow$ are the densities of states for up and down spin at the Fermi level and F is a function that describe the non-conservation of conduction electron spin. If the domain wall width is comparable to or smaller than the electron wavelength, the spin should be conserved (F = 1) in the conduction process and the BMR is given only by degree of polarization [520,521].

The BMR is a result of the spin-dependent scattering (SDS) of electrons across the nano or point contact from a ferromagnetically aligned state (low resistance) to an antiferromagnetically aligned state (high resistance) in an applied field. In bulk ferromagnets, Cabrera and Falicov (1974) [520] and later Tatara and Fukuyana (1997) [521] have shown that the SDS by domain walls is negligible, owing to adiabatic nature of electron transport across a wall, which is typically of order of several tens of nanometers wide. The SDS, nature of domain walls [522] and geometry of nanocontacts [523] play a key role in the observed huge BMR effect but the cause of such huge resistance change is not fully understood. Recently a plausible mechanism for the observed large BMR was given by Tagirov et al. (2002) [524] on the basis of spin splitting of quantized conduction states.

## 6. Envisaged Applications of Manganites

Perovskite manganites have a large potential for applications based on their various physical and chemical properties [232,525,526]. The magnetic field sensitivity of the transport properties, the strong metal insulator transition at the Curie temperature, the electric field polarizability of the material and its subsequent effect on the transport properties, the half metallicity of the electronic bands, etc., are properties of the rare earth manganites that could be exploited in a variety of devices. Based on the properties, a number of device approaches are being explored and few of them are described below.

(1) The magnetoresistance of manganites might be used in magnetic sensors, magnetoresistive read heads, and magnetoresistive random access memory. Magnetic sensors can be made from either thin films or single crystals and can be used to sense the magnitude of a magnetic field in one or several directions by choosing the right crystal form and de-magnetizing factor. A good low field magnetoresistive response however, can be obtained in manganite samples with a high density of grain boundaries and in tunnel spin valve structures. One of the first working devices of this kind was constructed by Sun et al. (1996) [500]. It consists of two layers of ferromagnetic $La_{0.67}Ca_{0.33}MnO_3$, separated by a $SrTiO_3$ spacer layer, and shows a resistance decrease by a factor of 2 in a field of less than 20 mT. The main disadvantage of devices based on grain-boundary magnetoresistance or on ferromagnetic tunneling junctions is that large magnetic field sensitivities are only achieved at temperatures below 200 K.

(2) The electric field effect has also been observed in mangnaites. Here the top layer can be paramagnetic, such as STO [527], or a ferroelectric layer, such as PZT ($PbZr_{0.2}Ti_{0.8}O_3$), and the bottom layer is a CMR material, but the changes are more profound in the case of PZT where only 3% change in the channel resistance is measured over a period of 45 min at room temperature which makes this material attractive for nonvolatile ferroelectric field-effect device applications [528,529].

(3) The large temperature coefficient of resistance (TCR, calculated as *(1/R)(dR/dT)*) just below the resistivity peak makes these CMR materials interesting for use in bolometric detectors [530-532]. Bolometer is an instrument for detecting and measuring radiation. Indeed, the TCR can reach 15% per degree at 300K [533], which is about one order of magnitude larger than that of $VO_2$, the material commonly used in bolometers.

(4) Since the properties of the CMR materials are quite spectacular at reduced temperatures, i.e. below 100 K, so at these low temperatures, the combination of high-$T_C$ superconducting cuprates thin films and CMR manganites could lead to Hybrid HTSC–CMR structures [534,535] These HTSC–CMR structures not only lead to potentially new spin-injection devices but also may serve as a useful medium for understanding some of the forefront theoretical ideas.

(5) Chemical applications include catalysis such as catalysts for automobile exhausts, oxygen sensors and solid electrolytes in fuel cells. The catalytic activity is associated with the $Mn^{3+}$-$Mn^{4+}$ mixed valence and the possibility of forming oxygen vacancies in the solid [536,537].

## 7. Summary and Future Prospects:

In recent past, intense research have been carried out on various aspects of magnetoresistance. Although the studies are focused on both, various details covering the underlying physics as well as on applied aspects, it is the former, which has received most of the attention. This review embodies various aspects of colossal magnetoresistance (CMR) in mangnaites from its genesis through recent studies to the several efforts made for making the phenomenon intelligible. Many salient features such as Double Exchange, Jahn Teller effect, Charge/Orbital/Spin Ordering, Phase Separation etc have recently emerged and these have been described in this review.

In this review we have paid particular emphasis on low field induced colossal magnetoresistnce. This is an important area from the point of view of applied aspects. We have extensively described and discussed the low field magnetotransport behaviour of various doped perovskite manganite systems such as polycrystalline $La_{0.67}Ca_{0.33}MnO_3$ films, Ag admixed $La_{0.67}Ca_{0.33}MnO_3$ films, polycrystalline ($La_{0.7}Ca_{0.2}Ba_{0.1}MnO_3$) and epitaxial ($La_{0.67}Ca_{0.33}MnO_3$) films on different substrates, nanophasic $La_{0.7}Ca_{0.3}MnO_3$, manganites-polymer composites ($La_{0.7}Ba_{0.2}Sr_{0.1}MnO_3$–PMMA and $La_{0.67}Ca_{0.33}MnO_3$–PMMA) and double layered polycrystalline $La_{1.4}Ca_{1.6-x}Ba_xMn_2O_7$ (bulk) and $La_{1.4}Ca_{1.6}Mn_2O_7$ (films). An important motivation for all CMR studies is the potential for applications.

Firstly we have described and discussed magneto-transport properties of polycrystalline $La_{0.67}Ca_{0.33}MnO_3$ (LCMO) films deposited by spray pyrolysis on YSZ substrate. The $T_{IM}$ is ~195K and the maximum MR of ~ 18% in ~ 3 kOe has been obtained at 77K. This enhanced low field MR (LFMR) has been explained on the basis of domain rotation and weakening of inter-domain interaction in weak link (grain boundary) region. In another study we have admixed optimum concentration of Ag v.i.z. ~ 25 wt % in LCMO and found that for Ag free LCMO film $T_{IM}$ shifts towards lower value. The conduction noise in these Ag free LCMO film increases whereas in Ag–LCMO the value of $T_{IM}$, $Tc$ and conduction noise did not change even after several measurements. Silver segregation at the grain boundaries in Ag–LCMO polycrystalline film seems to be responsible for improving the characteristics of Ag–LCMO films. For applications mangnaite films are relevant therefore we have investigated the effect of substrate-induced strain on the CMR characteristics of polycrystalline $La_{0.7}Ca_{0.2}Ba_{0.1}MnO_3$ films deposited on LAO, STO, ALO and YSZ substrates. The lattice strain for these thick films (~500nm) gets relaxed and leads to creation of disorders dominantly by increase in the grain boundary density. The $T_{IM}$ for films on LAO and STO is higher as compared to those on ALO and YSZ. For the films grown on LAO and STO which have lower amount of disorder (due to low mismatch) have low MR (~ 2-3%) and largerstrain leads to high value of MR (~ 7-9 %) as in the case of ALO and YSZ at 77k and in 1.5 kOe applied magnetic field. We have also studied substrate effect on magnetotransport properties of epitaxial films of $La_{0.7}Ca_{0.3}MnO_3$ on LAO, STO and MgO substrates. As the films are thicker (~ 200nm), the lattice strain gets relaxed and affects the magneto-transport properties by generating various defects e.g. dislocations, stacking faults etc. The film on MgO has maximum lattice relaxation (6.411) and hence shows reduced $T_{IM}$ and $T_C$ as compared to minimum lattice relaxation for film on STO (0.384). By applying Mott`s VRH transport model to these films, we have found that as the lattice relaxation increases the localization length decreases which in turn decrease the hopping distance, leading to reduction in $T_{IM}$ and $T_C$.

In another study we have investigated the effect of sintering temperature on microstructure and low field magneto-transport properties of polycrystalline nanophasic $La_{0.7}Ca_{0.3}MnO_3$ (LCMO) bulk, which have been synthesized by low temperature polymeric precursor route. Microstructure reveals that particle size also decreases from ~200 nm (~ 1000°C) to ~50 nm (~ 700°C). Both the transition temperatures $T_C$ and $T_{IM}$ shift towards lower temperatures as the particle size decreases (due to lowering of sintering temperature). There is only a slight decrease in $T_C$ (from 274K to 256K) whereas $T_{IM}$ decreases considerably from 267K to 138K as the sintering temperature is lowerd from 1000 to 700 °C. Also, we have observed a huge low field MR alongwith intrinsic MR. It has been found that low field MR increases as the sintering temperature (particle size) decreases but at the same time peak (intrinsic) MR decreases. This enhanced LFMR for small size particles is due to increased spin polarized tunneling behavior at lower temperatures. In another study we have described and discussed the effect of PMMA on low field magnetotransport in LBSMO–PMMA composite. The metal-like transition observed at ~150K in the pure LBSMO sample vanishes in the composites. The resistivity shows huge increments as a consequence of PMMA addition. Despite the increase in the resistivity by almost three orders of magnitude from virgin LBSMO to 10wt% PMMA admixed LBSMO composite, the MR shows fair amount of enhancement in the low field magneto-resistance at low temperatures (T < 125 K). These results show that spin polarized tunneling is slightly enhanced as a result of PMMA admixture. In continuation to this we have also investigated the (LCMO)1-x(PMMA)x, (x = 0.0, 10, 20, 35 and 50 wt %) composites synthesized by a unique polymeric sol-gel route that facilitates the LCMO grain growth in the presence of the polymer. All the samples are single phasic and nano-structured, and a small decrease in the average grain diameter as a consequence of composite formation is evidenced by XRD and SEM results. The PM-FM transition is reduced moderately. The resistivity increases nearly by two orders of magnitude and the insulator-metal transition temperature $T_{IM}$ is severely suppressed when the PMMA concentration is increased up to 50 wt %. The intrinsic component of MR is reduced while the LFMR is observed to increase gradually with the polymer concentration and at 77 K it shows a relative enhancement of nearly 35 % at a magnetic field of~3.6 kOe. The observed variation in the magneto-transport parameters have been explained in terms of the magnetic disorder created due to PMMA admixture to LCMO.

Another study deals with the synthesis and magneto-transport characterization of new manganite type CMR material, $La_{0.7}Hg_{0.3}MnO_3$. The synthesis has been carried out through a solid-state reaction route consisting of the formation of $La_{0.7}MnO_3$ followed by diffusion of Hg leading to $La_{0.7}Hg_{0.3}MnO_3$. The sample shows characteristic $T_{IM}$ at 227 K and $T_C$ at 264.59K. The room temperature MR is too small but it steadily increases as the temperature is decreased. e.g. MRs at 227K and 77 K are 3.41 and 9.05%, respectively, in an applied field of 1.5 kOe.

We have also investigated magneto-transport properties of the double layer manganite $La_{1.4}Ca_{1.6-x}Ba_xMn_2O_7$ (x=0.0, 0.2, 0.4 and 0.6). It has been found that the lattice parameter, $T_{IM}$ and $T_C$ increases with increasing Ba content. The increase in $T_C$ is due to the suppression of the JT distortion due to substitution of larger Ba cations. The difference between $T_C$ and $T_{IM}$ is due to the increased anisotropy of the FM exchange interaction as a consequence of Ba doping. There is a significant increase (~30 %) in the low temperature LFMR for the Ba doped samples. At 77 K and $H_{dc}$ = 0.6 kOe the MR values are 10 and 13.2 % respectively for x=0.0 and 0.6. In addition, we have studied the LFMR of polycrystalline films of $La_{1.4}Ca_{1.6}Mn_2O_7$. The films are single phasic having $T_C$ of ~ 107 K and $T_{IM}$ ~ 55 K. The transport mechanism above $T_C$ is of Mott's VRH type. These DLCMO films exhibit reasonable LFMR and at 77 K, the MR is ~5 % at 0.6 kOe and ~13 % at 3 kOe.

For applications apparently low (in the range Oe) and not high (covering Tesla) magnetic field produced CMR are relevant. The initial flurry of activity was brought about by reports of large MR at high field and low temperature. Recently, through a sequence of ingenious experiments on manganite perovskites, it is now possible to achieve 10 % MR at 2 kOe at 247 K, and there is every reason to expect this figure of merit to increase. They are to challenge GMR in the near future. Finally, there have been efforts to find compounds with features similar to the perovskites, and these searches have led to the synthesis and development of materials; the pyrochlores, chalcogenides, ruthenates, diluted magnetic semiconductors and various types of structures (e.g. tunneling junctions and nanocontacts). Besides large MR other aspects for example, issues involving power consumption, noise and compatibility with established fabrication methods also form pre-requisites for device application. These devices may lead to an exciting new generation of applications e.g. memory, read-on devices and sensors. However, intensive efforts are still required, by material science researchers as well as by device developers, to achieve significant low field room temperature magnetoresistance.

## Acknowledgements

One of the author (PKS) acknowledges CSIR, New Delhi for the award of SRF. Authors wish to thank Professors A R Verma, C N R Rao, Vikram Kumar, S B Ogale, J Kumar, R.S. Tiwari, Dr Kishan Lal and Dr. N. Khare for valuable discussions and suggestions. We will like to put our sincere thanks to Prof. T V Ramakrishnan and Prof. A K Raychaudhari for valuable discussions and a critical reading of the manuscript. This work was supported financially by UGC and CSIR, New Delhi

| Sample | Lattice parameters (Å) |
| --- | --- |
| Bulk LCBMO | a=3.892 |
| F1 (LAO/LCBMO) | a=5.479, b=5.470, c=7.804 |
| F2 (STO/LCBMO) | a=5.503, b=5.444, c=7.789 |
| F3 (ALO/LCBMO) | a=5.480, b=5.433, c=7.786 |
| F4 (YSZ/LCBMO) | a=5.505, b=5.456, c=7.789 |

**Table 1:** Lattice parameters of various films calculated from the XRD data. F1 to F4 are the LCBMO films deposited on different substrates.

| Sample | $T_{IM}$ (K) | $T_C$ (K) | % MR at 77K in 150 mT |
| --- | --- | --- | --- |
| Bulk LCBMO | 283 | 287 | 9.17 |
| F1 (LAO/LCBMO) | 259 | 283 | 3.06 |
| F2 (STO/LCBMO) | 264 | 285 | 2.12 |
| F3 (ALO/LCBMO) | 160 | 245 | 9.36 |
| F4 (YSZ/LCBMO) | 167 | 243 | 7.01 |

**Table 2:** The insulator-metal ($T_{IM}$), paramagnetic-ferromagnetic (Tc) transition temperatures and MR at 77K (in 150 mT applied magnetic field) for bulk LCBMO and F1, F2, F3 and F4 films.

| Characteristics | STO (3.905 Å) | LAO (3.821 Å) | MgO (4.216 Å) |
|---|---|---|---|
| Out of plane lattice constant | 3.878 Å | 3.889 Å | 3.899 Å |
| Lattice Relaxation L = $(a_F - a_S) \times 100/a_S$ | -0.384 | 3.057 | -6.411 |
| $T_C$ | 245 K | 220 K | 186 K |
| $T_{IM}$ | 243 K | 217 K | 191 K |
| ρ at 300K and at $T_{IM}$ | 36 mΩ-cm, 72 mΩ-cm | 31 mΩ-cm, 113 mΩ-cm | 38 mΩ-cm, 275 mΩ-cm |
| $\rho(T_{IM})/\rho(300\ K)$ | 2 | 3.65 | 7.25 |
| M R % (0.5T) (At T~$T_C$, $T_{IM}$ and 80K) | 2.5, 3.5, 1.4 | 7.1, 11.2, 0.3 | 13.9, 21, 3.5 |
| M R % (1.0T) (At T~$T_C$, $T_{IM}$ and 80K) | 9.1, 12.5, 2.1 | 16.5, 5.9, 4.4 | 25.7, 38, 4.7 |
| $MR_{max.}$ (0.5 T, 1.0 T) | 6.9, 17.6 | 22.5, 41 | 26, 44 |
| T ($MR_{max.}$) (0.5 T, 1.0 T) | 220, 225 K | 195, 200 K | 175, 180 K |
| $T_0$ | $1.242 \times 10^7$ K | $2.251 \times 10^7$ K | $3.830 \times 10^7$ K |

**Table 3:** The various characteristics parameters of epitaxial films of $La_{0.7}Ca_{0.3}MnO_3$ on $SrTiO_3$ (001), $LaAlO_3$ (001) and MgO (001) substrates.

| Sintering Temp. (°C) | Lattice parameters | | | Cell volume (Å$^3$) | Crystallite size(nm) XRD | Particle size(nm) SEM |
|---|---|---|---|---|---|---|
| | a(Å) | b(Å) | c(Å) | | | |
| 600 | 5.468 | 5.530 | 7.839 | 237.082 | ~29 | ~35 |
| 700 | 5.465 | 5.512 | 7.854 | 236.934 | ~32 | ~50 |
| 800 | 5.460 | 5.494 | 7.895 | 236.876 | ~35 | ~75 |
| 900 | 5.474 | 5.501 | 7.784 | 234.404 | ~40 | ~125 |
| 1000 | 5.473 | 5.494 | 7.762 | 233.420 | ~62 | ~200 |

**Table 4:** Lattice parameters, cell volumes, crystallite sizes (XRD) and particle sizes (SEM) of the samples sintered at different temperatures.

| Sintering Temp.(°C) Ts | $T_C$(K) | $T_{IM}$(K) | LFMR (%) at 1.0 T | |
|---|---|---|---|---|
| | | | 80K | 150K |
| 600 | 256 | 138 | -- | -- |
| 700 | 268 | 179 | 21.19 | 14.75 |
| 800 | 270 | 210 | 20.32 | 13.99 |
| 900 | 272 | 240 | 19.54 | 12.90 |
| 1000 | 274 | 267 | 18.94 | 11.54 |

**Table 5:** Paramagnetic-ferromagnetic transition temperature ($T_C$), insulator-metal transition temperature ($T_{IM}$) and low field magnetoresistance (LFMR) at the temperatures 80K and 150K for the field of 10 kG for all the samples.

| | a (Å) | b (Å) | c (Å) | $T_c$(K) | $T_{IM}$(K) | MR (%) at 150 mT |
|---|---|---|---|---|---|---|
| LHMO | 5.5183 | 5.6383 | 7.5368 | 264.59 | 227.13 | 0.16 (295K), 3.41 ($T_{IM}$), 9.05 (77K) |
| LMO | 5.5008 | 5.5469 | 7.8044 | 245.61 | 246.76 | 0.09 (295K), 2.89 ($T_{IM}$), 10.66 (77K) |

**Table 6:** The lattice parameters (a,b and c) the ferromagnetic temperature ($T_c$), the insulator-metal transition temperature ($T_{IM}$) and magnetoresistance for $La_{0.7}Hg_{0.3}MnO_3$ (LHMO and $La_{0.7}MnO_3$ (LMO) bulk samples.

| Ba (x) | $<r_A>$ (Å) | a (Å) | c (Å) | c/a | $T_{IM}$ (K) | $T_c$ (K) | MR% at 77 K | |
|---|---|---|---|---|---|---|---|---|
| | | | | | | | 30 mT | 300 mT |
| 0 | 1.1977 | 3.852 | 19.276 | 5.004 | 125 | 135 | 6.5 | 18.5 |
| 0.2 | 1.2171 | 3.868 | 19.306 | 4.991 | 131 | 144 | 6.8 | 19.1 |
| 0.4 | 1.2364 | 3.882 | 19.315 | 4.976 | 133 | 163 | 8.1 | 20.7 |
| 0.6 | 1.2557 | 3.889 | 19.322 | 4.968 | 136 | 185 | 10.5 | 23.8 |

**Table 7:** The lattice parameters (a and c), the insulator-metal transition temperature ($T_{IM}$), the ferromagnetic transition temperature ($T_c$) and magnetoresistance for bulk $La_{1.4}Ca_{1.6-x}Ba_xMn_2O_7$ (x=0.0, 0.2, 0.4, 0.6) s